\documentclass[12pt,english]{article}
\usepackage[T1]{fontenc}
\usepackage[latin1]{inputenc}
\usepackage{subfigure}
\usepackage{graphicx}
\usepackage{rotating}
\usepackage{hyperref}
\usepackage{amssymb}

\providecommand{\tabularnewline}{\\}

\makeatletter

\usepackage{babel}
\makeatother
\usepackage{geometry}
\usepackage{babel}
\usepackage{graphics}
\def\beq{\begin{equation}}
\def\eeq{\end{equation}}

\def\beqn{\begin{eqnarray}}
\def\eeqn{\end{eqnarray}}
\def\ba{\begin{eqnarray}}
\def\ea{\end{eqnarray}}

\def\atp{\frac{\alpha_s(Q^2)}{2\pi}}

\def\slash#1{#1\hskip-6pt/\hskip6pt}

\newcommand{\beqa}{\begin{eqnarray}}
\newcommand{\eeqa}{\end{eqnarray}}

\makeatletter

\providecommand{\LyX}{L\kern-.1667em\lower.25em\hbox{Y}\kern-.125emX\@}

\makeatother
\begin{document}
\rightline{LTH-679}
\begin{center}
\vspace{1.cm}
{\bf \large NNLO Logarithmic Expansions and Exact Solutions
of the DGLAP Equations from $x$-Space: New Algorithms for Precision Studies at the LHC}

\vspace{2cm}
 {\large \bf $^{1}$Alessandro Cafarella, $^{2,3}$Claudio Corian\`{o} and $^{2}$Marco Guzzi\\}
\vspace{1.cm}
{\it $^1$Department of Physics, University of Crete, 71003 Heraklion, Greece\\}
\vspace{0.5cm}
{\it $^2$Dipartimento di Fisica, Universit\`{a} di Lecce and INFN sezione
di Lecce\\
Via per Arnesano, 73100 Lecce, Italy\\}
\vspace{0.5cm}
{\it $^3$Department of Mathematical Sciences, University of Liverpool\\
Liverpool L69 3BX, UK\\}
\vspace{0.5cm}
{\it cafarell@physics.uoc.gr, claudio.coriano@le.infn.it, marco.guzzi@le.infn.it\\}
\end{center}
\vspace{0.5 cm}
\begin{abstract}

A NNLO analysis of certain logarithmic expansions, developed for precision studies of the evolution of the QCD parton
distributions (pdf) at the Large Hadron Collider, is presented.
We elaborate on their relations to all the solutions of the DGLAP equations that have been hitherto obtained from Mellin space, to which are equivalent.
Exact expansions, equivalent to exact solutions of the equations, are constructed in the non-singlet sector. The algorithmic features of our approach are also emphasized, since this method allows
to obtain numerical solutions of the evolution equations with the same accuracy of other methods, based on Mellin space, and of brute force methods,
which solve the equations by finite differences. The implementation of our analysis allows to compare with existing
benchmarks for the evolution of the pdf's, useful for applications at the LHC, and to extend them significantly in a systematic fashion, especially when solutions that retain logarithmic corrections
only of a certain accuracy are searched for.


\end{abstract}
\newpage
\section{Introduction}

Precision studies of some hadronic processes in the perturbative regime are going to be very important
in order to confirm the validity of
the mechanism of mass generation in the Standard Model
at the new collider, the LHC. This program involves a rather complex analysis of the
QCD background, with the corresponding radiative corrections taken into account to higher orders.
Studies of these corrections for specific processes
have been performed by various groups,
at a level of accuracy which has reached the next-to-next-to-leading order (NNLO) 
in $\alpha_s$, the QCD coupling constant. The quantification of the impact of these corrections
requires the determination of the hard scattering of the partonic cross sections up to
order $\alpha_s^3$, with the matrix of the
anomalous dimensions of the DGLAP kernels determined at the same perturbative order.

The study of the evolution of the parton distributions is  critical for the success of
this program and may include both a NNLO analysis and, possibly,
a resummation of the large logs which may appear in certain kinematic regions of specific
processes \cite{Sterman}. The questions that we address in our work concern the types of approximations which 
are involved when we try to solve the DGLAP equations to higher orders and the differences among the various methods proposed for their solution.

The clarification of these issues is important, since a chosen method has a direct impact on the structure of the 
evolution codes and on their phenomenological predictions. 
We address these questions by going over a discussions 
of these methods and, in particular, we compare those based in Mellin space and 
the analogous ones based in $x$-space. Mellin methods have been the most popular and have been implemented
up to NLO and, very recently, also at NNLO \cite{Vogt3}.
We remark that $x$-space methods based on logarithmic expansions have never been thoroughly justified in the previous literature 
even at NLO, in the case of the QCD parton distributions (pdf's) \cite{cafacor,nostri,gordon}. 

We fill this gap and present 
exact proofs of the equivalence of these methods - in the case of the evolution of the QCD pdf's - 
extending a proof which had been outlined by Rossi \cite{Rossi} at LO and by 
Da Luz Vieira and Storrow \cite{Storrow} at NLO in their 
study of the parton distributions of the photon. 

In more recent times, these studies on the pdf's of the photon have triggered 
similar studies also for the QCD pdf's. The result of these efforts was the proposal of new 
expansions for the quark and gluon parton distributions \cite{cafacor,nostri,gordon} which had to capture the logarithmic 
behaviour of the solution up to NLO. Evidence of the consistency of the ansatz was, 
in part, based on a comparative study of the generic structures of the logarithms that appear in the solution
using Mellin moments, since the same logarithms of the coupling could be reobtained by recursion relations.

In this work we are going to clarify - using the exact solutions of the corresponding recursion relations - 
the role of these previous expansions and present their generalizations. 
In particular we will show that they can be extended to retain higher logarithmic corrections and how they
can be made exact. It is shown that by a suitable extension of this analysis, all
that has been known so far in moment space can be reobtained directly from $x$-space.
In the photon case the Da Luz Vieira-Storrow solution \cite{Storrow} can be now understood 
simply as a {\em first truncated ansatz} of the general truncated solutions that we analize. However, 
the analysis presented here is limited to the QCD pdf's, 
while a similar study of the pdf's of the photon will be discussed by us in a forthcoming paper.

We introduce the notion of first, second, and so on, truncated solutions of the DGLAP, 
that correct the LO behaviour, from which the nature of the expansions and the contributions retained 
by the various approximations will appear quite clearly. 

Our work is organized as follows. After defining our conventions, we bring in a simple example
that shows how a non-singlet LO solution of the DGLAP is obtained by an $x$-space ansatz. 
Then we move to NLO and introduce the notion of truncated logarithmic solutions at this order, 
moving afterwards to define exact recursive solutions from $x$-space. In all these cases, we show that 
these solutions contain exactly 
the same information of those obtained in Mellin space, to which they turn out to be equivalent.  
The same analysis is then extended to NNLO. Our approach lays at the foundation of a numerical 
method -based on $x$-space - that solves the NNLO DGLAP with great accuracy down to very small-$x$ $(10^{-5})$. 
The method, therefore, not only does not suffer from the usual well known
inaccuracy of $x$-space based approaches at small-$x$ values \cite{CorianoSavkli}, but is, from our viewpoint, 
a complementary way to look at the evolution of the pdf's in an extremely simple fashion. We conclude with some 
comments concerning the timely issue of defining benchmarks for the evolution of the pdf's, obtained by 
comparing solutions extracted by Mellin methods against those derived from our approach, in particular 
for those solutions which retain accuracy of a given order in $\alpha_s$ ($O(\alpha_s)$ accurate 
solutions), relevant for precise determination of certain NNLO observables at the LHC.

\section{Definitions and conventions}

Before we start our analysis it is convenient to define
here our notations and conventions that we will use in the rest of the paper.

We introduce the 3-loop evolution of the coupling via its $\beta$-function
\begin{equation}
\beta(\alpha_{s})\equiv\frac{\partial\alpha_{s}(Q^{2})}{\partial\log Q^{2}},
\label{eq:beta_def}
\end{equation}
and its three-loop expansion is
\begin{equation}
\beta(\alpha_{s})=-\frac{\beta_{0}}{4\pi}\alpha_{s}^{2}-\frac{\beta_{1}}{16\pi^{2}}\alpha_{s}^{3}-
\frac{\beta_{2}}{64\pi^{3}}\alpha_{s}^{4}+O(\alpha_{s}^{5}),
\label{eq:beta_exp}
\end{equation}
where
\ba
&&\beta_{0}=\frac{11}{3}N_{C}-\frac{4}{3}T_{f},\nonumber\\
&&\beta_{1}=\frac{34}{3}N_{C}^{2}-\frac{10}{3}N_{C}n_{f}-2C_{F}n_{f},\nonumber\\
&&\beta_{2}=\frac{2857}{54}N_{C}^{3}+2C_{F}^{2}T_{f}-\frac{205}{9}C_{F}N_{C}T_{f}-
\frac{1415}{27}N_{C}^{2}T_{f}+\frac{44}{9}C_{F}T_{f}^{2}+\frac{158}{27}N_{C}T_{f}^{2},\nonumber\\
\ea
are the coefficients of the beta function. In particular, 
$\beta_2$ \cite{betafunction1,betafunction2} and $\beta_3$ \cite{betafunction3} are in the $\overline{MS}$ scheme.
We have set
\begin{equation}
N_{C}=3,\qquad C_{F}=\frac{N_{C}^{2}-1}{2N_{C}}=\frac{4}{3},\qquad T_{f}=T_{R}n_{f}=
\frac{1}{2}n_{f},
\end{equation}
where $N_{C}$ is the number of colors, $n_{f}$ is the number of
active flavors, that is fixed by the number of quarks with $m_{q}\leq Q$.
One can obtain either an exact or an accurate (truncated)
solution of this equation.
An exact solution includes higher order effects in $\alpha_s$, while a
truncated solution retains contributions only up to a given (fixed) order in a certain expansion parameter. 
The structure of the NLO exact solution of the RGE for the coupling is well known
and relates $\alpha_s(\mu_1^2)$ in terms of $\alpha_s(\mu_2^2)$
via an implicit solution
\ba
\label{implicit}
\frac{1}{a_s(\mu_1^2)}=\frac{1}{a_s(\mu_2^2)}
+\beta_0 \ln \left(\frac{\mu_1^2}{\mu_2^2} \right)
-b_1\ln\left\{\frac{a_s(\mu_1^2) \, [ 1 + b_1 a_s(\mu_2^2) ]}
{a_s(\mu_2^2) \, [ 1 + b_1 a_s(\mu_1^2) ]} \right\},
\ea
where $a_s(\mu^2)=\alpha_s(\mu^2)/(4\pi)$.
The truncated solution is obtained by expanding up to a given order in a small
variable
\ba
\label{alphas}
\alpha_s(\mu_1^2)=\alpha_s(\mu_2^2)-\left[\frac{\alpha_s^2(\mu_2^2)}{4\pi}
+\frac{\alpha_s^3(\mu_2^2)}{(4\pi)^2}(-\beta_0^2 L^2+\beta_1 L)\right],
\ea
where the $\mu_1^2$ dependence is shifted into the factor $L=\ln(\mu_1^2/\mu_2^2)
$, and we have used a $\beta$-function expanded up to NLO, 
involving $\beta_0$ and $\beta_1$. Exact solutions of the RGE for the running coupling are not
available (analytically) beyond NLO, while they can be obtained numerically.
Truncated solutions instead can be obtained quite easily, for instance
expanding in terms of the logarithm of a specific scale ($\Lambda$)
\begin{equation}
\alpha_{s}(Q^{2})=\frac{4\pi}{\beta_{0}L_{\Lambda}}\left\{ 1-\frac{\beta_{1}}{\beta_{0}^{2}}
\frac{\log L_{\Lambda}}{L_{\Lambda}}+
\frac{1}{\beta_{0}^{3}L_{\Lambda}^{2}}\left[\frac{\beta_{1}^{2}}{\beta_{0}}
\left(\log^{2}L_{\Lambda}-\log L_{\Lambda}-1\right)+
\beta_{2}\right]+O\left(\frac{1}{L_{\Lambda}^{3}}\right)\right\},
\label{eq:alpha_s_nnlo}
\end{equation}
where
\begin{equation}
L_{\Lambda}=\log\frac{Q^{2}}{\Lambda_{\overline{MS}}^{2}},
\end{equation}

and where $\Lambda_{\overline{MS}}^{(n_{f})}$ is calculated using the known
value of $\alpha_{s}(m_{Z})$ and imposing the continuity of $\alpha_{s}$
at the thresholds identified by the quark masses.

\section{General Issues}
For an integro-differential equation of DGLAP type, which is defined in a perturbative fashion, 
the kernel $P(x)$ is known  perturbatively up to the first few orders
in $\alpha_s$, approximations which are commonly known as LO, NLO, NNLO.

The equation is of the form
\ba
\frac{\partial f(x,Q^2)}{\partial \ln{Q^2}}=P(x,Q^2)\otimes f(x,Q^2),
\label{dglap}
\ea
with
\beq
a(x)\otimes b(x)\equiv \int_0^1\frac{dy}{y} a(y) b(x/y),
\eeq
and the expansion of the kernel at LO, for instance,  is given by
\ba
\label{kernLO}
&&{P(x,Q^2)}^{LO}=\left(\atp\right) P^{(0)}(x).
\ea
In the case of QCD one equation is scalar, termed non-singlet, the other equation involves 2-by-2 matrices,
the singlet. In other cases, for instance in supersymmetric QCD, both the singlet and the
non-singlet equations have a matrix structure \cite{Coriano}. Except for the LO case, exact analytic solutions of
the singlet equations are not known. However, various methods are available in order to
obtain a numerical solution with a good accuracy. These methods are of two types: 
{\em brute force} approaches based in $x$-space and those based on the inversion of the Mellin moments.

A brute force method involves a numerical solution of the PDE based on finite differences schemes.
One can easily find a stability scheme in which the differential 
operator on the left-hand-side of the equation gets replaced by its finite difference expression.
This method has the advantage that it allows to obtain the so called ``exact'' solution of the
equation at a given order (LO, NLO, NNLO). The only approximation involved in this numerical solution
comes from the perturbative expansion of the kernels. Solutions of this type are not
accurate to the working order of the expansion of the kernels,
since they retain higher order terms in $\alpha_s$. 
\footnote{It is common, however, to refer to
these solutions as to the ``exact'' ones, though they have no better status than the accurate (truncated) ones. 
In principle, large cancelations between contributions of 
higher order in the perturbative expansion of the kernels beyond NNLO, which are not available, 
and the known contributions, could take place at higher orders and this possibility 
remains unaccounted for in these ``exact'' solution. 
The term ``exact'', though being a misnomer, is however wide spread 
in the context of perturbative applications and 
for this reason we will use it throughout our work.} A short-come of brute force methods is the lack of an ansatz 
for the solution, which could instead be quite useful in order to understand the role of the retained perturbative logarithms. 
The use of Mellin inversion allows to extract, in the non singlet case, the exact solution quite immediately up to NNLO. 

The Mellin moments are defined as

\begin{equation}
a(N)=\int_{0}^{1}a(x)x^{N-1}\textrm{d}x
\end{equation}
and the basic advantage of working in moment space is to reduce the convolution product $\otimes$
into an ordinary product. For instance, at leading order we obtain
the LO DGLAP equation (\ref{dglap}) in moment space
\begin{equation}
\frac{\partial f(N,\alpha_{s})}{\partial\alpha_{s}}
=-\frac{\left(\frac{\alpha_{s}}{2\pi}\right)
P^{(0)}(N)}{\frac{\beta_{0}}{4\pi}\alpha_{s}^{2}}f(N,\alpha_{s}),
\label{dg1}
\end{equation}
which is solved by
\begin{equation}
f(N,\alpha_s)=f(N,\alpha_{0})\left(
\frac{\alpha_s}{\alpha_{0}}\right)^{-\frac{2P^{(0)}(N)}{\beta_{0}}}
=f(N,\alpha_{0})\exp\left\{-\frac{2P^{(0)}(N)}{\beta_{0}}
\log\left(\frac{\alpha_s}{\alpha_{0}}\right)\right\},
\label{solLO}
\end{equation}
where we have used the notation $\alpha\equiv\alpha(Q^{2})$ and $\alpha_{0}\equiv\alpha(Q_{0}^{2})$.
At this point, to contruct the solution in $x$-space, we need to perform a numerical
inversion of the moments, following a contour in the complex plane. This method is widely used in the numerical construction of the solutions and various optimization of this technique have been proposed 
\cite{kosower}.
We will show below how one can find solutions of any desired accuracy 
by using a set of recursion relations without the need of using numerical inversion of the 
Mellin moments.

\subsection{The logarithmic ansatz in LO}
To illustrate how the logarithmic expansion works and why it can reproduce the same solutions obtained from
moment space, it is convenient, for simplicity, to work at LO.
We try, in the ansatz, to organize the logarithmic behaviour of the solution
in terms of $\alpha_s$ and its logarithmic powers, times some
scale-invariant functions $A_n(x)$, which depend only on Bjorken $x$. As we are going to see, this re-arrangement of 
the scale dependent terms is rather general for evolution equations in QCD.
The number of the scale-invariant functions $A_n$ is actually infinite, and they are obtained recursively from a given initial condition.

The expansion that summarizes the logarithmic behaviour of the solution at LO is chosen
of the form
\beq
{f(x,Q^2)}^{LO}=\sum_{n=0}^{\infty}\frac{A_{n}(x)}{n!}
\left[\ln{\left(\frac{\alpha_s(Q^2)}{\alpha_s(Q^2_0)}\right)}\right]^{n}.
\label{an1}
\eeq

To determine $A_{n}(x)$ for every $n$ we introduce the simplified notation
\begin{equation}
L \equiv\log\frac{\alpha_{s}(Q^{2})}{\alpha_{s}(Q_{0}^{2})},
\end{equation}
and insert our ansatz (\ref{an1}) into the DGLAP equation
together with the LO expansion of the $\beta$-function to get

\begin{equation}
-\sum_{n=0}^{\infty}\frac{A_{n+1}}{n!}L^{n}
\frac{\beta_{0}}{4\pi}\alpha_{s}=\sum_{n=0}^{\infty}\frac{L^{n}}{n!}
\frac{\alpha_{s}}{2\pi}P^{(0)}\otimes A_{n}.
\end{equation}
Equating term by term in powers of $L$ we find the recursion relation
\begin{equation}
A_{n+1}=-\frac{2}{\beta_{0}}P^{(0)}\otimes A_{n}.
\label{LOrecursion}
\end{equation}
At this point we need to show that these recursion relations can be solved in terms of some
initial condition and that they reproduce the exact LO solution in moment space. This can be done by
taking Mellin moments of the recursion relations and solving the chain of these relations in terms of
the initial condition $A_0(x)$.
At LO the solution of (\ref{LOrecursion}) in moment space is simply given by  
\begin{equation}
A_{n}(N)=\left(-\frac{2}{\beta_{0}}P^{(0)}\right)^n q {(N,\alpha_s(Q_0^2))},
\end{equation}
having imposed the initial condition $A_0=q(x,\alpha_s(Q_0^2))$. 
At this point we plug in this solution into (\ref{an1}) to obtain 
\begin{equation}
f(N,Q^{2})=\sum_{n=0}^{\infty}\frac{A_{n}(N)}{n!}\log^{n}
\frac{\alpha_s(Q^{2})}{\alpha_s(Q_{0}^{2})},
\label{eq:LOansatz}
\end{equation}

which clearly coincides with (\ref{solLO}), after a simple expansion of the latter
\begin{equation}
f(N,\alpha_s)=f(N,\alpha_{0})\sum_{n=0}^{\infty}
\left\{\frac{1}{n!}\left[-\frac{2P^{(0)}(N)}{\beta_{0}}\right]^{n}
\log^{n}\left(\frac{\alpha_s}{\alpha_{0}}\right)\right\}.
\end{equation}
Notice that this non-singlet solution is an exact one. In the singlet case the same approach will succeed 
at the same order and there is no need to introduce truncated solution at this order. As expected, 
however, things will 
get more involved at higher orders, especially in the singlet case. 

The strategy that we follow in order to construct solutions of the DGLAP
equations is all contained in this trivial example, and we can summarize
our systematic search of logarithmic solutions at any order as follows: we

1) define the logarithmic ansatz up to a certain perturbative order and we insert it
into the DGLAP equation, appropriately expanded at that order;

2) derive recurrency relations for the scale invariant coefficients of the expansion;

3) take the Mellin moments of the recurrency relations and
{\em solve} them in terms of the moments of the initial conditions;

4) we show, finally, that the solution of the recursion relations, so obtained, is {\em exactly}
the solution of the original evolution equation firstly given in moment space.
This approach is sufficient to solve all the equations at any desired order of accuracy in the strong 
coupling, 
as we are going to show in the following sections. Ultimately, the success of the logarithmic ansatz lays on the fact that the solution
of the DGLAP equations in QCD resums only logarithms of the coupling constant.

\section{Truncated solution at NLO. Non-singlet}

The extension of our procedure to NLO (non-singlet) is more involved, but also in this
case proofs of consistency of the logarithmic ansatz can be formulated. 
However, before starting our technical analysis, we define the notion of  ``truncated solutions'' of the DGLAP
equations, expanding our preliminary discussion of the previous sections. We start with some definitions. 

A {\em truncated solution} retains only contributions up to a certain order in the expansion in the coupling.
We could define a $1$-$st$ truncated solution, a $2$-$nd$ truncated 
solution and so on. The sequence of truncated
solutions is expected to converge toward the exact solution of the DGLAP as the number of truncates 
increases. This can be done at any order in the 
expansion of the DGLAP kernels (NLO,NNLO,NNNLO,...). For instance, at NLO, we can build an exact solution
in moment space (this is true only in the non-singlet case) but we can also build the sequence of
truncated solutions. It is convenient to illustrate 
the kind of approximations which are involved in order 
to obtain these solutions and for this reason we try to detail the derivations.

Let's consider the NLO non-singlet DGLAP equation, written directly in moment space
\beq
\frac{\partial f(N,\alpha_{s})}{\partial\alpha_{s}}=
-\frac{\left(\frac{\alpha_{s}}{2\pi}\right)P^{(0)}(N)+
\left(\frac{\alpha_{s}}{2\pi}\right)^{2}P^{(1)}(N)}
{\frac{\beta_{0}}{4\pi}\alpha_{s}^{2}
+\frac{\beta_{1}}{16\pi^{2}}\alpha_{s}^{3}}f(N,\alpha_{s}),
\label{nontrunc}
\end{equation}
and search for its exact solution, which is given by
\begin{eqnarray}
f(N,\alpha_s) & = & f(N,\alpha_{0})\left(
\frac{\alpha_s}{\alpha_{0}}\right)^{-\frac{2P^{(0)}(N)}{\beta_{0}}}
\left(\frac{4\pi\beta_{0}+\alpha_s\beta_{1}}{4\pi\beta_{0}+
\alpha_{0}\beta_{1}}\right)^{\frac{2P^{(0)}(N)}{\beta_{0}}-
\frac{4P^{(1)}(N)}{\beta_{1}}}.
\label{exactsol}
\eeqa
Notice that equation (\ref{nontrunc}) is the exact NLO equation. In particular we have preserved the
structure of the right-hand side, that involves both the beta function and the NLO kernels and is
given as a ratio of two polynomials in $\alpha_s$

\beq
\frac{P^{NLO}(x,\alpha_s)}{\beta^{NLO}(\alpha_s)},
\eeq
where
\beq
\label{kernNLO}
{P(x,Q^2)}^{NLO}=\left(\atp\right) P^{(0)}(x)+\left(\atp\right)^2 P^{(1)}(x)
\eeq
is the NLO kernel. The factorization of the LO solution from the NLO equation can be obtained 
expanding the ratio $P/\beta$ in $\alpha_s$, which allows the factorization of a $1/\alpha_s$ 
contribution. Equivalently, one can redefine the integral of the solution in moment space 
by subtraction of the LO part
\beq
\int_{\alpha_0}^{\alpha_s}d\alpha \left(\frac{P^{NLO}(x,\alpha)}{\beta^{NLO}(\alpha)} - \frac{{P}_{LO}(\alpha)}
{\beta_{LO}(\alpha)}\right).
\label{evolint}
\eeq
Denoting by $b_1=\beta_1/\beta_0$, the truncated differential
equation can be written as
\ba
\frac{\partial f(N,\alpha_s)}{\partial \alpha_s}=-\frac{2}{\beta_0 \alpha_s}
\left[P^{(0)}(N)+\atp \left(P^{(1)}-\frac{b_1}{2} P^{(0)}\right)\right]f(N,\alpha_s),
\label{tron}
\ea
which has the solution
\ba
f(N,\alpha_s)=\left[\frac{\alpha_s}{\alpha_0}\right]^{-\frac{2 P^{(0)}}{\beta_0}}
\times \exp{\left\{\frac{\left(\alpha_s-\alpha_0\right)}{\pi\beta_0}\left(
\frac{b_1}{2} P^{(0)}-P^{(1)}\right)\right\}}f(N,\alpha_0).
\label{TTR}
\ea
Notice that this solution of the truncated equation, exactly as in the exact solution (\ref{exactsol}),
contains as a factor the LO solution and therefore can be rewritten in the form
\ba
f(N,\alpha_s)=\exp{\left\{\frac{\left(\alpha_s-\alpha_0\right)}{\pi\beta_0}\left(
\frac{b_1}{2} P^{(0)}-P^{(1)}\right)\right\}}f^{LO}(N,\alpha_s),
\label{expansol}
\ea
where $f^{LO}(N,\alpha_s)$ is given by
\ba
f^{LO}(N,\alpha_s)=\left[\frac{\alpha_s}{\alpha_0}\right]^{-\frac{2 P^{(0)}}{\beta_0}}f(N,\alpha_0).
\ea
Eq. (\ref{expansol}) exemplifies a typical mathematical encounter in the search of solutions of PDE's 
of a certain accuracy: if we allow a perturbative expansion of the defining equation arrested at a given order,
the solution, however, is still affected
by higher order terms in the expansion parameter (in our case $\alpha_s$). 
To identify the expansion which converges to (\ref{expansol}) proceeds as follows. We start 
from the $1$-$st$ truncated solution.

Expanding (\ref{expansol}) to first order around the LO solution we obtain

\ba
\label{trunc1}
f(N,\alpha_s)=f^{LO}(N,\alpha_s)\times
\left\{1+\frac{(\alpha_s-\alpha_0)}{\pi\beta_0}\left(\frac{b_1}{2} P^{(0)}-P^{(1)}\right)\right\},
\ea
which is the expression of the $1$-$st$ truncated solution, accurate at order $\alpha_s$.
One can already see from (\ref{trunc1}) that the ansatz which we are looking for should involve a double expansion in two values of the coupling
constant: $\alpha_s$ and $\alpha_0$. This point will be made more clear below. 
For this reason 
we are naturally lead to study the logarithmic expansion
\beq
{f(x,Q^2)}^{NLO}=\sum_{n=0}^{\infty}\frac{A_{n}(x)}{n!}
\left[\ln{\left(\frac{\alpha_s}{\alpha_0}\right)}\right]^{n}
+\alpha_s\sum_{n=0}^{\infty}\frac{B_{n}(x)}{n!}
\left[\ln{\left(\frac{\alpha_s}{\alpha_0}\right)}\right]^{n},\,
\label{logexp}
\eeq
which is the obvious generalization of the analogous LO expansion (\ref{an1}).

Inserting this ansatz in the NLO DGLAP equation, we derive the following recursion relations
for $A_{n}$ and $B_{n}$
\ba
&&A_{n+1}=-\frac{2}{\beta_0}P^{(0)}(x)\otimes A_{n}(x),
\nonumber\\
&&B_{n+1}=-B_{n}(x)-\frac{\beta_1}{4\pi\beta_0}A_{n+1}(x)
-\frac{2}{\beta_0}P^{(0)}(x)\otimes B_{n}(x)-\frac{1}{\pi\beta_0}P^{(1)}(x)\otimes A_{n}(x),
\nonumber\\
\label{rr}
\ea
together with the initial condition
\ba
f(x,Q^2_{0})=A_0+\alpha_s(Q_0^2)B_0.
\ea

At this point we need to prove that the recursion relations (\ref{rr}) reproduce in moment
space (\ref{trunc1}).
To do so we rewrite the recursion relations in Mellin-space
\ba
&&A_{n+1}(N)=-\frac{2}{\beta_0}P^{(0)}(N)A_{n}(N),\nonumber\\
&&B_{n+1}(N)=-B_{n}(N)-\frac{\beta_1}{4\pi\beta_0}A_{n+1}(N)
-\frac{2}{\beta_0}P^{(0)}(N)B_{n}(N)-
\frac{1}{\pi\beta_0}P^{(1)}(N)A_{n}(N),\nonumber\\
\ea
and search for their solution over $n$.
After solving these relations with respect to $A_{0}$ and $B_{0}$,
it is simple to realize that our ansatz (\ref{logexp}) exactly
reproduces the truncated solution (\ref{trunc1}) only if the condition $B_{0}=0$
is satisfied. In fact, denoting by
\ba
&&R_0=-\frac{2}{\beta_0}P^{(0)}(N),\nonumber\\
&&R_1=\left(\frac{b_1}{2\pi\beta_0}P^{(0)}-
\frac{1}{\pi\beta_0}P^{(1)}\right),
\ea
the recursive coefficients, we can rewrite the recursion relations as
\ba
&&A_{n+1}=R_0 A_n,\nonumber\\
&&B_{n+1}=(R_0-1)B_n + R_1 A_n .
\label{rec1}
\ea
Then, observing that
\ba
&&A_n=R_0^{n}A_0,\nonumber\\
&&B_1=(R_0-1)B_0+R_1 A_0,\nonumber\\
&&B_2=(R_0-1)^2 B_0+R_1 A_0(2R_0-1),\nonumber\\
&&B_3=(R_0-1)^3 B_0+R_1 A_0\left[(2R_0-1)(R_0-1)+R_0^2\right],\nonumber\\
&&\vdots\nonumber\\
\ea
we identify the structure of the $n_{th}$ iterate in close form
\beq
B_n=(R_0-1)^n B_0+R_1 A_0\left[R_0^n-(R_0-1)^n\right].\nonumber
\label{rec2}
\eeq
Substituting the expressions for $A_n$ and $B_n$ so obtained in terms of $A_0$ and $B_0$
in the initial ansatz, and summing the logarithms (a procedure that we call ``exponentiation ``) we obtain
\beqa
\sum_{n=0}^\infty  \frac{A_n(N)}{n!} L^n &=& A_0\left(\frac{\alpha_s}{\alpha_0}\right)^{R_0},
\nonumber \\
\sum_{n=0}^\infty  \alpha_s \frac{B_n(N)}{n!} L^n &=& \sum_{n=0}^\infty
\alpha_s \frac{1}{n!}\left\{(R_0-1)^n B_0+R_1 A_0\left[R_0^n-(R_0-1)^n\right]\right\} \nonumber \\
&=&
\alpha_s B_0
\left(\frac{\alpha_s}{\alpha_0}\right)^{R_0-1}+
\alpha_s R_1 A_0\left(\frac{\alpha_s}{\alpha_0}\right)^{R_0}-
\alpha_s R_1 A_0\left(\frac{\alpha_s}{\alpha_0}\right)^{R_0-1},
\nonumber \\
\eeqa
expression that can be rewritten as
\ba
f(N,\alpha_s)=A_0\left(\frac{\alpha_s}{\alpha_0}\right)^{R_0} +\alpha_s B_0
\left(\frac{\alpha_s}{\alpha_0}\right)^{R_0-1}+
\alpha_s R_1 A_0\left(\frac{\alpha_s}{\alpha_0}\right)^{R_0}-
\alpha_s R_1 A_0\left(\frac{\alpha_s}{\alpha_0}\right)^{R_0-1}.
\ea
This expression after a simple rearrangement becomes 
\ba
f(N,\alpha_s)=\left(\frac{\alpha_s}{\alpha_0}\right)^{R_0}\left[1+
\left(\alpha_s-\alpha_0\right)R_1\right]f(N,Q^2_0)-
\left(\frac{\alpha_s}{\alpha_0}\right)^{R_0}\left(\alpha_s-\alpha_0\right)
(R_1\alpha_0 B_0).
\ea
This solution in moment space exactly coincides with the truncated solution (\ref{trunc1}) if we impose 
the condition $B_0=0$.
 It is clear that the solution gets organized in the form of a double expansion in the
two variables $\alpha_s$ and $\alpha_0$. While $\alpha_s$ appears explicitely
in the ansatz (\ref{logexp}), $\alpha_0$ appears only after the logarithmic summation
and the factorization of the leading order solution.
An obvious question to ask is how should we modify our ansatz if we want to reproduce
the exact solution of the truncated DGLAP equation in moment space, given by eq.~(\ref{TTR}).
The answer comes from a simple extension of our recursive method.

\subsection{Higher order truncated solutions}

We start by expanding the solution of the truncated equation (\ref{expansol}),
whose exponential factor is approximated by its double expansion in 
$\alpha_s$ and $\alpha_0$ to second order, thereby identifying the approximate solution 
\beqa
f(N,\alpha_s) &=&\exp{\left\{\frac{\left(\alpha_s-\alpha_0\right)}{\pi\beta_0}\left(
\frac{b_1}{2} P^{(0)}-P^{(1)}\right)\right\}}\times f^{LO}(N,\alpha_s)\nonumber \\
&\simeq& \left(\frac{\alpha_s}{\alpha_{0}}\right)^{R_0}
\left[1- R_1 (\alpha_0-\alpha_s)
+\frac{1}{2} R_1^2(\alpha_0-\alpha_s)^2
+R_1(\alpha_0^2 -\alpha_s^2)\frac{b_1}{8\pi}\right]f(N,\alpha_{0}).
\label{newapp}
\nonumber\\
\ea
To generate this solution with the recursive method it is sufficient to introduce 
the higher order (2nd order)  ansatz 
\ba
\label{anss1}
\tilde{f}(x,\alpha_s)=\sum_{n=0}^{+\infty}\frac{L^n}{n!}\left[
A_n(x) +\alpha_s B_n(x)+\alpha_s^2 C_n(x)\right],
\ea
where we have included some new coefficients $C_n(x)$ that will take care of the higher order 
terms we aim to include.
Inserted into the NLO DGLAP equation, this ansatz
generates an appropriate chain of recursion relations
\ba
\label{}
&&A_{n+1}(x)=-\frac{2}{\beta_0}P^{(0)}(x)\otimes A_{n}(x),
\nonumber\\
&&B_{n+1}(x)=-B_{n}(x)-\frac{2}{\beta_0}P^{(0)}(x)\otimes B_{n}(x)
-\frac{b_1}{(4\pi)}A_{n+1}(x)-\frac{1}{\pi\beta_0}P^{(1)}(x)\otimes A_{n}(x),
\nonumber\\
&&C_{n+1}(x)=-2 C_{n}(x)-\frac{2}{\beta_0}P^{(0)}(x)\otimes C_{n}(x)
-\frac{b_1}{(4\pi)}B_{n+1}(x)-\frac{b_1}{(4\pi)}B_{n}(x)
-\frac{1}{\pi\beta_0}P^{(1)}(x)\otimes B_{n}(x),
\nonumber\\
\ea
that we solve by going to Mellin space and obtain
\ba
\label{highNLO}
&&A_n=R_0^n A_0,\nonumber\\
&&B_n=R_1\left[R_0^n-(R_0-1)^n\right]A_0,
\nonumber\\
&&C_n=\left[\frac{1}{2}\left(R_0 -2\right)^n -\left(R_0 -1\right)^n
\right]R_1^2 A_0 + \nonumber\\
&&\hspace{1cm}\frac{1}{2}R_1^2 R_0^n A_0 +
\frac{1}{8\pi} R_1 b_1 \left(R_0 -2\right)^n A_0
-\frac{1}{8\pi}R_1 R_0^n b_1 A_0,
\ea
where the initial conditions are $\tilde{f}(N,\alpha_0)=A_0$ and $B_0=C_0=0$. It is a trivial exercise
to show that the solution of the recursion relation, inserted into (\ref{anss1}), coincides with (\ref{newapp}), after exponentiation.

Capturing more and more logs of the truncated logarithmic equation at this point is as easy as never before.
We can consider, for instance, a higher order ansatz accurate to $O(\alpha_s^3)$
for the NLO non-singlet solution
\ba
\label{anss2}
\tilde{f}(x,\alpha_s)=\sum_{n=0}^{\infty}\frac{L^n}{n!}\left[
A_n(x) +\alpha_s B_n(x)+\alpha_s^2 C_n(x)+\alpha_s^3 D_n(x)\right],
\ea
that generates four independent recursion relations. The relations for $A_{n+1}$,
$B_{n+1}$, $C_{n+1}$ are, for this extension, the same as in the previous case and are listed
in (\ref{highNLO}). Hence, we are left with an additional relation for the
$D_{n+1}$ coefficient which reads
\ba
D_{n+1}(x)=-3D_{n}(x)-\frac{2}{\beta_0}P^{(0)}\otimes D_{n}(x)
-\frac{b_1}{(4\pi)}C_{n+1}(x)-\frac{b_1}{(2\pi)}C_{n}(x)
-\frac{1}{\pi\beta_0}P^{(1)}\otimes C_{n}(x).
\nonumber\\
\ea
These are solved in Mellin space with respect to
$A_0,B_0,C_0,D_0$ (with the condition $B_0=C_0=D_0=0$). We obtain
\ba
&&A_n=R_0^n A_0,\nonumber\\
&&B_n=R_1\left[R_0^n- (R_0-1)^n\right]A_0,
\nonumber\\
&&C_n=\left[\frac{1}{2}\left(R_0 -2\right)^n -\left(R_0 -1\right)^n
\right]R_1^2 A_0 + \nonumber\\
&&\hspace{1cm}\frac{1}{2}R_1^2 R_0^n A_0 +
\frac{1}{8\pi} R_1 b_1 \left(R_0 -2\right)^n A_0
-\frac{1}{8\pi}R_1 R_0^n b_1 A_0,\nonumber\\
&&D_n=\left[-\frac{1}{6}\left(R_0 -3\right)^n
+\frac{1}{2}\left(R_0 -2\right)^n
-\frac{1}{2}\left(R_0 -1\right)^n
+\frac{1}{6}R_0^n \right]R_1^3 A_0\nonumber\\
&&\hspace{1cm}\left[-\frac{1}{8\pi}\left(R_0 -3\right)^n b_1
+\frac{1}{8\pi}\left(R_0 -2\right)^n b_1
+\frac{1}{8\pi}\left(R_0 -1\right)^n b_1
-\frac{1}{8\pi}R_0^n b_1\right]R_1^2 A_0\nonumber\\
&&\hspace{1cm}\left[-\frac{1}{48\pi^2}\left(R_0 -3\right)^n b_1^2
+\frac{1}{48\pi^2}R_0^n b_1^2\right]R_1 A_0.
\label{chain2}
\ea
Exponentiating we have
\ba
\label{hanss2}
&&\tilde{f}(x,\alpha_s)=
\left\{1+\alpha_s\left(1-\frac{\alpha_0}{\alpha_s}\right)R_1\right\}
A_0\left(\frac{\alpha_s}{\alpha_0}\right)^{R_0}
\nonumber\\
&&\hspace{1.5cm}+\alpha_s^2\left\{\left[\frac{1}{2}\left(\frac{\alpha_0^2}{\alpha_s^2}
-2~\frac{\alpha_0}{\alpha_s}+1\right)R_1^2
+\frac{b_1}{8\pi}\frac{\alpha_0^2}{\alpha_s^2}R_1
-\frac{b_1}{8\pi}R_1\right]\right\}A_0\left(\frac{\alpha_s}{\alpha_0}\right)^{R_0}
\nonumber\\
&&\hspace{1.5cm}+\alpha_s^3\left\{\left(-\frac{1}{6}\frac{\alpha_0^3}{\alpha_s^3}
+\frac{1}{2}\frac{\alpha_0^2}{\alpha_s^2}-\frac{1}{2}\frac{\alpha_0}{\alpha_s}
+\frac{1}{6}\right)R_1^3+\left(-\frac{b_1}{8\pi}\frac{\alpha_0^3}{\alpha_s^3}
+\frac{b_1}{8\pi}\frac{\alpha_0^2}{\alpha_s^2}
+\frac{b_1}{8\pi}\frac{\alpha_0}{\alpha_s}-\frac{b_1}{8\pi}\right)R_1^2
\right.\nonumber\\
&&\hspace{2.5cm}\left.+\left(\frac{b_1^2}{48\pi^2}\frac{\alpha_0^3}{\alpha_s^3}
+\frac{b_1^2}{48\pi^2}\right)R_1\right\}A_0\left(\frac{\alpha_s}{\alpha_0}\right)^{R_0}\,,
\ea
which is the solution of the truncated
equation computed with an $O(\alpha_s^3)$ accuracy.

\section{Non-singlet truncated solutions at NNLO}

The generalization of the method that takes to the truncated solutions
at NNLO is more involved, but to show the equivalence of these solutions to those in Mellin space
one proceeds as for the lower orders. As we have already pointed out, one has first to expand the ratio $P/\beta$ at a certain
order in $\alpha_s$, then solve the equation in moment space - solution that will bring in automatically
higher powers of $\alpha_s$ - and then reconstruct this solution via iterates.

At NNLO the kernels are given by
\ba
\label{ansatzeNNLO}
&&{P(x,Q^2)}^{NNLO}=\left(\atp\right) P^{(0)}(x)+\left(\atp\right)^2 P^{(1)}(x)
+\left(\atp\right)^3P^{(2)}(x),
\nonumber\\
\ea
and the equation in Mellin-space is given by
\ba
\frac{\partial f(N,\alpha_s)}{\partial\alpha_s}&=& \frac{P^{NNLO}(N)}{\beta^{NNLO}}f(N,\alpha_s).
\ea
We search for solutions of this equation of a given accuracy in $\alpha_s$ and for this purpose
we truncate the evolution integral of the ratio $P/\beta$ to $O(\alpha_s^2)$. 
This is the first order at which the $P^{(2)}$ component of the kernels appear. As we are going to see, 
this will generate the first truncate for the NNLO case. Therefore, while the first truncate at NLO 
is of $O(\alpha_s)$, the first truncate at NNLO is of $O(\alpha_s^2)$.
We obtain  
\beq
I_{NNLO}=\int_{\alpha_0}^{\alpha_s}d\alpha \left(\frac{P_{NNLO}(x,\alpha)}{\beta_{NNLO}(\alpha)}
- \frac{{P}_{LO}(\alpha)}{\beta_{LO}(\alpha)}\right)\approx
-R_1\alpha_0 -\frac{1}{2}R_2\alpha_0^2+R_1\alpha_s+\frac{1}{2}R_2\alpha_s^2.
\label{Intevol1}
\eeq

At this retained accuracy of the evolution integral,
the exact solution of the corresponding (truncated) DGLAP equation can be found, in moment space, as in (\ref{ltrunc})
\ba
&&f(N,\alpha_s)=f(N,\alpha_0)\left(\frac{\alpha_s}{\alpha_0}\right)^{-2 \frac{P^{(0)}}{\beta_0}} \left\{1 + \left(\alpha_s-\alpha_0\right)\left[-\frac{P^{(1)}}{\pi\beta_0}
+\frac{P^{(0)}\beta_1}{2\pi\beta_0^2}\right] \right.\nonumber\\
&&\left.\hspace{1.6cm}+\alpha_s^2\left[\frac{{P^{(1)}}^2}{2\pi^2\beta_0^2}
-\frac{P^{(2)}}{4\pi^2\beta_0}-\frac{P^{(0)}P^{(1)}\beta_1}{2\pi^2\beta_0^2}
+\frac{P^{(1)}\beta_1}{8\pi^2\beta_0^2}+\frac{{P^{(0)}}^2\beta_1^2}{8\pi^2\beta_0^4}
-\frac{P^{(0)}\beta_1^2}{16\pi^2\beta_0^3}+\frac{P^{(0)}\beta_2}{16\pi^2\beta_0^2}
\right]\right.\nonumber\\
&&\left.\hspace{1.6cm}+\alpha_0^2\left[\frac{{P^{(1)}}^2}{2\pi^2\beta_0^2}
+\frac{P^{(2)}}{4\pi^2\beta_0}-\frac{P^{(0)}P^{(1)}\beta_1}{2\pi^2\beta_0^2}
-\frac{P^{(1)}\beta_1}{8\pi^2\beta_0^2}+\frac{{P^{(0)}}^2\beta_1^2}{8\pi^2\beta_0^4}
+\frac{P^{(0)}\beta_1^2}{16\pi^2\beta_0^3}-\frac{P^{(0)}\beta_2}{16\pi^2\beta_0^2}
\right]\right.\nonumber\\
&&\left.\hspace{1.6cm}+\alpha_0\alpha_s\left[-\frac{{P^{(1)}}^2}{\pi^2\beta_0^2}
+\frac{P^{(0)}P^{(1)}\beta_1}{\pi^2\beta_0^3}-
\frac{{P^{(0)}}^2\beta_1^2}{4\pi^2\beta_0^4}\right]\right\}\,.
\label{para}
\ea
At this order this solution coincides with the exact NNLO solution of the DGLAP
equation, obtained from an exact evaluation of the integral (\ref{Intevol1}), followed by a 
double epansion in the couplings. Therefore, similarly to (\ref{newapp}), 
the solution is organized effectively as a double expansion
in $\alpha_s$ and $\alpha_0$. This approach remains valid also in the singlet case, when the equations assume a matrix form. As we have already pointed
out above, all the known solutions of the singlet equations in moment space are obtained after a truncation of the corresponding PDE, having retained a given accuracy of the ratio $P/\beta$. For this reason, and to compare with the
previous literature, it is convenient to rewrite (\ref{para}) in a form that parallels the analogous
singlet result \cite{ellis}. It is not difficult to perform the match of our result with that previous one, which takes the form 
\cite{ellis}, \cite{Vogt3}, \cite{Buras}
\ba
f(N,\alpha_s)&=&U(N,\alpha_s)f_{LO}(N,\alpha_s,\alpha_0)U^{-1}(N,\alpha_0)
\nonumber\\
&=&\left[1+\sum_{\kappa=1}^{+\infty}U_{\kappa}(N)\alpha_s^{\kappa}\right]
f_{LO}(N,\alpha_s,\alpha_0)\left[1+\sum_{\kappa=1}^{+\infty}U_{\kappa}(N)\alpha_0^{\kappa}\right]^{-1},
\ea
which becomes, after some manipulations 
\ba
\label{solNNLO}
&&f(N,\alpha_s)=\left(\frac{\alpha_s}{\alpha_0}\right)^{-\frac{2}{\beta_0}P^{(0)}}
\left[1+\left(\alpha_s-\alpha_0\right)U_1(N) +\alpha_s^2 U_2(N)\right.\nonumber\\
&&\hspace{4.5cm}\left.-\alpha_s\alpha_0 U_1^2(N)
+\alpha_0^2\left(U_1^2(N)-U_2(N)\right)\right]f(N,\alpha_0)\,,\nonumber\\
\ea
where the functions $U_{i}(N)$ are defined as
\ba
&&U_1(N)=\frac{1}{\pi\beta_0}\left[\frac{b_1 P^{(0)}(N)}{2}-P^{(1)}(N)\right]
\equiv R_1(N),\nonumber\\
&&U_2(N)=\frac{1}{2}\left[R_1^2(N)-R_2(N)\right],\nonumber\\
&&R_2(N)=\left[\frac{P^{(2)}(N)}{2\pi^2\beta_0}+
\frac{b_1}{4\pi}R_1(N)+\frac{b_2}{(4\pi)^2}R_0(N)\right],\nonumber\\
&&R_0(N)=-\frac{2}{\beta_0}P^{(0)}(N),
\,
\ea
where $\beta_1/\beta_0=b_1$, $\beta_2/\beta_0=b_2$. 

We intend to show rigorously that this solution is generated by a simple logarithmic ansatz
arrested at a specific order. 
For this purpose we simplify (\ref{solNNLO}) obtaining 

\ba
\label{arrangedNNLO}
&&f(N,Q^2)=\left(\frac{\alpha_s}{\alpha_0}\right)^{R_0(N)}
\left[1-R_1(N)\alpha_0+\frac{1}{2}R_1^2(N)\alpha_0^2+
\frac{1}{2}R_2(N)\alpha_0^2+R_1(N)\alpha_s\right.\nonumber\\
&&\hspace{4cm}\left.-R_1^2(N)\alpha_s\alpha_0+\frac{1}{2}R_1^2(N)\alpha_s^2-
\frac{1}{2}R_2(N)\alpha_s^2\right]f(N,Q^2_0),\nonumber\\
\ea
and the ansatz that captures its logarithmic behaviour can be easily
found and is given by
\ba
\label{nnloans}
&&{f(x,Q^2)}^{NNLO}=\sum_{n=0}^{\infty}\frac{A_{n}(x)}{n!}
\left[\ln{\left(\frac{\alpha_s(Q^2)}{\alpha_s(Q^2_0)}\right)}\right]^{n}
+\alpha_s(Q^2)\sum_{n=0}^{\infty}\frac{B_{n}(x)}{n!}
\left[\ln{\left(\frac{\alpha_s(Q^2)}{\alpha_s(Q^2_0)}\right)}\right]^{n}\nonumber\\
&&\hspace{3cm}+\alpha_s^2(Q^2)\sum_{n=0}^{\infty}\frac{C_{n}(x)}{n!}
\left[\ln{\left(\frac{\alpha_s(Q^2)}{\alpha_s(Q^2_0)}\right)}\right]^{n}\,.
\ea
Setting the initial conditions as
\ba
f(x,Q^2_{0})=A_0(x)+\alpha_0 B_0(x) + \alpha_0^2 C_0(x),
\ea
and introducing the 3-loop expansion of the $\beta$-function, we derive the following recursion relations

\ba
\label{NNLOrecurrence}
&&A_{n+1}(x)=-\frac{2}{\beta_0}P^{(0)}(x)\otimes A_{n}(x),
\nonumber\\
&&B_{n+1}(x)=-B_{n}(x)-\frac{\beta_1}{4\pi\beta_0}A_{n+1}(x)
-\frac{2}{\beta_0}P^{(0)}(x)\otimes B_{n}(x)-\frac{1}{\pi\beta_0}P^{(1)}(x)\otimes A_{n}(x),
\nonumber\\
&&C_{n+1}(x)=-2C_{n}(x)-\frac{\beta_{1}}{4\pi\beta_{0}}B_{n}(x)-
\frac{\beta_{1}}{4\pi\beta_{0}}B_{n+1}(x)-
\frac{\beta_{2}}{16\pi^{2}\beta_{0}}A_{n+1}(x)\nonumber \\
&&\hspace{2cm} -\frac{2}{\beta_{0}}P^{(0)}(x)\otimes C_{n}(x)-
\frac{1}{\pi\beta_{0}}P^{(1)}(x)\otimes B_{n}(x)\nonumber \\
&&\hspace{2cm}-\frac{1}{2\pi^{2}\beta_{0}}P^{(2)}(x)\otimes A_{n}(x).
\ea

We need to show that the solution of the NNLO recursion relations
reproduces (\ref{solNNLO}) in Mellin-space, once we have chosen appropriate initial conditions for
$A_0(N)$, $B_0(N)$ and $C_0(N)$ \footnote{It can be shown that the infinite set of recursion relations have internal symmetries 
and different choices of initial conditions can bring to the same solution. The choice that we make in our analysis is the simplest one.}. 

At NLO we have already seen that $B_0(N)$ has to vanish for any $N$,
i.e. $B_0(x)=0$, and we try to impose the same condition on $C_0(N)$.
In this case we obtain the recursion relation for the moments
\ba
\label{NNLOrec_nonsing}
&&C_{n+1}(N)=-2 C_{n}(N)-\frac{\beta_{1}}{4\pi\beta_{0}}B_{n}(N)-
\frac{\beta_{1}}{4\pi\beta_{0}}B_{n+1}(N)-
\frac{\beta_{2}}{16\pi^{2}\beta_{0}}A_{n+1}(N)\nonumber \\
&&\hspace{2cm} -\frac{2}{\beta_{0}}P^{(0)}(N) C_{n}(N)-
\frac{1}{\pi\beta_{0}}P^{(1)}(N)  B_{n}(N)\nonumber \\
&&\hspace{2cm}-\frac{1}{2\pi^{2}\beta_{0}}P^{(2)}(N)  A_{n}(N),
\ea
which combined with the relations for $A_n(N)$ and $B_n(N)$
give
\ba
&&C_{n}=\left\{-R_1^2 \left(R_0-1\right)^n -\frac{1}{2}R_2 R_0^{n}
+\frac{1}{2}\left[\left(R_0-2\right)^n R_1^2+
\left(R_0-2\right)^n R_2+R_1^2 R_0^n\right]\right\}A_0,\nonumber\\
&&B_{n}= \left[R_0^n-\left(R_0-1\right)^n \right]R_1 A_0,\nonumber\\
&&A_{n}=R_0^n A_0\,,
\ea
where the $N$ dependence in the coefficients $R_i$ has been
suppressed for simplicity.
The solution is determined exactly as in (\ref{rec1})
and (\ref{rec2}) and can be easily brought to the form
\beqa
f(N,Q^2)=\frac{1}{2}\left(\frac{\alpha_s}{\alpha_0}\right)^{R_0(N)}
\left[2-2 R_1 (\alpha_s -\alpha_0)+R_1^2(\alpha_s-\alpha_0)^2+
R_2(\alpha_0^2-\alpha_s^2)\right]A_0,
\ea
which is the result quoted in eq.~(\ref{arrangedNNLO}) with $A_0=f(N,\alpha_0)$.
We have therefore identified the
correct logarithmic expansion at NNLO that solves in $x$-space the
DGLAP equation with an accuracy of order $\alpha_s^2$.

\section{Generalizations to all orders : exact solutions of truncated equations built 
recursively}

We have seen that the order of approximation in $\alpha_s$ of the truncated solutions is in direct 
correspondence with the order of the approximation used in the computation of the integral on the right-hand-side of the evolution equation (\ref{evolint}).
This issues is particularly important in the singlet case, as we are going to investigate next,
since any singlet solution involves a truncation. We have also seen that all the known solutions
obtained in moment space can be
easily reobtained from a logarithmic ansatz and therefore there is complete equivalence between the two
approaches. We will also seen that the structure of the ansatz is insensitive to whether the equations
that we intend to solve are of matrix forms or are scalar equations, since the ansatz and the
recursion relations are linear in the unknown matrix coefficients (for the singlet) and generate in both cases 
the same recursion relations. 

We pause here and try to describe the patterns that we have investigated in some generality.
We work at a generic order $N^mLO$, with m=1 denoting the NLO, $m=2$ the NNLO and so on.
We have already seen that
one can factorize the LO solution, having defined the evolution integral
\beq
I_{N^mLO}(\alpha_s,\alpha_0)=\int_{\alpha_0}^{\alpha_s}d\alpha
 \left(\frac{P_{N^mLO}(x,\alpha)}{\beta_{N^mLO}(\alpha)}
-\frac{{P}_{LO}(\alpha)}
{\beta_{LO}(\alpha)}\right),
\label{Intevolm}
\eeq
and the exact solution can be formally written as
\ba
f(N,\alpha_s)=f^{LO}(N,\alpha_s)\times e^{I_{N^mLO}(\alpha_s,\alpha_0)}.
\ea
A Taylor expansion of the integrand in the (\ref{Intevolm})
around $\alpha_s=0$ at order $\kappa=(m-1)$ gives, in moment space,
\ba
&&\left(\frac{P_{N^mLO}(N,\alpha_s)}{\beta_{N^mLO}(\alpha_s)}-\frac{{P}_{LO}(\alpha_s)}
{\beta_{LO}(\alpha_s)}\right)\approx R_1(P^{(0)},P^{(1)},N)+
R_2(P^{(0)},P^{(1)},P^{(2)},N)\alpha_s
\nonumber\\
&&\hspace{5cm}
+R_3(P^{(0)},P^{(1)},P^{(3)},N)\alpha_s^2
\nonumber\\
&&\hspace{5cm}
+\dots+R_{\kappa+1}(P^{(0)},P^{(1)},\dots,P^{(m)},N)\alpha_s^{\kappa},\,
\ea
which at NNLO becomes
\ba
&&\left(\frac{P_{NNLO}(N,\alpha_s)}{\beta_{NNLO}(\alpha_s)}-\frac{{P}_{LO}(\alpha_s)}
{\beta_{LO}(\alpha_s)}\right)\approx R_1(P^{(0)},P^{(1)},N)+
R_2(P^{(0)},P^{(1)},P^{(2)},N)\alpha_s.
\ea
Thus, integrating between $\alpha_s$ and $\alpha_0$ we, obtain the
following expression for $I_{N^mLO}(\alpha_s,\alpha_0)$ at 
$O(\alpha_s^{\kappa})$
\ba
I_{N^mLO}^{(\kappa)}(\alpha_s,\alpha_0)=-R_1\alpha_0 -\frac{1}{2}R_2\alpha_0^2-
\dots -\frac{1}{\kappa}R_{\kappa}\alpha_0^{\kappa}+R_1\alpha_s
+\frac{1}{2}R_2\alpha_s^2+\dots+\frac{1}{\kappa}R_{\kappa}\alpha_s^{\kappa},
\ea
where the $(P^{(0)},P^{(1)},\dots,P^{(m)},N)$ dependence in the $R_{\kappa}$
coefficients has been omitted.
To summarize: in order to solve the
$\kappa$-truncated version of the N$^{m}$LO DGLAP equation
\ba
\frac{\partial f(N,\alpha_{s})}{\partial\alpha_{s}}=
\frac{P_{\rm{N}^{m}\rm{LO}}(N,\alpha_s)}{\beta_{\rm{N}^{m}\rm{LO}}(\alpha_s)}f(N,\alpha_{s}),
\ea
which is obtained by a Taylor expansion - around $\alpha_s=0$ -
of the ratio $P_{\rm{N}^{m}\rm{LO}}(N,\alpha_s)/\beta_{\rm{N}^{m}\rm{LO}}(\alpha_s)$, we need to solve the equation
\ba
\label{trunc_eq}
\frac{\partial f(N,\alpha_{s})}{\partial\alpha_{s}}=\frac{1}{\alpha_s}\left[R_0+
\alpha_s R_1+\alpha_s^2 R_2+\dots+\alpha_s^{\kappa} R_{\kappa}\right]f(N,\alpha_{s}),
\ea
where the coefficients $R_{\kappa}$ have a dependence on $P^{(0)}$ and $P^{(1)}$
in the NLO case, and on $P^{(0)}$, $P^{(1)}$ and $P^{(2)}$ in the NNLO case.
Eq. (\ref{trunc_eq}) admits an exact solution of the form
\ba
f(N,\alpha_{s})=\left(\frac{\alpha_s}{\alpha_0}\right)^{R_0}\exp\left\{R_1\left(\alpha_s-\alpha_0\right)+
\frac{1}{2}R_2\left(\alpha_s^2-\alpha_0^2\right)+\dots +
\frac{1}{\kappa}R_{\kappa}\left(\alpha_s^{\kappa}-\alpha_0^{\kappa}\right)\right\}f(N,\alpha_0),
\nonumber\\
\ea
having factorized the LO solution.

At this stage, the Taylor expansion of the exponential
around $(\alpha_s,\alpha_0)=(0,0)$ generates an expanded solution of the form
\beqa
f_{N^mLO}(N,\alpha_s) &\approx& f_{LO}(N,\alpha_0) e^{I_{N^mLO}^{(\kappa)}}\nonumber \\
&=& f_{LO}(N,\alpha_0) \left(1 + I_{N^mLO}^{(\kappa)} + \frac{1}{2!}\left(I_{N^mLO}^{(\kappa)}\right)^2
+ \dots \right),
\label{ltrunc}
\eeqa

which can be also written as
\ba
&&f_{LO}(N,\alpha_0) \left(1 + I_{N^mLO}^{(\kappa)} + \frac{1}{2!}\left(I_{N^mLO}^{(\kappa)}\right)^2
+ \dots \right)= \nonumber\\
&&f^{LO}(N,\alpha_s)\times\left[c_0 + \alpha_s \left(c_{1,0} + c_{1,1}\alpha_0
+ c_{1,2}\alpha_0^2 + \dots + c_{1,\kappa-1}\alpha_0^{\kappa-1}+c_{1,\kappa}\alpha_0^{\kappa}\right)
\right. \nonumber \\
&&\hspace{3cm} \left.+ \alpha_s^2 \left( c_{2,0} + c_{2,1}\alpha_0 + c_{2,2}\alpha_0^2
+ \dots + c_{2,\kappa}\alpha_0^{\kappa}\right) + \dots c_{\kappa,0}\alpha_s^{\kappa} + \cdots\right],
\nonumber \\
\ea
where the coefficients $c_{ij}$ are defined in moment space.
The $\kappa$-th truncated solution of the equation above 
\beq
f_{N^m LO}(N,\alpha_s)\big|_{O(\alpha_s^{\kappa})}=f^{LO}(N,\alpha_s)
\left({\sum_{i,j=0}}^{i+j \leq \kappa} \alpha_s^i \alpha_0^j c_{i,j}\right),
\eeq
is therefore accurate to $O(\alpha_s^{\kappa})$, and clearly does not retain all the powers of the coupling constant which are, instead, part of (\ref{ltrunc}). However, as far as we are interested in an accurate solution of order $\kappa$, we can reobtain exactly the same expression from $x$-space using the ansatz

\beq
f_{N^mLO}(x,\alpha_s)\big|_{O(\alpha_s^{\kappa'})}=
\sum_{n=0}^\infty\left(A^0_n(x) + \alpha_s A^1_n(x)+ \alpha_s^2 A^2_n(x) + \dots +
\alpha_s^{\kappa'} A^{\kappa'}_n(x)\right)
\left[\ln{\left(\frac{\alpha_s(Q^2)}{\alpha_s(Q^2_0)}\right)}\right]^{n},
\label{truncatedseries}
\eeq
which can be correctly defined to be a {\em truncated solution of order $\kappa'$} of the $(\kappa)$-truncated equation.
Since we have truncated the evolution integral at order $\kappa$, this is also the maximum order at
which the truncated logarithmic expansion (\ref{truncatedseries}) coincides with the exact solution of the
full equation. This corresponds to the 
choice $\kappa'=\kappa$. Notice, however, that 
the number of coefficients $A_n^{\kappa'}$ that one introduces in the ansatz is unrelated to 
$\kappa$ and can be {\em larger} than this specific value. This implies that 
we obtain an improved accuracy as we let $\kappa'$ in (\ref{truncatedseries}) grow. 

If we choose the accuracy of the evolution integral to be $\kappa$, while 
sending the index $\kappa'$
in the logarithmic expansion of (\ref{truncatedseries}) 
to infinity, then the ansatz that accompanies this choice becomes 
\beq
f_{N^mLO}(N,\alpha_s)=\sum_{n=0}^\infty\left(\sum_{l=0}^{\infty}
\alpha_s^{l} A_n^{l}(x)\right)
\left[\ln{\left(\frac{\alpha_s(Q^2)}{\alpha_s(Q^2_0)}\right)}\right]^{n},
\label{truncatedseries1}
\eeq
and converges to the exact solution of the (order $\kappa$) truncated equation (\ref{ltrunc}). 
Obviously, this exact solution starts to differ from the exact
solution of the exact DGLAP equation at $O(\alpha_s^{\kappa+1})$. Also in this case, as before, one should notice that the double expansion in
$\alpha_s$ and $\alpha_0$ of the exact solution of the $(\kappa)$-truncated equation 
can be reobtained after using our exponentiation, and not before. We remark, if not obvious, that 
the recursion relations, in this case, 
need to be solved at the chosen order $\kappa'$, as widely shown in the examples discussed before.

We remark that, as done for the LO, we could also factorize the NLO solution and determine the $N^mLO$ solution using
the integral
\beq
I_{N^mLO}=\int_{\alpha_0}^{\alpha_s}d\alpha \left(\frac{P_{N^mLO}(x,\alpha)}{\beta_{N^mLO}(\alpha)} 
- \frac{{P}_{NLO}(\alpha)}{\beta_{NLO}(\alpha)}\right),
\label{Intevol2}
\eeq
and then restart the previous procedure. Obviously, the two approaches imply a resummation of the logarithmic behaviour
of the pdf's in the two cases. 

It is convenient to summarize what we have achieved up to now.
A truncation of the evolution integral introduces an approximation in the search for
solutions, which is controlled by the accuracy ($\kappa$)
in the expansion of the same integral. The
exact solution of the corresponding truncated equation, as we have seen from
the previous examples, involves all the powers of $\alpha_s$ and $\alpha_0$ and, obviously,
a further expansion around
the point $\alpha_s=\alpha_0=0$ is needed in order to identify a set of truncated solutions which can be
reobtained by a logarithmic ansatz. This is possible because of 
the property of analiticity of the solution. 
Therefore two types of truncations
are involved in the approximation of the solutions: 1) truncation of the equation and
2) truncation ($\kappa'$) of the corresponding solution. In the non-singlet
case, which is particularly simple, one can therefore identify a wide choice 
of solutions (by varying $\kappa$ and $\kappa'$) that retain
higher order effects in quite different fashions. Previous studies of the evolution 
using an ansatz a la Rossi-Storrow \cite{Rossi,Storrow}, borrowed from the pdf's of the photon, 
were therefore quite limited in accuracy. Our generalized procedure is the logical step forward 
in order to equate the accuracy of solutions obtained in moment space to those in $x$-space, without having to rely on purely numerical brute force methods.

 \footnote{In \cite{Vogt3} a flag variable called
IMODEV allows to switch among the exact solution (IMODEV=1), the exact solution of
the $O(\alpha_s^2)$ truncated equation (IMODEV=2). A third option (IMODEV=0)
involves at NLO and at NNLO $O(\alpha_s)$ and $O(\alpha_s^2)$, respectively, truncated ansatz that,
in our cases, are reconstructed logarithmically.}

\subsection{Recursion relations beyond NNLO and for all $\kappa$'s}

In the actual numerical implementations, if we intend to use a generic truncate of the non-singlet equation
(the result is actually true also for the singlet), it is convenient
to work with implementations of the
recursion relations that are valid at any order. In fact it is not practical to rederive them at each new
order $\kappa'$ of approximation. Next we are going to show how to do it, identifying generic
relations that are easier to implement numerically.

The expression to all orders of the DGLAP kernels is given by
\ba
&&P(N,\alpha_s)=\sum_{l=0}^{\infty}\left(\frac{\alpha_s}{2\pi}\right)^{l+1} P^{(l)}(N)
\nonumber\\
&&\frac{\partial \alpha_s}{\partial \ln{Q^2}}=-\sum_{i=0}^{\infty}\alpha_s^{i+2}
\frac{\beta_i}{(4\pi)^{i+1}},
\ea
and the equation in Mellin-space in Melin space is given by
\ba
\label{tot1}
\frac{\partial f(N,\alpha_s)}{\partial \alpha_s}=-\frac{
\sum_{l=0}^{\infty}\left(\frac{\alpha_s}{2\pi}\right)^{l+1} P^{(l)}(N)}
{\sum_{i=0}^{\infty}\alpha_s^{i+2}
\frac{\beta_i}{(4\pi)^{i+1}}}f(N,\alpha_s),
\ea
whose exact solution can be formally written as
\ba
f(N,\alpha_s)=\left(\frac{\alpha_s}{\alpha_0}\right)^{-\frac{2}{\beta_0}P^{(0)}}
e^{{\cal F}(\alpha_s,\alpha_0,P^{(0)},P^{(1)},P^{(2)},...,
\beta_0,\beta_1,...)}f(N,\alpha_0)\,,
\ea
where ${\cal F}$ is obtained from the evolution integral and whose specific form is not relevant 
at this point.

We will use the notation $(\vec{P},\vec{\beta})$ to indicate the sequence of 
components of the kernels and the coefficients of the $\beta$-function 
$(P^{(0)},P^{(1)},P^{(2)},...,\beta_0,\beta_1,...)$.

Then, the Taylor expansion around $\alpha_s=\alpha_0$ of the solution is formally given by 
\ba
e^{{\cal F}(\alpha_s,\alpha_0,\vec{P},\vec{\beta})}=\sum_{n=0}^{\infty}
\Phi_n(\partial{\cal F},\partial^2{\cal F},...,\partial^n{\cal F})
|_{\alpha_s=\alpha_0}(\alpha_s-\alpha_0)^n\frac{1}{n!}
\ea
for an appropriate $\Phi_n(\partial{\cal F},\partial^2{\cal F},...,\partial^n{\cal F})$.
$\Phi_n$ is a function that depends over all the partial derivative obtained
by the Taylor expansion. Since it is
calculated for $\alpha_s=\alpha_0$, it has a parametric dependence only on $\alpha_0$,
and we can perform a further expansion around the value $\alpha_0=0$ obtaining
\ba
&&e^{{\cal F}(\alpha_s,\alpha_0,\vec{P},\vec{\beta})}=
\sum_{m=0}^{\infty}
\frac{\alpha_0^{m}}{m!}
\frac{{\partial}^m}{{\partial \alpha_0}^m}
\left[\sum_{n=0}^{\infty}
\Phi_n(\alpha_0,\vec{P},\vec{\beta})\left(\alpha_s-\alpha_0\right)^n
\frac{1}{n!}\right]_{\alpha_0=0}.\,
\ea
 This expression can always be arranged and simplified as follows
\ba
&&e^{{\cal F}(\alpha_s,\alpha_0,\vec{P},\vec{\beta})}=
\sum_{n=0}^{\infty}\left[\frac{\alpha_s^{n}}{n!}\Phi_n(0,\vec{P},\vec{\beta})
+\left(-n\frac{\alpha_s^{n-1}}{n!}\Phi_n(0,\vec{P},\vec{\beta})
+\frac{\alpha_s^{n}}{n!}\partial
\Phi_n(\alpha_0,\vec{P},\vec{\beta})\bigg|_{\alpha_0=0}\right)\alpha_0
+\right.\nonumber\\
&&\hspace{3cm}\left.\left(\frac{1}{2}\frac{(n-1)n\alpha_s^{n-2}}{n!}
\Phi_n(0,\vec{P},\vec{\beta})-n\frac{\alpha_s^{n-1}}{n!}
\partial\Phi_n(\alpha_0,\vec{P},\vec{\beta})\bigg|_{\alpha_0=0}
\right.\right.\nonumber\\
&&\hspace{3cm}\left.\left.
+\frac{1}{2}\alpha_s^n\partial^2\Phi_n(\alpha_0,\vec{P},\vec{\beta})
\bigg|_{\alpha_0=0}\right)\alpha_0^2+(\dots)\alpha_0^3+\dots\right]
\nonumber\\
&&\hspace{2cm}=\left[
\left(1+\alpha_s{\cal \xi}_{1}^{(0)}(\vec{P},\vec{\beta})+\dots
+\alpha_s^n{\cal \xi}_{n}^{(0)}(\vec{P},\vec{\beta})\right)
\right.\nonumber\\
&&\hspace{3cm}\left.
+\alpha_0\left({\cal \xi}_{0}^{(1)}(\vec{P},\vec{\beta})
+\alpha_s{\cal \xi}_{1}^{(1)}(\vec{P},\vec{\beta})
+\alpha_s^2{\cal \xi}_{2}^{(1)}(\vec{P},\vec{\beta})+\dots
+\alpha_s^n{\cal \xi}_{n}^{(1)}(\vec{P},\vec{\beta})\right)
\right.\nonumber\\
&&\hspace{3cm}\left.
+\alpha_0^2\left({\cal \xi}_{0}^{(2)}(\vec{P},\vec{\beta})
+\alpha_s{\cal \xi}_{1}^{(2)}(\vec{P},\vec{\beta})
+\alpha_s^2{\cal \xi}_{2}^{(2)}(\vec{P},\vec{\beta})+\dots
+\alpha_s^n{\cal \xi}_{n}^{(2)}(\vec{P},\vec{\beta})\right)
\right.\nonumber\\
&&\hspace{3cm}\left.
\vdots
\right.\nonumber\\
&&\hspace{3cm}\left.
+\alpha_0^m\left({\cal \xi}_{0}^{(m)}(\vec{P},\vec{\beta})
+\alpha_s{\cal \xi}_{1}^{(m)}(\vec{P},\vec{\beta})
+\alpha_s^2{\cal \xi}_{2}^{(m)}(\vec{P},\vec{\beta})+\dots+
\alpha_s^n{\cal \xi}_{n}^{(m)}(\vec{P},\vec{\beta})\right)
\right]\nonumber\\
&&\hspace{2cm}=\sum_{n=0}^{\infty}\sum_{m=0}^{\infty}
\alpha_0^{m}\alpha_s^{n}{\cal \xi}_{n}^{(m)}(\vec{P},\vec{\beta}),
\nonumber\\
\ea
where we are formally absorbing all the dependence on both the kernels $\vec{P}$
and the $\beta$-function $\vec{\beta}$, coming from
the functions $\partial^m\Phi_n$ calculated at the point $\alpha_0=0$,
in the coefficients ${\cal \xi}_{n}^{(m)}$. 
Finally, we can reorganize the solution to all orders as

\ba
\label{mar1}
f(N,\alpha_s)=\left(\frac{\alpha_s}{\alpha_0}\right)^{-\frac{2}{\beta_0}P^{(0)}}
f(N,\alpha_0)\sum_{n=0}^{\infty}\sum_{m=0}^{\infty}
\alpha_0^m\alpha_s^{n}{\cal \xi}_{n}^{(m)}(\vec{P},\vec{\beta}).
\ea
With the help of the general notation
\ba
&&\vec{P}_{NLO}=(P^{(0)},P^{(1)}),\nonumber\\
&&\vec{P}_{NNLO}=(P^{(0)},P^{(1)},P^{(2)}),\nonumber\\
&&\vec{\beta}_{NLO}=(\beta_{0},\beta_{1}),\nonumber\\
&&\vec{\beta}_{NNLO}=(\beta_{0},\beta_{1},\beta_{2})\,,
\ea
we can try to indentify, by a formal reasoning, the exact solution up to a fixed - but generic -
perturbative order of the expansion of the
kernels.

To obtain the NLO/NNLO exact solution it is sufficient to
take as null the components $(P^{(2)},P^{(3)},...)$ and
$(\beta_2,\beta_3,...)$ for the NLO and $(P^{(3)},...)$ and
$(\beta_3,...)$ for the NNLO case.
Since eq.(\ref{mar1}) contains all the powers of $\alpha_0 \alpha_s$ up to
$\alpha_0^m \alpha_s^n$ (i.e. it is a polynomial expression of order
$\alpha^{n+m}$), if we aim at an accuracy of order $\alpha_s^{\kappa}$,
we have to arrange the exact expanded solution as
\ba
\label{mar2}
f(N,\alpha_s)=\left(\frac{\alpha_s}{\alpha_0}\right)^{-\frac{2}{\beta_0}P^{(0)}}
f(N,\alpha_0)\sum_{n=0}^{\kappa}\sum_{j=0}^{\infty}\alpha_0^j
\alpha_s^{n-j} {\cal \xi}_{n}^{(j)}(\vec{P},\vec{\beta})+ O(\alpha_s^{\kappa}),
\ea
with $O(\alpha_s^{\kappa})$ indicating all the
higher-order terms containing powers of the type
$\alpha_0\alpha_s^{\kappa}+\dots+\alpha_0^{\kappa}\alpha_s^{\kappa}$.
Hence, the index $\kappa$ represents the order at which we truncate the
solution.

As a natural generalization of the cases discussed in the previous sections, we introduce
the higher-order ansatz ($\kappa$-truncated solution)
\ba
\tilde{f}(N,\alpha_s)=\sum_{n=0}^{\infty}\left[\sum_{m=0}^{\kappa}
\frac{O_{n}^m(N)}{n!}\alpha_s^m \right]\log^n\left(\frac{\alpha_s}{\alpha_0}\right)
\label{higher}
\ea
that reproduces the exact solution (\ref{mar2}) expanded at order $\kappa$.

In fact, inserting this last ansatz in (\ref{tot1}) we generate a generic 
chain of recursion relations of the form

\ba
\label{chain1}
&&O_{n+1}^0 (N)=F^0 (O_n^0(P^{(0)},\beta_0)),\nonumber\\
&&O_{n+1}^1 (N)=F^1 (O_n^0,O_{n+1}^0,O_n^1,\vec{P},\vec{\beta}),\nonumber\\
&&O_{n+1}^2 (N)=F^2 (O_n^0,O_n^1,O_n^2,O_{n+1}^0,O_{n+1}^1,\vec{P},\vec{\beta}),\nonumber\\
&&\vdots\nonumber\\
&&O_{n+1}^{\kappa} (N)=F^{\kappa}(O_n^0,...,O_n^{\kappa},O_{n+1}^0,...,
O_{n+1}^{\kappa-1},\vec{P},\vec{\beta})\,.
\ea

A deeper look at the explicit structure of (\ref{chain1}),
for the NLO non-singlet case, reveals the following structures for the generic iterates

\ba
\label{genrecnlo}
&&O_{n+1}^{0}(N)=-\frac{2}{\beta_{0}}\left[P^{(0)}(N)\, O_{n}^{0}(N)\right],
\nonumber\\
&&O_{n+1}^{\kappa}(N)= -\frac{2}{\beta_{0}}\left[P^{(0)}\, O_{n}^{\kappa}\right](N)
-\frac{1}{\pi\beta_{0}}\left[P^{(1)}(N)\, O_{n}^{\kappa-1}(N)\right]
\nonumber \\
&&\hspace{2cm}-\frac{\beta_{1}}{4\pi\beta_{0}}
O_{n+1}^{\kappa-1}(N)-\kappa O_{n}^{\kappa}(N)-(\kappa-1)
\frac{\beta_{1}}{4\pi\beta_{0}} O_{n}^{\kappa-1}(N),\,
\ea
at NLO, while for the NNLO case we obtain
\ba
\label{genrecnnlo}
&&O_{n+1}^{0}(N)=-\frac{2}{\beta_{0}}\left[P^{(0)}(N)\, O_{n}^{0}(N)\right],
\nonumber\\
&& O_{n+1}^{1}(N)= -\frac{2}{\beta_{0}}\left[P^{(0)}(N)\, O_{n}^{1}(N)\right]
-\frac{1}{\pi\beta_{0}}\left[P^{(1)}(N)\, O_{n}^{0}(N)\right]\nonumber \\
&&\hspace{2cm} -\frac{\beta_{1}}{4\pi\beta_{0}}O_{n+1}^{0}(N)-O_{n}^{1}(N),
\nonumber\\
&& O_{n+1}^{\kappa}(N) =  -\frac{2}{\beta_{0}}\left[P^{(0)}
(N)\, O_{n}^{\kappa}(N)\right]
-\frac{1}{\pi\beta_{0}}\left[P^{(1)}(N)\, O_{n}^{\kappa-1}(N)\right]\nonumber \\
&&\hspace{2cm} -\frac{1}{2\pi^{2}\beta_{0}}
\left[P^{(2)}(N)\, O_{n}^{\kappa-2}(N)\right]\nonumber \\
&&\hspace{2cm} -\frac{\beta_{1}}{4\pi\beta_{0}}O_{n+1}^{\kappa-1}(N)
-\frac{\beta_{2}}{16\pi^{2}\beta_{0}} O_{n+1}^{\kappa-2}(N)\nonumber \\
&&\hspace{2cm} -\kappa O_{n}^{\kappa}(N)-(\kappa-1)
\frac{\beta_{1}}{4\pi\beta_{0}} O_{n}^{\kappa-1}(N)-(\kappa-2)
\frac{\beta_{2}}{16\pi^{2}\beta_{0}} O_{n}^{\kappa-2}(N).
\ea
Hence, one is able to determine the structure of the $\kappa$-$th$
recursion relation when the $\kappa=0$ and $\kappa=1$ cases are known.
This property is very useful from the computational point of view.
\footnote{The relations (\ref{genrecnlo}) and (\ref{genrecnnlo}) hold
also in the NLO/NNLO singlet case and can be generalized to any
perturbative order in the expansion of the kernels.}

Passing to the resolution of the recursion relations in moment space, we get the
formal expansion of each $O_{n}^{\kappa}(N)$
in terms of the initial condition $O_{0}^0(N)$, which reads
\ba
\tilde{f}(N,\alpha_0)=O_{0}^0(N)+\slash{0}\alpha_0+\slash{0}\alpha_0^2+...
+\slash{0}\alpha_0^{\kappa}\,.
\ea
Here we have set to zero all the higher order terms, as a natural 
generalization of $B_0=C_0=0$..., according to what has been discussed above.

These relations can be solved as we have shown in previous examples,
and the generic structure of their solution can be identified.
If we define $R_0=-\frac{2}{\beta_0}P^{(0)}$, the expressions of all the
$O_{n}^{\kappa}(N)$ in terms of $\tilde{f}(N,\alpha_0)\equiv\tilde{f}_0$ become
\ba
&&O_{n}^0 (N)=R_0^n\tilde{f}_0\nonumber\\
&&O_{n}^1 (N)=G^1 (R_0^n,(R_0 -1)^n,\vec{P},\vec{\beta})
\tilde{f}_0,\nonumber\\
&&O_{n}^2 (N)=G^2 (R_0^n,(R_0 -1)^n,(R_0 -2)^n,\vec{P},\vec{\beta})
\tilde{f}_0,\nonumber\\
&&\vdots\nonumber\\
&&O_{n}^{\kappa} (N)=G^{\kappa} (R_0^n,(R_0 -1)^n,(R_0 -2)^n,...,(R_0 -{\kappa})^n,\vec{P},\vec{\beta})
\tilde{f}_0\,,
\ea
and in particular, by an explicit calculation of $O_{n}^{\kappa}(N)$, one can work out
the structure of these special functions $G^m$. For instance, we get for $m=\kappa$ the expression
\ba
\label{Gform}
G^{\kappa} (R_0^n,(R_0 -1)^n,(R_0 -2)^n,...,(R_0 -{\kappa})^n,\vec{P},\vec{\beta})=\sum_{j=0}^{\kappa}
(R_0 -j)^n \xi_{\kappa}^{(j)}(\vec{P},\vec{\beta})\,,
\ea
for suitable coefficients $\xi_{\kappa}^{(j)}$.
Substituting the $O_{n}^{m}(N)$ functions with $m=0,\dots\kappa$  in the higher-order ansatz
(\ref{higher}) and performing our exponentiation, we get an expression of the form
\ba
&&\tilde{f}(N,\alpha_s)=G^{0}\Bigg(\left(\frac{\alpha_s}{\alpha_0}\right)^{R_0}\Bigg)
+\alpha_s \,G^{1}\Bigg(\left(\frac{\alpha_s}{\alpha_0}\right)^{R_0},
\left(\frac{\alpha_s}{\alpha_0}\right)^{R_0-1}\Bigg)
\nonumber\\
&&\hspace{2cm}
+\alpha_s^2 \,G^{2}\Bigg(\left(\frac{\alpha_s}{\alpha_0}\right)^{R_0},
\left(\frac{\alpha_s}{\alpha_0}\right)^{R_0-1},\left(\frac{\alpha_s}{\alpha_0}\right)^{R_0-2}\Bigg)
\nonumber\\
&&\hspace{2cm}
+\dots \alpha_s^{\kappa} \,G^{\kappa}\Bigg(\left(\frac{\alpha_s}{\alpha_0}\right)^{R_0},
\left(\frac{\alpha_s}{\alpha_0}\right)^{R_0-1},\left(\frac{\alpha_s}{\alpha_0}\right)^{R_0-2},\dots,
\left(\frac{\alpha_s}{\alpha_0}\right)^{R_0-\kappa}\Bigg)\,,
\nonumber\\
\ea
which can be written as follows by the use of eq.~(\ref{Gform})
\ba
\tilde{f}(N,\alpha_s)=\left(\frac{\alpha_s}{\alpha_0}\right)^{R_0}\tilde{f}(N,\alpha_0)
\sum_{m=0}^{\kappa}\sum_{j=0}^{m} \alpha_s^{m-j}\alpha_0^{j}~\xi_{m}^{(j)}(\vec{P},\vec{\beta})\,.
\ea
This is the exact solution expanded up to $O(\alpha_s^\kappa)$ in accuracy.

\section{The Search for the exact non-singlet NLO solution}
We have shown in the previous sections how to construct exact solutions of truncated equations
using logarithmic expansions. We have also shown the equivalence of these approaches with the Mellin method, since the recursion relations for the unknown
coefficient functions
of the expansions can be solved to all orders and so reproduce the solution in Mellin space
of the truncated equation. The question that we want to address in this section is whether we
can search for exact solutions of the exact (untruncated) equations as well.
These solutions are known exactly in the non singlet case up to NLO. It is not difficult also to obtain the
exact NNLO solution in Mellin space, and we will reconstruct the same solutions 
using modified recursion relations. 
The expansions that we will be using at NLO are logarithmic and solve the untruncated equation. 
The NNLO case, instead, will be treated in a following section, where, again, we will use 
recursion relations to build the exact solution but with a non-logarithmic ansatz \footnote{In PEGASUS 
\cite{Vogt3} the NNLO non-singlet solution is built by truncation in Mellin space of the evolution equation,
while the NLO solution is implemented as an exact solution.}.

The exact NLO non-singlet solution has been given in eq.~(\ref{exactsol}).
The identification of an expansion that allows to reconstruct in moment space eq.~(\ref{exactsol}) follows quite naturally once the typical
properties of the convolution product $\otimes$ are identified.
For this purpose we define the serie of convolution products
\beqa
e^{F_A P_A(x)\otimes} &\equiv& \sum_{n=0}^\infty \frac{F_A^n}{n!}\left(P_A(x)\otimes\right)^n
\ea
that acts on a given initial function as
\beqa
e^{F_A P_A(x)\otimes}\phi(x)&=&
\left( \delta(1-x)\otimes + F_A P_A(x)\otimes + \frac{1}{2!}F_A^2 P_A\otimes P_A\otimes
+ \dots\right)\phi(x) \nonumber \\
&=&
\phi(x) +  F_A \left(P_A \otimes \phi\right) + \frac{1}{2!}F_A^2 \left(P_A\otimes P_A\otimes \right)\phi(x)
\,+ \dots.
\eeqa

The functions $F_A$ and $F_B$ are parametrically dependent on any other variable except the variable $x$.
The proof of the associativity, distributivity and commutativity of the $\otimes$ product is easily obtained after mapping these products in Mellin space. For instance, for generic functions $a(x)$ $b(x)$ and $c(x)$ for which the $\otimes$ product is a regular function one has
\beqa
\mathcal{M}\left[\left(a\otimes b\right)\otimes c\right](N)&=&\mathcal{M}\left[ a\otimes \left(b\otimes c\right)\right](N) \nonumber \\
&=& a(N) B(N) C(N),
 \eeqa
where $\mathcal{M}$ denotes the Mellin transform and $N$ is the moment variable.
Also one obtains
\beq
e^{F_A P_A(x)\otimes}\,e^{F_B P_B(x)\otimes}\,\phi(x)=e^{\left(F_A P_A(x) + F_B P_B(x)\right)\otimes}\,\phi(x),
\label{expo}
\eeq
and
\beq
\mathcal{M}\left[e^{M a(x)\otimes}\,\phi\right](N)=e^{M a(N)}\phi(N),
\eeq
with $M$ $x$-independent,
since both left-hand-side and right-hand-side of (\ref{expo}) can be mapped to the same function in Mellin space.
Notice that the role of the identity in $\otimes$-space is taken by the function $\delta(1-x)$. We will also use the notation
\beq
\left( \sum_{n=0}^\infty {A'}_n(x) F_A^n\right)_\otimes \phi(x)\equiv \left( A_0(x)\otimes + F_A A_1(x)\otimes + \dots\right)\phi(x),
\eeq
where the ${A'}_n(x)$ and the $A_n(x)$ capture the operatorial and the functional expansion 
- respectively - and are trivially related
\beq
A'_n(x)\otimes \phi(x)= A_n(x).
\eeq
To identify the $x$-space ansatz we rewrite (\ref{exactsol}) as

\beqa
f(N,\alpha)& = & f(N,\alpha_{0})e^{a(N)L}e^{b(N)M}\nonumber \\
& = & f(N,\alpha_{0})\left(\sum_{n=0}^{\infty}
\frac{a(N)^{n}}{n!}L^{n}\right)\left(\sum_{m=0}^{\infty}
\frac{b(N)^{m}}{m!}M^{m}\right),
\label{labelE}
\end{eqnarray}
where we have introduced the notations

\begin{eqnarray}
L & = & \log\frac{\alpha_s}{\alpha_{0}},\nonumber \\
M & = & \log\frac{4\pi\beta_{0}+\alpha_s\beta_{1}}{4\pi\beta_{0}+\alpha_{0}\beta_{1}},\nonumber \\
a(N) & = & -\frac{2P^{(0)}(N)}{\beta_{0}},\nonumber \\
b(N) & = & \frac{2P^{(0)}(N)}{\beta_{0}}-\frac{4P^{(1)}(N)}{\beta_{1}}.
\label{labelA}
\end{eqnarray}
Our ansatz for the exact solution in $x$-space is chosen of the form
\begin{eqnarray}
f(x,Q^{2}) & = & \left(\sum_{n=0}^{\infty}\frac{A'_{n}(x)}{n!}L^{n}\right)_\otimes
\left(\sum_{m=0}^{\infty}\frac{B'_{m}(x)}{m!}M^{m}\right)_\otimes f(x,Q_0^2)\nonumber \\
& = & \sum_{s=0}^{\infty}\sum_{n=0}^{s}L^{n}M^{s-n} \frac{A'_{n}(x)\otimes B'_{s-n}(x)}{n!(s-n)!}
\otimes f(x,Q_0^2)\nonumber \\
& = & \sum_{s=0}^{\infty}\sum_{n=0}^{s}\frac{C_{n}^{s}(x)}{n!(s-n)!}L^{n}M^{s-n},
\label{eq:NLOansatz}
\end{eqnarray}
where in the first step we have turned the product of two series into
a single series of a combined exponent $s=n+m$, and in the last step
we have introduced the functions
\begin{equation}
C_{n}^{s}(x)=A'_{n}(x)\otimes B'_{s-n}(x)\otimes f(x,Q_0^2),\qquad(n\leq s).
\end{equation}
Setting $Q=Q_{0}$ in (\ref{eq:NLOansatz}) we get the initial condition $A'_0(x)=B'_0(x)=\delta(1-x)$ or,
equivalently,
\begin{equation}
f(x,Q_{0}^{2})=C_{0}^{0}(x).
\end{equation}
Inserting the ansatz (\ref{eq:NLOansatz}) into the NLO DGLAP equation, with the expressions of
the kernel and beta function included at the corresponding order, we obtain the identity
\begin{eqnarray}
\sum_{s=0}^{\infty}\sum_{n=0}^{s}\left\{\left(-\frac{\beta_{0}}{4\pi}
\alpha-\frac{\beta_{1}}{16\pi^{2}}\alpha^{2}\right)C_{n+1}^{s+1}-
\frac{\beta_{1}}{16\pi^{2}}\alpha^{2}C_{n}^{s+1}\right\}
\frac{L^{n}M^{s-n}}{n!(s-n)!}\nonumber \\
=\sum_{s=0}^{\infty}\sum_{n=0}^{s}\left\{ \frac{\alpha}{2\pi}P^{(0)}
\otimes C_{n}^{s}+\frac{\alpha^{2}}{4\pi^{2}}P^{(1)}\otimes C_{n}^{s}\right\}
\frac{L^{n}M^{s-n}}{n!(s-n)!}.
\end{eqnarray}
Equating term by term the coefficients of $\alpha$ and
$\alpha^{2}$, we find from this identity the new exact recursion relations
\begin{eqnarray}
C_{n+1}^{s+1} & = & -\frac{2}{\beta_{0}}P^{(0)}\otimes C_{n}^{s},\\
C_{n}^{s+1} & = & -C_{n+1}^{s+1}-\frac{4}{\beta_{1}}P^{(1)}\otimes C_{n}^{s},
\end{eqnarray}
or, equivalently,  
\begin{eqnarray}
C_{n}^{s} & = & -\frac{2}{\beta_{0}}P^{(0)}\otimes C_{n-1}^{s-1}
\label{eq:NLO_rec_diagonale},\\
C_{n}^{s} & = & -C_{n+1}^{s}-\frac{4}{\beta_{1}}P^{(1)}\otimes C_{n}^{s-1}.
\label{eq:NLO_rec_verticale}
\end{eqnarray}
Notice that although the coefficients $C_n^s$ are convolution products of two functions,
the recursion relations do not let these products appear explicitely.
These relations just written down allow to compute all the coefficients $C_{n}^{s}$
$(n\leq s)$ up to a chosen $s$ starting from $C_{0}^{0}$, which is
given by the initial conditions. In particular eq.~(\ref{eq:NLO_rec_diagonale}) allows us to move along the diagonal
arrow according to the diagram reproduced in Table \ref{cap:schemaNLO};
eq.~(\ref{eq:NLO_rec_verticale}) instead allows us to compute a coefficient
in the table once we know the coefficients at its right and the coefficient
above it (horizontal and vertical arrows).
To compute $C_{n}^{s}$ there is a certain freedom, as illustrated
in the diagram.  For instance, to determine $C_{s}^{s}$
we can only use (\ref{eq:NLO_rec_diagonale}),
and for the coefficients $C_{0}^{s}$ we can only use (\ref{eq:NLO_rec_verticale}).
For all the other coefficients one can prove that using (\ref{eq:NLO_rec_diagonale})
or (\ref{eq:NLO_rec_verticale}) brings to the same determination of the coefficients, 
and in our numerical studies we
have chosen to implement (\ref{eq:NLO_rec_diagonale}), being this relation 
less time consuming since it involves $P^{(0)}$ instead of $P^{(1)}$.

The recursion relations defining the iterated solution can be solved as follows.
From the first relation (\ref{eq:NLO_rec_diagonale}), keeping the $s$-index fixed, we have
\ba
\label{exasolve1}
&&C_{n}^{s} = -\frac{2}{\beta_{0}}P^{(0)}\otimes C_{n-1}^{s-1}\Rightarrow
\nonumber\\
&&C_{n}^{s} = \left[-\frac{2}{\beta_{0}}P^{(0)}\right]^{n}\otimes C_{0}^{s-n-1}\,,
\ea
then, since the second relation (\ref{eq:NLO_rec_verticale}) also holds for $n=0$,
we can write (using eq. (\ref{exasolve1}))
\ba
&&C_{n}^{s}=-C_{n+1}^{s}-\frac{4}{\beta_{1}}P^{(1)}\otimes C_{n}^{s-1} \Rightarrow
\nonumber\\
&&C_{0}^{s}=\left[\frac{2}{\beta_0}P^{(0)}-\frac{4}{\beta_{1}}P^{(1)}\right]
\otimes C_{0}^{s-1}\Rightarrow
\nonumber\\
&&C_{0}^{s}=\left[\frac{2}{\beta_0}P^{(0)}-\frac{4}{\beta_{1}}P^{(1)}\right]^{s}
\otimes C_{0}^{0}.
\ea
Finally, inserting the above relation in (\ref{exasolve1}) we can write
\ba
C_{n}^{s} =\left[\frac{2}{\beta_0}P^{(0)}\right]^{n}\otimes
\left[\frac{2}{\beta_0}P^{(0)}-\frac{4}{\beta_{1}}P^{(1)}\right]^{s-n}
\otimes C_{0}^{0}\,,
\ea
which is the solution we have been searching for. The last step in the proof consists in
taking the Mellin transform of this operatorial solution and summing the corresponding series

\ba
f(N,\alpha_s)&=&\sum_{s=0}^{\infty}\sum_{n=0}^{s}\frac{C_{n}^{s}(N)}{n!(s-n)!}L^{n}M^{s-n}
\nonumber\\
&=&\sum_{s=0}^{\infty}\sum_{n=0}^{s}\frac{L^{n}M^{s-n}}{n!(s-n)!}
\left[\frac{2}{\beta_0}P^{(0)}\right]^{n}
\left[\frac{2}{\beta_0}P^{(0)}-\frac{4}{\beta_{1}}P^{(1)}\right]^{s-n}
 C_{0}^{0}(N),\,
\ea
that after summation gives  
\ba
f(N,\alpha_s)=e^{-\frac{2}{\beta_0}P^{(0)}(N)\log\left(\frac{\alpha_s}{\alpha_0}\right)}
\exp\left\{\left[\frac{2}{\beta_0}P^{(0)}(N)-\frac{4}{\beta_{1}}P^{(1)}(N)\right]
\log{\left(\frac{4\pi\beta_0+\alpha_s\beta_1}{4\pi\beta_0+\alpha_0\beta_1}\right)}\right\}
C_{0}^{0}(N),
\nonumber\\
\ea
which is exactly the expression in eq. (\ref{labelE}).
Hence, it is obvious that the exact solution of the DGLAP 
equation (\ref{nontrunc}) can be written in $x$-space as
\ba
f(x,\alpha_s(Q^2))=e^{-\log\left(\frac{\alpha_s}{\alpha_0}\right)\frac{2}{\beta_0}P^{(0)}(x)\otimes}
e^{\log{\left(\frac{4\pi\beta_0+\alpha_s\beta_1}{4\pi\beta_0+\alpha_0\beta_1}\right)}
\left[\frac{2}{\beta_0}P^{(0)}(x)-\frac{4}{\beta_{1}}P^{(1)}(x)\right]\otimes}C_{0}^{0}(x),
\nonumber\\
\ea
therefore proving that the ansatz (\ref{eq:NLOansatz}) reproduces the exact solution of the
NLO DGLAP equation from $x$-space.

A second version of the same ansatz for the NLO exact solution can be built using a
factorization of the NLO DGLAP equation. This strategy is analogous to the method of
factorization for ordinary PDE's. For this purpose we define a modified LO DGLAP equation,
involving $\beta^{NLO}$

\begin{equation}
\frac{\partial \tilde{f}_{LO}(x,\alpha_{s})}{\partial\alpha_{s}}
=\left(\frac{\alpha_{s}}{2\pi \beta^{NLO}}\right)
P^{(0)}(x)\otimes \tilde{f}_{LO}(x,\alpha_{s}),
\end{equation}

whose solution is given by
\beqa
\tilde{f}_{LO}(x,\alpha)&=& e^{M \left(\frac{2 P^{(0)}}{\beta_0}\right)\otimes} f_{LO}(x,\alpha)
\nonumber \\
f_{LO}(x,\alpha)&=& e^{L \left(\frac{-2 P^{(0)}}{\beta_0}\right)\otimes} f(x,\alpha_0),
\eeqa
where we have introduced the ordinary LO solution $f_{LO}$, expressed in terms of a typical initial
condition $f(x,\alpha_0)$, and the NLO recursion relations can be obtained from the expansion
\beq
f_{NLO}(x,\alpha)=\left(\sum_{n=0}^{\infty}\frac{B_{n}(x)}{n!}M^{n}\right)_\otimes 
\tilde{f}_{LO}(x,\alpha).
\label{nlosolve}
\eeq

Inserting this relation into (\ref{nontrunc}) we obtain the recursion relations
\beqa
B_{n+1}&=& \left(-\frac{4}{\beta_1} P^{(1)}\right)\otimes B_n \nonumber \\
B_0(x) &=&\delta(1-x)\,,
\eeqa
which is solved in moment space by
\beqa
B_n(N)&=&\left(-\frac{4}{\beta_1} P^{(1)}\right)^n B_0(N)\nonumber \\
B_0(N)&=&1.
\eeqa
The solution eq.~(\ref{nlosolve}) can be re-expressed in the form

\beqa
f_{NLO}(x,\alpha) &=& e^{M \left(-\frac{1}{4 \beta_1} P^{(1)}\right)\otimes}e^{M \left(\frac{2 P^{(0)}}{\beta_0}\right)\otimes}e^{L \left(\frac{-2 P^{(0)}}{\beta_0}\right)\otimes} f(x,\alpha_0) \nonumber \\
&=& e^{M \left(-\frac{1}{4 \beta_1} P^{(1)} +\frac{2 P^{(0)}}{\beta_0}\right)\otimes}e^{L \left(\frac{-2 P^{(0)}}{\beta_0}\right)\otimes} f(x,\alpha_0)
\eeqa
which agrees with (\ref{labelE}) once $a(N)$ and $b(N)$ have been defined as in (\ref{labelA}).

We have therefore proved that the exact NLO
solution of the DGLAP equation can be described by an exact ansatz. Since the ansatz is built
by inspection, it is obvious that one needs to know the solution in moment space in order
to reconstruct the coefficients. Though the recursive scheme used to construct the solution
in $x$-space is more complex compared to the recursion relations for the truncated solution,
its numerical implementation is still very stable and very precise, reaching the same level of
accuracy of the traditional methods based on the inversion of the Mellin
moments.

\begin{table}
\begin{center}$\begin{array}{ccccccccc}
C_{0}^{0}\\
\downarrow & \searrow\\
C_{0}^{1} & \leftarrow & C_{1}^{1}\\
\downarrow & \searrow &  & \searrow\\
C_{0}^{2} & \leftarrow & C_{1}^{2} &  & C_{2}^{2}\\
\downarrow & \searrow &  & \searrow &  & \searrow\\
C_{0}^{3} & \leftarrow & C_{1}^{3} &  & C_{2}^{3} &  & C_{3}^{3}\\
\downarrow & \searrow &  & \searrow &  & \searrow &  & \searrow\\
\ldots &  & \ldots &  & \ldots &  & \ldots &  & \ldots\end{array}$\end{center}

\caption{Schematic representation of the procedure followed to compute each
coefficient $C_{n}^{s}$.\label{cap:schemaNLO}}
\end{table}

\section{Finding the exact non-singlet  NNLO solution}

To identify the NNLO exact solution we proceed similarly to the NLO case and start
from the DGLAP equation in moment space at the corresponding perturbative order

\beq
\frac{\partial f(N,\alpha_{s})}{\partial\alpha_{s}}=
-\frac{\left(\frac{\alpha_{s}}{2\pi}\right)P^{(0)}(N)+
\left(\frac{\alpha_{s}}{2\pi}\right)^{2}P^{(1)}(N)+
\left(\frac{\alpha_{s}}{2\pi}\right)^{3}P^{(2)}(N)}{\frac{\beta_{0}}{4\pi}
\alpha_{s}^{2}+\frac{\beta_{1}}{16\pi^{2}}\alpha_{s}^{3}+
\frac{\beta_{2}}{64\pi^{3}}\alpha_{s}^{4}}f(N,\alpha_{s}).
\label{eqnn}
\end{equation}
After a separation of variables, all the new logarithmic/non logarithmic and dependences
come from the integral
\beq
\int_{\alpha_s(Q_0^2)}^{\alpha_s(Q^2)} d\alpha \frac{P^{NNLO}(\alpha_s)}{\beta^{NNLO}(\alpha)},
\label{integer}
\eeq
and the solution of (\ref{eqnn}) is
\begin{eqnarray}
f(N,\alpha) & = & f(N,\alpha_{0})e^{a(N)\mathcal{L}}e^{b(N)\mathcal{M}}e^{c(N)\mathcal{Q}}\nonumber \\
& = & f(N,\alpha_{0})\left(\sum_{n=0}^{\infty}\frac{a(N)^{n}}{n!}\mathcal{L}^{n}\right)
\left(\sum_{m=0}^{\infty}\frac{b(N)^{m}}{m!}\mathcal{M}^{m}\right)\left(\sum_{p=0}^{\infty}
\frac{c(N)^{p}}{p!}\mathcal{Q}^{p}\right),
\label{nnlotest}
\end{eqnarray}
where we have defined
\begin{eqnarray}
\mathcal{L} & = & \log\frac{\alpha}{\alpha_{0}},\\
\mathcal{M} & = & \log\frac{16\pi^{2}\beta_{0}+4\pi\alpha\beta_{1}
+\alpha^{2}\beta_{2}}{16\pi^{2}\beta_{0}+4\pi\alpha_{0}\beta_{1}+\alpha_{0}^{2}\beta_{2}},\\
\mathcal{Q} & = & \frac{1}{\sqrt{4\beta_{0}\beta_{2}-\beta_{1}^{2}}}
\arctan\frac{2\pi(\alpha-\alpha_{0})\sqrt{4\beta_{0}\beta_{2}-
\beta_{1}^{2}}}{2\pi(8\pi\beta_{0}+(\alpha+\alpha_{0})\beta_{1})+
\alpha\alpha_{0}\beta_{2}},\\
a(N) & = & -\frac{2P^{(0)}(N)}{\beta_{0}},\\
b(N) & = & \frac{P^{(0)}(N)}{\beta_{0}}-\frac{4P^{(2)}(N)}{\beta_{2}},\\
c(N) & = & \frac{2\beta_{1}}{\beta_{0}}P^{(0)}(N)-8P^{(1)}(N)
+\frac{8\beta_{1}}{\beta_{2}}P^{(2)}(N).
\end{eqnarray}
Notice that for $n_{f}=6$ the solution has a branch point since
$4\beta_{0}\beta_{2}-\beta_{1}^{2}<0$. If we increase $n_f$ as we
step up in the factorization scale then, for $n_f=6$,
$\mathcal{Q}$ is replaced by its analytic continuation
\begin{equation}
\mathcal{Q}=\frac{1}{\sqrt{\beta_{1}^{2}-4\beta_{0}\beta_{2}}}
\textrm{arctanh}\frac{2\pi(\alpha-\alpha_{0})\sqrt{\beta_{1}^{2}
-4\beta_{0}\beta_{2}}}{2\pi(8\pi\beta_{0}+(\alpha+
\alpha_{0})\beta_{1})+\alpha\alpha_{0}\beta_{2}}.
\end{equation}
Eq.~(\ref{nnlotest}) incorporates all the nontrivial dependence on the coupling
constant $\alpha_s$ (now determined at 3-loop level)
into $\mathcal{L}, \mathcal{M}$ and $\mathcal{Q}$.

As a side remark we emphasize that it is also possible to obtain various  NNLO exact recursion
relations using the formalism of the convolution series introduced above.
For this purpose it is convenient to define suitable operatorial expressions, for instance
\beqa
\mathcal{E}_1 &\equiv& e^{\int_{\alpha_0}^{\alpha_s} d\alpha
 \frac{P^{NLO}(x,\alpha)}{\beta^{NNLO}(\alpha)}_\otimes}, \nonumber \\
\mathcal{E}_2 &\equiv& e^{\int_{\alpha_0}^{\alpha_s} d\alpha
 \left(\frac{\alpha}{2 \pi}\right)^3 \frac{P^{(2)}(x,\alpha)}{\beta^{NNLO}(\alpha)}_\otimes},
\label{seq1}
\ea

which are manipulated under the prescription that the integral in $\alpha$ is evaluated before that any
convolution product acts on the initial conditions. The re-arrangement of these operatorial
expressions is therefore quite simple and one can use simple identities such as
\beqa
J_0 &=&\int_{\alpha_0}^{\alpha_s} d\alpha
 \left(\frac{\alpha}{2 \pi}\right) \frac{P^{(0)}(x,\alpha)}{\beta^{NNLO}(\alpha)}_\otimes
= 2\frac{\beta_1}{\beta_0}\mathcal{Q}P^{(0)}_\otimes -\frac{2}{\beta_0}\mathcal{L} P^{(0)}_\otimes
+ \frac{1}{\beta_0}\mathcal{M}P^{(0)}_\otimes, \nonumber\\
J_1 &=&\int_{\alpha_0}^{\alpha_s} d\alpha
 \left(\frac{\alpha}{2 \pi}\right)^2 \frac{P^{(1)}(x,\alpha)}{\beta^{NNLO}(\alpha)}_\otimes
= - 8 \mathcal{Q} P^{(1)}\otimes, \nonumber \\
J_2&=&\int_{\alpha_0}^{\alpha_s} d\alpha
 \left(\frac{\alpha}{2 \pi}\right)^3 \frac{P^{(2)}(x,\alpha)}{\beta^{NNLO}(\alpha)}_\otimes =
\left(-\frac{4}{\beta_2}\mathcal{M} + 8 \frac{\beta_1}{\beta_2}\mathcal{Q}\right) P^{(2)}\otimes,
\nonumber \\
J_{NNLO}&=&\int_{\alpha_0}^{\alpha_s} d\alpha
 \frac{P^{NLO}(x,\alpha)}{\beta^{NNLO}(\alpha)}_\otimes \nonumber \\
&=&  \mathcal{Q}\left(2 \frac{\beta_1}{\beta_0} P^{(0)}\otimes - 8 P^{(1)}\otimes \right)
-\frac{2}{\beta_0}\mathcal{L}P^{(0)}\otimes + \frac{1}{\beta_0}\mathcal{M} P^{(0)}\otimes,
\nonumber \\
\ea
to build the NNLO exact solution using a suitable recursive algorithm. For instace, using
(\ref{seq1})
one can build an intermediate solution of the equation
\beq
\frac{\partial \tilde{f}_{NLO}(x,\alpha_s)}{\partial \alpha_s}=
\frac{P^{NLO}}{\beta^{NNLO}}\tilde{f}_{NLO}(x,\alpha_s)
\eeq
given by
\beqa
\tilde{f}_{NLO} &=& \mathcal{E}_1 f(x,\alpha_0),
\eeqa

and then constructs with a second recursion the exact solution
\beqa
f(x,\alpha_s) &=& \mathcal{E}_2 \tilde{f}_{NLO}.
\ea
A straightforward approach, however, remains the one described in the previous section, 
that we are going now to extend to NNLO.
In this case, in the choice of the recursion relations, 
one is bound to equate 3 independent logarithmic powers of 
$\mathcal{L}$, $\mathcal{M}$ and $\mathcal{Q}$ that appear in the symmetric ansazt

\begin{eqnarray}
f(x,Q^{2}) & = & \left(\sum_{n=0}^{\infty}\frac{A'_{n}(x)}{n!}\mathcal{L}^{n}\right)_\otimes
\left(\sum_{m=0}^{\infty}\frac{B'_{m}(x)}{m!}\mathcal{M}^{m}\right)_\otimes
\left(\sum_{p=0}^{\infty}\frac{C'_{p}(x)}{p!}\mathcal{Q}^{p}\right)_\otimes f(x,Q_0^2)\nonumber \\
& = & \sum_{s=0}^{\infty}\sum_{t=0}^{s}\sum_{n=0}^{t}\frac{A'_{n}(x)\otimes
B'_{t-n}(x)\otimes {C'_{s-t}(x)}}{n!(t-n)!(s-t)!}\otimes f(x,Q_0^2)\,\mathcal{L}^{n}\mathcal{M}^{t-n}\mathcal{Q}^{s-t}\nonumber \\
& = & \sum_{s=0}^{\infty}\sum_{t=0}^{s}\sum_{n=0}^{t}
\frac{ D_{t,n}^{s}(x)}{n!(t-n)!(s-t)!}\mathcal{L}^{n}\mathcal{M}^{t-n}\mathcal{Q}^{s-t},
\label{eq:NNLOansatz}
\end{eqnarray}
and where
\beq
D_{t,n}^{s}(x)= A'_{n}(x)\otimes
B'_{t-n}(x)\otimes C'_{s-t}(x)\otimes f(x,Q_0^2).
\eeq
The ansatz is clearly identified quite simply by inspection, once 
the structure of the solution in moment space (\ref{nnlotest}) is known explicitely.
In (\ref{eq:NNLOansatz}) we have at a first step re-arranged the product of the three series
into a single series with a given total exponent $s=n+m+p$, and we have
introduced an index $t=n+m$. The triple-indexed function $D_{t,n}^{s}(x)$ can be defined also as
an ordinary product
\begin{equation}
D_{t,n}^{s}(x)=A_{n}(x)\otimes B_{t-n}(x)\otimes C_{s-t}(x),\qquad(n\leq t\leq s),
\end{equation}
where we have absorbed the $\otimes$ operator into the definition of $A$, $B$ and $C$,
\beq
A(x) B(x) C(x)=A'(x)\otimes \bigg[\Big(B'(x)\otimes \Big(C'(x)\otimes f(x,Q_0^2)\Big)\Big)\bigg].
\eeq
Setting $Q=Q_{0}$ in (\ref{eq:NNLOansatz}) we get the initial condition
\begin{equation}
f(x,Q_{0}^{2})=D_{0,0}^{0}(x).
\end{equation}
Inserting the ansatz (\ref{eq:NNLOansatz}) into the 3-loop DGLAP equation
together with the beta function determined at the same order and equating
the coefficients of $\alpha$, $\alpha^{2}$ and $\alpha^{3}$, we
find the recursion relations satisfied by the unknown coefficients $D_{t,n}^{s}(x)$
\begin{eqnarray}
D_{t+1,n+1}^{s+1} & = & -\frac{2}{\beta_{0}}P^{(0)}\otimes D_{t,n}^{s},\\
D_{t+1,n}^{s+1} & = & -\frac{1}{2}D_{t+1,n+1}^{s+1}-
\frac{4}{\beta_{2}}P^{(2)}\otimes D_{t,n}^{s},\\
D_{t,n}^{s+1} & = & -2\beta_{1}\left(D_{t+1,n}^{s+1}+
D_{t+1,n+1}^{s+1}\right)-8P^{(1)}\otimes D_{t,n}^{s},
\end{eqnarray}
or equivalently
\begin{eqnarray}
D_{t,n}^{s} & = & -\frac{2}{\beta_{0}}P^{(0)}
\otimes D_{t-1,n-1}^{s-1}
\label{eq:NNLO_rec_diagonale},\\
D_{t,n}^{s} & = & -\frac{1}{2}D_{t,n+1}^{s}-
\frac{4}{\beta_{2}}P^{(2)}\otimes D_{t-1,n}^{s-1}
\label{eq:NNLO_rec_verticale},\\
D_{t,n}^{s} & = & -2\beta_{1}\left(D_{t+1,n}^{s}+
D_{t+1,n+1}^{s}\right)-8P^{(1)}\otimes D_{t,n}^{s-1}.
\label{eq:NNLO_rec_orizzontale}
\end{eqnarray}
In the computation of a given coefficient $D_{t,n}^{s}$, if more
than one recursion relation is allowed to determine that specific coefficient,
we will choose to implement the less time consuming path,
i.e.~in the order (\ref{eq:NNLO_rec_diagonale}),
(\ref{eq:NNLO_rec_orizzontale}) and (\ref{eq:NNLO_rec_verticale}).
At a fixed integer $s$ we proceed as follows: we

\begin{enumerate}
\item compute all the coefficients $D_{t,n}^{s}$ with $n\neq0$ using (\ref{eq:NNLO_rec_diagonale});
\item compute the coefficient $D_{s,0}^{s}$ using (\ref{eq:NNLO_rec_verticale});
\item compute the coefficient $D_{t,0}^{s}$ with $t\neq s$ using (\ref{eq:NNLO_rec_orizzontale}),
in decreasing order in $t$.
\end{enumerate}
This computational strategy is exemplified in the diagram in Table \ref{cap:schemaNNLO}
for $s=4$, where the various paths are highlighted.

\begin{table}
\begin{center}$\begin{array}{ccccccccc}
 &  &  &  &  &  &  &  & \underline{D_{4,4}^{4}}\\
\\ &  &  &  &  &  & \underline{D_{3,3}^{4}} &  & \underline{D_{4,3}^{4}}\\
\\ &  &  &  & \underline{D_{2,2}^{4}} &  & \underline{D_{3,2}^{4}} &  &
\underline{D_{4,2}^{4}}\\
\\ &  & \underline{D_{1,1}^{4}} &  & \underline{D_{2,1}^{4}} &  &
\underline{D_{3,1}^{4}} &  & \underline{D_{4,1}^{4}}\\
 & \swarrow &  & \swarrow &  & \swarrow &  & \swarrow & \downarrow\\
D_{0,0}^{4} & \longleftarrow & D_{1,0}^{4} & \longleftarrow & D_{2,0}^{4} &
\longleftarrow & D_{3,0}^{4} & \longleftarrow & D_{4,0}^{4}\end{array}$
\end{center}

\caption{Schematic representation of the procedure followed to compute each
coefficient $D_{t,n}^{s}$ for $s=4$. The underlined coefficients are
computed via eq.~(\ref{eq:NNLO_rec_diagonale}).\label{cap:schemaNNLO}}
\end{table}

Following a procedure similar to the one used for the NLO case, we can solve
the recursion relations for the NNLO ansatz with the initial
conditions $D_{0,0}^{0}(x)$.
Solving the relations (\ref{eq:NNLO_rec_diagonale}-\ref{eq:NNLO_rec_orizzontale}), 
we obtain the chain conditions

\ba
&&D_{t,n}^{s}=-\frac{2}{\beta_{0}}P^{(0)}
\otimes D_{t-1,n-1}^{s-1}\Rightarrow
\nonumber\\
&&D_{t,n}^{s}= \left[-\frac{2}{\beta_{0}}P^{(0)}\right]^{n}\otimes D_{t-n,0}^{s-n}.
\ea
Then, from the second relation we get the additional ones
\ba
&&D_{t,n}^{s}=-\frac{1}{2}D_{t,n+1}^{s}-
\frac{4}{\beta_{2}}P^{(2)}\otimes D_{t-1,n}^{s-1}\Rightarrow
\nonumber\\
&&D_{t-n,0}^{s-n}=\left[\frac{P^{(0)}}{\beta_{0}}
-\frac{4P^{(2)}}{\beta_2}\right]^{t-n}\otimes D_{0,0}^{s-t-2n}.
\ea
From the last relation we also obtain the relations 
\ba
&&D_{t,n}^{s}=-2\beta_{1}\left(D_{t+1,n}^{s}+
D_{t+1,n+1}^{s}\right)-8P^{(1)}\otimes D_{t,n}^{s-1}\Rightarrow
\nonumber\\
&&D_{0,0}^{s-t-2n}=\left[-8 P^{(1)}+\frac{2\beta_1}{\beta_0}P^{(0)}
+\frac{8\beta_1}{\beta_2}P^{(2)}\right]\otimes D_{0,0}^{0}\,,
\ea
which solve the recursion relations in x-space in terms of the initial condition $D_{0,0}^{0}$.
Finally, the explicit expression of the $D_{t,n}^{s}$ coefficient
will be given by 
\ba
D_{t,n}^{s}(x)=\left[-\frac{2}{\beta_0}P^{(0)}\right]^{n}
\otimes\left[\frac{P^{(0)}}{\beta_0}-\frac{4P^{(2)}}{\beta_2}\right]^{t-n}
\otimes\left[-8 P^{(1)} +\frac{2\beta_1}{\beta_0}P^{(0)}+
8\frac{\beta_1}{\beta_2}P^{(2)}\right]^{s-t-2n}\otimes D_{0,0}^{0}(x).
\nonumber\\
\ea
The solution of the NNLO DGLAP equation reproduced by (\ref{eq:NNLOansatz})
in Mellin space will then be written in the form 
\ba
f(N,\alpha_s)&=&\sum_{s=0}^{\infty}\sum_{t=0}^{s}\sum_{n=0}^{t}
\frac{D_{t,n}^{s}(N)}{n!(t-n)!(s-t)!}\mathcal{L}^{n}\mathcal{M}^{t-n}\mathcal{Q}^{s-t}
\nonumber\\
&=&\sum_{s=0}^{\infty}\sum_{t=0}^{s}\sum_{n=0}^{t}
\frac{\mathcal{L}^{n}\mathcal{M}^{t-n}\mathcal{Q}^{s-t}}{n!(t-n)!(s-t)!}
\left[-\frac{2}{\beta_0}P^{(0)}\right]^{n}
\left[\frac{P^{(0)}(N)}{\beta_0}-\frac{4P^{(2)}(N)}{\beta_2}\right]^{t-n}
\nonumber\\
&\times&\left[-8 P^{(1)}(N) +\frac{2\beta_1}{\beta_0}P^{(0)}(N)
+ 8\frac{\beta_1}{\beta_2}P^{(2)}(N)\right]^{s-t-2n}D_{0,0}^{0}(N)\,,
\ea
which is equivalent to
\ba
&&f(N,\alpha_s)= e^{\left[-\frac{2}{\beta_0}P^{(0)}\right]
\log{\left(\frac{\alpha_s}{\alpha_0}\right)}}
\exp\left\{\left[\frac{P^{(0)}(N)}{\beta_{0}}-\frac{4P^{(2)}(N)}{\beta_{2}}\right]
\log\frac{16\pi^{2}\beta_{0}+4\pi\alpha_s\beta_{1}
+\alpha_s^{2}\beta_{2}}{16\pi^{2}\beta_{0}+4\pi\alpha_{0}\beta_{1}+\alpha_{0}^{2}\beta_{2}}
\right\}\times
\nonumber\\
&&\hspace{2cm}\exp\left\{\left[\frac{2\beta_{1}}{\beta_{0}}P^{(0)}(N)-8P^{(1)}(N)
+\frac{8\beta_{1}}{\beta_{2}}P^{(2)}(N)\right]\times
\right.\nonumber\\
&&\left.\hspace{2cm}\Bigg(\frac{1}{\sqrt{4\beta_{0}\beta_{2}-\beta_{1}^{2}}}
\arctan\frac{2\pi(\alpha_s-\alpha_{0})\sqrt{4\beta_{0}\beta_{2}-
\beta_{1}^{2}}}{2\pi(8\pi\beta_{0}+(\alpha_s+\alpha_{0})\beta_{1})+
\alpha_s\alpha_{0}\beta_{2}}\Bigg)
\right\}D_{0,0}^{0}(N),
\ea
and it reproduces the result in (\ref{nnlotest}).
In $x$-space the above solution can be simply written as
\ba
&&f(x,\alpha_s(Q^2))= e^{\left[\log{\left(\frac{\alpha_s}{\alpha_0}\right)}
-\frac{2}{\beta_0}P^{(0)}\right]\otimes}
\exp\left\{\log\frac{16\pi^{2}\beta_{0}+4\pi\alpha_s\beta_{1}
+\alpha_s^{2}\beta_{2}}{16\pi^{2}\beta_{0}+4\pi\alpha_{0}\beta_{1}+\alpha_{0}^{2}\beta_{2}}
\left[\frac{P^{(0)}(N)}{\beta_{0}}-\frac{4P^{(2)}(N)}{\beta_{2}}\right]\otimes\right\}
\nonumber\\
&&\hspace{2cm}\exp\left\{
\Bigg(\frac{1}{\sqrt{4\beta_{0}\beta_{2}-\beta_{1}^{2}}}
\arctan\frac{2\pi(\alpha_s-\alpha_{0})\sqrt{4\beta_{0}\beta_{2}-
\beta_{1}^{2}}}{2\pi(8\pi\beta_{0}+(\alpha_s+\alpha_{0})\beta_{1})+
\alpha_s\alpha_{0}\beta_{2}}\Bigg)\right.\nonumber\\
&&\left.\hspace{2cm}
\left[\frac{2\beta_{1}}{\beta_{0}}P^{(0)}(N)-8P^{(1)}(N)
+\frac{8\beta_{1}}{\beta_{2}}P^{(2)}(N)\right]\otimes\right\}D_{0,0}^{0}(x).
\ea
We have shown how to obtain exact NNLO solutions of the non-singlet equations using recursion relations. 
It is clear that the solution shown above conceals all the logarithms of the coupling constant 
into more complicated functions of $\alpha_s$ and therefore 
it performs an intrinsic resummation of 
all these contributions, as obvious, being the exact solution of the non-singlet 
equation at NNLO. 
A numerical implementation of the recursion relations associated to these new 
functions of the coupling constants, in this case, is no 
different from the previous cases, when only functions of the form 
$\log(\alpha/\alpha_0)$ have been considered, but with a faster convergence rate.

\section{Truncated solutions at LO and NNLO in the singlet case}

The proof of the existence of a valid logarithmic ansatz that reproduces the
truncated solution of the singlet DGLAP equation at NLO is far
more involved compared to the non singlet
case. Before we proceed with this discussion, it is important to clarify
some points regarding some known results concerning these equations in moment space.
First of all, as we have widely remarked before, there are no exact solutions
of the singlet equations in moment space beyond those known at LO, 
due to the matrix structure of the equations. 
Therefore, it is no surprise that there is no logarithmic ansatz that can't do better
than to reproduce the truncated solution, since only these ones are available
analytically in moment space.
If we knew the structure of the exact solution in moment space we could 
construct an ansatz that would generate by recursion relations all the moments of that solution, 
following the same strategy outlined for the non-singlet equation. Therefore,
inverting numerically the equations for the moments has no advantage whatsoever compared 
to the numerical implementation of the logarithmic series using the algorithm that we have developed here. 
However, we can arbitrarily improve the logarithmic series in order to capture higher 
order contributions in the truncated solution, a feature that can be very appealing 
for phenomenological purposes.

The proof that a suitable logarithmic ansatz reproduces the truncated
solution of the moments of the singlet pdf's at NLO goes as follows.

\subsection{The exact solution at LO}

We start from the singlet matrix equation
\begin{equation}
\label{NLOsinglet1}
\frac{\partial}{\partial\log Q^{2}}\left(\begin{array}{c}
q^{(+)}(x,Q^{2})\\
g(x,Q^{2})\end{array}\right)=\left(\begin{array}{cc}
P_{qq}(x,\alpha_{s}(Q^{2})) & P_{qg}(x,\alpha_{s}(Q^{2}))\\
P_{gq}(x,\alpha_{s}(Q^{2})) & P_{gg}(x,\alpha_{s}(Q^{2}))
\end{array}\right)\otimes\left(\begin{array}{c}
q^{(+)}(x,Q^{2})\\
g(x,Q^{2})\end{array}\right),
\label{eq_singlet}
\end{equation}
whose well known LO solution in Mellin space can be easily identified
\ba
\vec{f}(N,\alpha_s)=\hat{L}(\alpha_s,\alpha_0,N)\vec{f}(N,\alpha_0),
\label{easily}
\ea
and where $\hat{L}(\alpha_s,\alpha_0,N)=
\left(\frac{\alpha_s}{\alpha_0}\right)^{\hat{R}_{0}(N)}$ is the evolution
operator.

Diagonalizing the $\hat{R}_{0}$ operator, in the equation above, we can write
the evolution operator $\hat{L}(\alpha_s,\alpha_0,N)$ as
\ba
\hat{L}(\alpha_s,\alpha_0,N)=
e_{+}\left(\frac{\alpha_s}{\alpha_0}\right)^{r_{+}}
+e_{-}\left(\frac{\alpha_s}{\alpha_0}\right)^{r_{-}},
\ea
where $r_{\pm}$ are the eigenvalues of the matrix
$\hat{R}_0=(-2/\beta_0) \hat{P}_0$ and $e_{+}$
and $e_{-}$ are projectors \cite{Petronzio1}, \cite{Petronzio2} defined as
\ba
e_{\pm}=\frac{1}{r_{\pm}-r_{\mp}}\left[\hat{R}_0-
r_{\mp}\hat{I}\right].
\ea
Since the $e_{\pm}$ are projection operators, the following properties
hold
\ba
e_+ e_+=e_+\hspace{1.5cm} e_- e_-=e_-\hspace{1.5cm} e_+ e_-=e_- e_+=0
\hspace{1.5cm} e_+ + e_-=1 \,.
\label{project}
\ea
Hence it is not difficult to see that
\ba
\hat{R}_{0}(N)=e_{+}r_{+}+e_{-}r_{-}\,.
\ea
It is important to note that one can write a solution of the singlet
DGLAP equation in a closed exponential form only at LO.

It is quite straightforward to reproduce this exact matrix solution at 
LO using a logarithmic expansion and the associated recursion relations. 
These are obtained from the ansatz (here written directly in moment space)
\ba
\label{RossiLO}
\vec{f}(N,\alpha_s)=\sum_{n=0}^{\infty}\frac{\vec{A}_{n}(N)}{n!}
\left[\ln{\left(\frac{\alpha_s}{\alpha_0}\right)}\right]^{n}\,,
\ea
subject to the initial condition
\ba
\vec{f}(N,\alpha_0)=\vec{A}_0(N).
\ea
Then the recursion relations become
\ba
\vec{A}_{n+1}(N)=-\frac{2}{\beta_0}\hat{P}^{(0)}(N)
\vec{A}_{n}(N)\equiv\hat{R}_{0}(N)\vec{A}_{n}(N),
\ea
and can be solved as
\ba
\vec{A}_{n}&=&\left[\hat{R}_{0}(N)\right]^n \vec{A}_{0}(N) \nonumber \\
&=&\left( e_+ r_+^n + e_- r_-^n\right) \vec{f}(N,\alpha_0),
\ea
having used eq.~(\ref{project}). Inserting this expression into eq.~(\ref{RossiLO}) we easily
obtain the relations
\ba
\vec{f}(N,\alpha_s)=\sum_{n=0}^{\infty}\frac{\left[\hat{R}_{0}(N)\right]^n}{n!}
\left[\ln{\left(\frac{\alpha_s}{\alpha_0}\right)}\right]^{n}\vec{A}_{0}(N)=
\left(\frac{\alpha_s}{\alpha_0}\right)^{\hat{R}_{0}(N)}\vec{f}(N,\alpha_0),
\ea
in agreement with eq.~(\ref{easily}).
\subsection{The standard NLO solution from moment space}
Moving to NLO, one can build a truncated
solution in moment space of eq.~(\ref{eq_singlet})
by a series expansion around the lowest order solution.

We start from the truncated version of the
vector equation (\ref{NLOsinglet1})
\ba
\frac{\partial{\vec{f}(N,\alpha_s)}}{\partial\alpha_s}=
-\frac{2}{\beta_0 \alpha_s}
\left[\hat{P}^{(0)}
+\frac{\alpha_s}{2\pi}\left(\hat{P}^{(1)}-
\frac{b_1}{2}\hat{P}^{(0)}\right)\right]\vec{f}(N,\alpha_s)\,,
\ea
that we re-express in the form
\ba
\label{NLOsinglet2}
\frac{\partial{\vec{f}(N,\alpha_s)}}{\partial\alpha_s} &=&
-\frac{1}{\alpha_s}\left[-\hat{R_0}+\alpha_s
\left(\frac{\hat{P}^{(1)}}{\pi\beta_0}+
\frac{\hat{R}_0 b_1}{4\pi}\right)\right]\vec{f}(N,\alpha_s) \nonumber \\
&=&
\frac{1}{\alpha_s}\left[\hat{R_0}+\alpha_s \hat{R}_1\right]
\vec{f}(N,\alpha_s),\, \nonumber\\
\ea
and with the $\hat{R_1}$ operator defined as
\ba
\hat{R_1}(N)=-\frac{1}{\pi}\left(\frac{b_1}{4}\hat{R}_0(N)+
\frac{\hat{P}^{(1)}(N)}{\beta_0}\right).
\ea
We use \cite{Petronzio1,Petronzio2} a truncated vector
solution of (\ref{NLOsinglet2}) - accurate at $O(\alpha_s)$ - of the form
\ba
\label{NLOans_sing1}
\vec{f}(N,\alpha_s)&=&\hat{U}(\alpha_s,N)\,\hat{L}(\alpha_s,\alpha_0,N)\,
\hat{U}^{-1}(\alpha_0,N)\,\vec{f}(N,\alpha_0)\nonumber\\
&=&\left[1+\alpha_s
\hat{U}_{1}(N)\right]
\hat{L}(\alpha_s,\alpha_0,N)\left[1+\alpha_0
\hat{U}_{1}(N)\right]^{-1}\vec{f}(N,\alpha_0)\,,\nonumber\\
\ea
where we have expanded in powers of $\alpha_s$ the operators
$\hat{U}(\alpha_s,N)$ and $\hat{U}^{-1}(\alpha_0,N)$.
Inserting (\ref{NLOans_sing1}) in eq.~(\ref{NLOsinglet2}),
we obtain the commutation relations involving the operators $\hat{U_1}$, $\hat{R}_0$ and $\hat{R}_1$
\ba
\label{com1}
\left[\hat{R}_0,\hat{U}_1\right]=\hat{U}_1-\hat{R}_1,
\ea
which appear in the solution in Mellin space \cite{Petronzio1,Petronzio2}.
Then, using the properties of the projection operators

\ba
\hat{U}_1=e_+\hat{U}_1 e_+ + e_+\hat{U}_1 e_- + e_-\hat{U}_1 e_+
+e_-\hat{U}_1 e_-, \,
\ea
and inserting this relation in the commutator (\ref{com1})
we easily derive the relation
\ba
\hat{U}_1=\left[e_+\hat{R}_1 e_+ + e_-\hat{R}_1 e_-\right]-
\frac{e_+\hat{R}_1 e_-}{r_+ -r_- -1}-\frac{e_-\hat{R}_1 e_+}{r_- -r_+ -1}.
\ea

Finally, expanding the term $\left[1+\alpha_0 \hat{U}_1\right]^{-1}$ in eq.~(\ref{NLOans_sing1}) we arrive at the solution \cite{Petronzio1,Petronzio2}
\ba
\label{NLOtrsolsin}
\vec{f}(N,\alpha_s)=\left[\hat{L}+\alpha_s\hat{U}_1\hat{L}
-\alpha_0\hat{L}\hat{U}_1\right]\vec{f}(N,\alpha_0)\,,
\ea
where the $(\alpha_s,\alpha_0,N)$ dependence has been dropped.
Such solution can be put in a more readable form as
\ba
\label{solution}
&&\vec{f}(N,\alpha_s)=\left\{
\left(\frac{\alpha_s}{\alpha_0}\right)^{r_+}\left[e_{+}
+(\alpha_s-\alpha_0)\,e_{+}\hat{R_1}e_{+}+\right.\right.\nonumber\\
&&\hspace{2cm}\left.\left.\left(\alpha_0-\alpha_s
\left(\frac{\alpha_s}{\alpha_0}\right)^{r_- -r_+}\right)
\frac{e_+\hat{R_1}e_{-}}{r_+ -r_- -1}\right]
+(+\leftrightarrow -)\right\}\vec{f}(N,\alpha_0)\,,
\ea
which can be called the {\em standard} NLO solution, having been introduced in the literature
about 20 years ago \cite{Petronzio2}. It is obvious that this solution is a
(first) truncated solution of the NLO singlet DGLAP equation, with the equation truncated
at the same order.

\subsection{Reobtaining the standard NLO solution using the logarithmic expansion}

Having worked out the well-known NLO singlet solution in moment space, our aim is
to show that the
same solution can be reconstructed using a logarithmic ansatz. This fills a gap in the 
previous literature on this types of ansatze for the QCD pdf's. 
To facilitate our duty, we stress once more that the type of recursion relations
obtained in the non-singlet and singlet cases are similar. 
In fact the matrix structure of the equations doesn't play any role 
in the derivation due to the linearity of the ansatz in the (vector) coefficient 
functions that appear in it. 

Our NLO singlet ansatz has the form
\ba
\label{ansatzevec}
{\vec{f}(x,Q^2)}^{NLO}=\sum_{n=0}^{\infty}\frac{\vec{A}_{n}(x)}{n!}
\left[\ln{\left(\frac{\alpha_s}{\alpha_0}\right)}\right]^{n}
+\alpha_s\sum_{n=0}^{\infty}\frac{\vec{B}_{n}(x)}{n!}
\left[\ln{\left(\frac{\alpha_s}{\alpha_0}\right)}\right]^{n}\,,
\ea

with $A_n$ and $B_n$ now being vectors involving the 
singlet components. The recursion relations are  
\ba
\label{vecrec}
&&\vec{A}_{n+1}(N)=\hat{R}_{0}(N)\vec{A}_{n}(N),\nonumber\\
&&\vec{B}_{n+1}(N)=-\vec{B}_{n}(N)-\frac{b_1}{4\pi}\vec{A}_{n+1}(N)
+\hat{R}_{0}(N)B_{n}(N)-
\frac{1}{\pi\beta_0}\hat{P}^{(1)}(N)\vec{A}_{n}(N),
\ea
subjected to the initial condition
\ba
\vec{f}(N,\alpha_0)=\vec{A}_0(N)+\alpha_0 \vec{B}_0(N).\,
\ea
The solution of (\ref{vecrec}) in moment space can be easily found and is given by
\ba
\label{vecrec1}
&&\vec{A}_{n+1}(N)=\left[e_+(r_+)^n +e_-(r_-)^n\right]\vec{A}_{0}(N),
\ea
while we can re-arrange the $\vec{B}_{n+1}(N)$ relation
into the form
\ba
&&\vec{B}_{n+1}(N)=\left(\hat{R}_0-1\right)\vec{B}_{n}(N)+
\hat{R}_1\hat{R}_0^n\vec{A}_{0}(N).
\label{brel1}
\ea
The last step to follow in order
to construct the truncated solution involves a projection of the recursion relations
(\ref{brel1}) in the basis of the projectors $\hat{e}_\pm$.
In this basis we separate the equations as 
\ba
&&\vec{B}_{n+1}=\left(e_+ r_+ +e_- r_- -1\right)\left[
\vec{B}_{n}^{++}+\vec{B}_{n}^{+-}+\vec{B}_{n}^{-+}+
\vec{B}_{n}^{--}\right]+\nonumber\\
&&\hspace{4.7cm}\left[\hat{R}_1^{++}+
\hat{R}_1^{+-}+\hat{R}_1^{-+}+\hat{R}_1^{--}\right]
\left(e_+ r_+^n +e_- r_-^n\right)\vec{A}_0,
\ea
where we have used the notation
\ba
&&e_{\pm}\vec{B}_{n}(N)e_{\pm}=\vec{B}_{n}^{\pm \pm}.
\ea
Then, using
\ba
&&\vec{B}_{n+1}^{++}=\left(r_+ -1\right)\vec{B}_{n}^{++}+
\hat{R}_1^{++}r_+^n\vec{A}_0,\nonumber\\
&&\vec{B}_{n+1}^{+-}=\left(r_+ -1\right)\vec{B}_{n}^{+-}+
\hat{R}_1^{+-}r_-^n\vec{A}_0,\nonumber\\
&&\vec{B}_{n+1}^{-+}=\left(r_- -1\right)\vec{B}_{n}^{-+}+
\hat{R}_1^{-+}r_+^n\vec{A}_0,\nonumber\\
&&\vec{B}_{n+1}^{--}=\left(r_- -1\right)\vec{B}_{n}^{--}+
\hat{R}_1^{--}r_-^n\vec{A}_0,\,
\ea
it is an easy task to verify that the solutions of the recursion relations
at NLO are given by
\ba
\label{Bproj}
&&\vec{B}_{n}^{++}=\left[r_+^n -(r_+ -1)^n \right]\hat{R}_1^{++}
\vec{A}_0,\nonumber\\
&&\vec{B}_{n}^{--}=\left[r_-^n -(r_- -1)^n \right]\hat{R}_1^{--}
\vec{A}_0,\nonumber\\
&&\vec{B}_{n}^{+-}=\left[-r_-^n +(r_+ -1)^n \right]
\frac{\hat{R}_1^{+-}}{r_+ -r_- -1}\vec{A}_0,\nonumber\\
&&\vec{B}_{n}^{-+}=\left[-r_+^n +(r_- -1)^n \right]
\frac{\hat{R}_1^{-+}}{r_- -r_+ -1}\vec{A}_0,\,\nonumber\\
\ea
where we have expressed the $n_{th}$ iterate in terms of the initial conditions,
and we have taken $\vec{B}_0=\vec{0}$.
Summing over all the projections, we arrive at the following expression for the
NLO truncated solution of the singlet parton distributions
\ba
\vec{f}(N,\alpha_s)=\sum_{n=0}^{\infty}\frac{L^n}{n!}
\left[\vec{A}_{n}^{++}+\vec{A}_{n}^{--}+\alpha_s
\left(\vec{B}_{n}^{++}+\vec{B}_{n}^{--}
+\vec{B}_{n}^{-+}+\vec{B}_{n}^{+-}\right)\right]\,,
\ea
which can be easily exponentiated to give
\ba
&&\vec{f}(N,\alpha_s)=e_{+}\vec{A}_0 \left(\frac{\alpha_s}{\alpha_0}\right)^{r_+}+
e_{-}\vec{A}_0 \left(\frac{\alpha_s}{\alpha_0}\right)^{r_-}+\nonumber\\
&&\hspace{1.5cm}\alpha_s\left\{
e_{+}\hat{R}_1e_{+}
\left(\frac{\alpha_s}{\alpha_0}\right)^{r_+}-
e_{+}\hat{R}_1e_{+}
\left(\frac{\alpha_s}{\alpha_0}\right)^{(r_+ -1)}+\right.\nonumber\\
&&\hspace{2cm}\left.
e_{-}\hat{R}_1e_{-}
\left(\frac{\alpha_s}{\alpha_0}\right)^{r_-}-
e_{-}\hat{R}_1e_{-}
\left(\frac{\alpha_s}{\alpha_0}\right)^{(r_- -1)}+\right.\nonumber\\
&&\hspace{2cm}\left.\frac{1}{(r_+ -r_- -1)}\left[
-e_{+}\hat{R}_1e_{-}
\left(\frac{\alpha_s}{\alpha_0}\right)^{r_-}+
e_{+}\hat{R}_1e_{-}
\left(\frac{\alpha_s}{\alpha_0}\right)^{(r_+ -1)}\right]+\right.\nonumber\\
&&\hspace{2cm}\left.\frac{1}{(r_- -r_+ -1)}\left[
-e_{-}\hat{R}_1e_{+}
\left(\frac{\alpha_s}{\alpha_0}\right)^{r_+}+
e_{-}\hat{R}_1e_{+}
\left(\frac{\alpha_s}{\alpha_0}\right)^{(r_- -1)}\right]\right\}\vec{A}_0\,.
\nonumber\\
\ea
Finally, organizing the various pieces we obtain exactly the solution in
eq.~(\ref{solution}). We have therefore shown that the logarithmic ansatz coincides with the 
solution of the singlet DGLAP equation at NLO known from the previous literature 
and reported in the previous section. It is intuitively obvious that we can 
build with this approach truncated solutions of higher orders improving on the standard 
solution (\ref{solution}) known from moment space, and we can do this with any accuracy. However, before discussing this point in one of the following sections, we want to show how the same strategy works at NNLO.

\subsection{Truncated Solution at NNLO}
At this point, to complete our investigation, we need to discuss the generalization 
of the procedure illustrated above to the
NNLO case. 
As usual, we start from a truncated version of eq.~(\ref{NLOsinglet1}), that at 
NNLO can be written as
\ba
\label{NNLOsinglet2}
\frac{\partial{\vec{f}(N,\alpha_s)}}{\partial\alpha_s}=
\frac{1}{\alpha_s}\left[\hat{R_0}+\alpha_s \hat{R}_1 +\alpha_s^2
\hat{R}_2\right]\vec{f}(N,\alpha_s),\,\nonumber\\
\ea
where
\ba
\hat{R}_2=-\frac{1}{\pi}\left(\frac{\hat{P^{(2)}}}{2\pi\beta_0}
+\frac{\hat{R}_1 b_1}{4}+\frac{\hat{R}_0 b_2}{16\pi}\right),
\ea
whose solution is expected to be of the form \cite{ellis}
\ba
\label{NNLOans_sing1}
\vec{f}(N,\alpha_s)=\left[1+\alpha_s\hat{U}_{1}(N)+\alpha_s^2 
\hat{U}_{2}(N)\right]\hat{L}(\alpha_s,\alpha_0,N)
\left[1+\alpha_0\hat{U}_{1}(N)+\alpha_0^2\hat{U}_{2}(N)\right]^{-1}
\vec{f}(N,\alpha_0),\,\nonumber\\
\ea
where  
\ba
&&\left[\hat{R}_0,\hat{U}_1\right]=\hat{U}_1-\hat{R}_1,\nonumber\\
&&\left[\hat{R}_0,\hat{U}_2\right]=-\hat{R}_2 -\hat{R}_1\hat{U}_1 +2\hat{U}_2.\,
\ea
Using the projectors in the $\pm$ subspaces, one can remove the commutators, obtaining
 \ba
&&\hat{U}_2^{++}=\frac{1}{2}\left[\hat{R}_1^{++}\hat{R}_1^{++}
+\hat{R}_2^{++}-\frac{\hat{R}_1^{+-}
\hat{R}_1^{-+}}{r_- -r_+ -1}\right],\nonumber\\
&&\hat{U}_2^{--}=\frac{1}{2}\left[\hat{R}_1^{--}\hat{R}_1^{--}
+\hat{R}_2^{--}-\frac{\hat{R}_1^{-+}
\hat{R}_1^{+-}}{r_+ -r_- -1}\right],\nonumber\\
&&\hat{U}_2^{+-}=\frac{1}{r_+ -r_- -2}\left[-\hat{R}_1^{+-}\hat{R}_1^{--}
-\hat{R}_2^{+-}+\frac{\hat{R}_1^{++}\hat{R}_1^{+-}}{r_+ -r_- -1}
\right],\nonumber\\
&&\hat{U}_2^{-+}=\frac{1}{r_- -r_+ -2}\left[-\hat{R}_1^{-+}\hat{R}_1^{++}
-\hat{R}_2^{-+}+\frac{\hat{R}_1^{--}\hat{R}_1^{-+}}{r_- -r_+ -1}
\right],\,
\ea
and the formal solution from Mellin space can be simplified to 
\ba
\label{NNLOtrsolsing}
&&\vec{f}(N,\alpha_s)=\left[\hat{L}+\alpha_s\hat{U}_1\hat{L}
-\alpha_0\hat{L}\hat{U}_1\right.\nonumber\\
&&\hspace{2cm}\left.+\alpha_s^2 \hat{U}_2\hat{L}
-\alpha_s\alpha_0\hat{U}_1\hat{L}\hat{U}_1
+\alpha_0^2\hat{L}\left(\hat{U}_1^2-\hat{U}_2\right)
\right]\vec{f}(N,\alpha_0)\,.
\ea

At this point we introduce our ($1$-$st$ truncated) logarithmic ansatz that is expected to reproduce 
(\ref{NNLOtrsolsing}). Now it includes also an infinite set of new coefficients $\vec{C_n}$, 
similar to the non-singlet NNLO case
\ba
\vec{f}(N,\alpha_s)=\sum_{n=0}^{\infty}\frac{L^n}{n!}\left[\vec{A_n} +\alpha_s\vec{B_n}+
\alpha_s^2\vec{C_n}\right].
\label{altroans}
\ea

Inserting the NNLO logarithmic ansatz into (\ref{NNLOsinglet2}), we obtain in moment space the
recursion relations
\ba
\label{vecrecNNLO}
&&\vec{A}_{n+1}=\hat{R}_{0}\vec{A}_{n},\nonumber\\
&&\vec{B}_{n+1}=\left(\hat{R}_0-1\right)\vec{B}_{n}+
\hat{R}_1\hat{R}_0^n\vec{A}_{0},\nonumber\\
&&\vec{C}_{n+1}=\left(\hat{R}_0-2\right)\vec{C}_{n}
-\frac{b_{1}}{4\pi}\left(\vec{B}_{n}+\vec{B}_{n+1}\right)
+\left[\frac{b_{1}}{4\pi}\hat{R}_{0}+\hat{R}_{1}\right]\vec{B}_{n},
\nonumber \\
&&\hspace{1cm}+\left[\hat{R}_2+\frac{b_1}{4\pi}\hat{R}_1\right]\hat{R}_0^n \vec{A}_0,
\ea
whose solution has to coincide with (\ref{NNLOtrsolsing}). Also in this case, as before,
we use the $e_{\pm}$ projectors and notice that the structure of the recursion relations for the
coefficients $A_n$ and $B_n$ remain the same as in NLO.
Therefore, the solutions of the recursion relations for $\vec{A}_n$ and $\vec{B}_n$ are still given by 
(\ref{vecrec1}) and (\ref{Bproj}). We then have to find only an explicit solution of the relations for the new 
coefficients $\vec{C}_{n+1}(N)$. 

These relations can be solved in terms of $\vec{A}_0$, $\vec{B}_0$ and
$\vec{C}_0$ with the help of (\ref{Bproj}).
Finally, taking $\vec{B}_0 =0$ and $\vec{C}_0 =0$ after a lengthy computation we obtain the explicit solutions 
for the projected components  
\ba
&&\vec{C}_{n}^{++}=-\frac{1}{2}\frac{\hat{R}_1^{+-}\hat{R}_1^{-+}}
{(r_+ -r_- -1)(r_- -r_+ -1)}\times\nonumber\\
&&\hspace{1cm}\left[2(r_- -1)^n -(r_+ -2)^n
-(r_+ -2)^n r_+ -r_+^n +r_+^{n+1}
+ r_- \left((r_+ -2)^n -r_+^n \right)\right]\vec{A}_0
\nonumber\\
&&\hspace{1cm}+\frac{1}{2}\hat{R}_1^{++}\hat{R}_1^{++}\left[
r_+^n -2(r_+ -1)^n +(r_+ -2)^n\right]\vec{A}_0
\nonumber\\
&&\hspace{1cm}+\frac{1}{2}\hat{R}_2^{++}\left[r_+^n
-(r_+ -2)^n\right]\vec{A}_0,
\nonumber\\
&&\vec{C}_{n}^{--}=-\frac{1}{2}\frac{\hat{R}_1^{-+}\hat{R}_1^{+-}}
{(r_+ -r_- -1)(r_- -r_+ -1)}\times\nonumber\\
&&\hspace{1cm}\left[2(r_+ -1)^n -(r_- -2)^n
-(r_- -2)^n r_- -r_+^n +r_+^{n+1}
+ r_+ \left((r_- -2)^n -r_-^n \right)\right]\vec{A}_0
\nonumber\\
&&\hspace{1cm}+\frac{1}{2}\hat{R}_1^{--}\hat{R}_1^{--}\left[
r_-^n -2(r_- -1)^n +(r_- -2)^n\right]\vec{A}_0
\nonumber\\
&&\hspace{1cm}+\frac{1}{2}\hat{R}_2^{--}\left[r_-^n
-(r_- -2)^n\right]\vec{A}_0,
\nonumber\\
&&\vec{C}_{n}^{+-}=\frac{1}{2+r_-^2 +r_-(3-2r_+)-3r_+ +r_+^2}
\left[\hat{R}_2^{+-}\left(r_-^n -(r_+-2)^n\right)\left(1+ r_- -r_+\right)
\right.\nonumber\\
&&\hspace{1cm}\left.-\hat{R}_1^{+-}\hat{R}_1^{--}\left(2(r_- -1)^n
+(r_--1)^n r_- -r_-^{n+1}\right.\right.\nonumber\\
&&\hspace{1cm}\left.\left.-(r_+ -2)^n +r_-^n(r_+ -1) -(r_- -1)^n r_+\right)
\right.\nonumber\\
&&\hspace{1cm}\left.+\hat{R}_1^{++}\hat{R}_1^{+-}
\left(r_-^n+(r_+ -2)^n +r_-(r_+ -2)^n
\right.\right.\nonumber\\
&&\hspace{1cm}\left.\left.-2(r_+ -1)^n -r_-(r_+ -1)^n
-r_+(r_+ -2)^n + r_+(r_+ -1)^n \right)
\right]\vec{A}_0,
\nonumber\\
&&\vec{C}_{n}^{-+}=\frac{1}{2+r_-^2+3r_+ +r_+^2 -r_-(3+2r_+)}
\left[\hat{R}_2^{-+}(r_- -r_+ -1)\left((r_- -2)^n -r_+^n\right)
\right.\nonumber\\
&&\hspace{1cm}\left.-\hat{R}_1^{--}\hat{R}_1^{-+}\left(
-(r_- -2)^n +2(r_- -1)^n +((r_- -2)^n -(r_- -1)^n)r_-
\right.\right.\nonumber\\
&&\hspace{1cm}\left.\left.-(r_- -2)^n r_+ +(r_- -1)^n r_+
-r_+^n\right)
\right.\nonumber\\
&&\hspace{1cm}\left.+\hat{R}_1^{-+}\hat{R}_1^{++}\left(
(r_- -2)^n -2(r_+ -1)^n +r_-(r_+ -1)^n\right.\right.\nonumber\\
&&\hspace{1cm}\left.\left.-(r_+ -1)^n r_+
+r_+^n -r_- r_+^n +r_+^{n+1}\right)
\right]\vec{A}_0.
\nonumber\\
\ea
Re-inserting these solutions into the NNLO ansatz (\ref{altroans}) and after exponentiation 
one can show explicitely that the logarithmic solution so obtained coincides
with (\ref{NNLOtrsolsing}). Details can be found in appendix A. 
In the practical implementations of these solutions, there are two obvious strategies
that can be followed. One consists in the implementation of the recursion relations 
as we have done in various cases above: a sufficiently large number of iterates will converge 
to the truncated solution (\ref{NNLOtrsolsing}). This is obtained by implementing eqs.~(\ref{vecrecNNLO}) 
and incorporating them into (\ref{altroans}). A second method consists in the direct computation of 
(\ref{NNLOtrsolsing}) in $x$-space which becomes 
\ba
\label{NNLOtrsolsingx}
&&\vec{f}(x,\alpha_s)=\left[\hat{L}+\alpha_s\hat{U}_1\otimes\hat{L}
-\alpha_0\hat{L}\otimes\hat{U}_1\right.\nonumber\\
&&\hspace{2cm}\left.+\alpha_s^2 \hat{U}_2\otimes\hat{L}
-\alpha_s\alpha_0\hat{U}_1\otimes\hat{L}\otimes \hat{U}_1
+\alpha_0^2\hat{L}\otimes \left(\hat{U}_1\otimes \hat{U}_1-\hat{U}_2\right)
\right]\otimes \vec{f}(x,\alpha_0),\nonumber \\
\ea
and with $\hat{L}$ now replaced by its operatorial $(\otimes)$ form 
\beq
\hat{L}\to e^{L^n R_0 \otimes}= \left(\sum_{n=0}^{\infty} \frac{R_0^n}{n!} L^n\right)_\otimes.
\eeq
An implementation of (\ref{NNLOtrsolsingx}) would reduce 
its numerical evaluation to that of a sequence of LO solutions built around ``artificial''
initial conditions given by
$\hat{U}_1\otimes f(x,\alpha_0)$, $\hat{U}_1\otimes \hat{U}_1\otimes f(x,\alpha_0)$ and so on.

\section{Higher order logarithmic approximation of the NNLO singlet solution}

The procedure studied in the previous section can be
generalized and applied to obtain solutions that retain higher order 
logarithmic contributions in the NLO/NNLO singlet cases.
The same procedure is also the one that
has been implemented in all the existing codes for the singlet:
one has to truncate the equation and then try to reach the exact solution by a sufficiently 
high number of iterates.
On the other hand, $x$-space (non brute force) implementations are, from this respect,
still lagging since they are only based
on the Rossi-Storrow formulation \cite{cafacor,gordon}, which we have analized thoroughly
and largely extended in this work.

Therefore, the only way at our disposal to reach from $x$-space the exact solution
is by using higher order truncates. This is of practical relevance since our algorithm 
allows to perform separate
checks between truncated solutions of arbitrary high orders built either from
Mellin or from $x$-space
\footnote{ The current benchmarks available at NNLO are limited to exact solutions and 
do not involve comparisons
between truncated solutions}. We recall, if not obvious, that in the analysis
of hadronic processes the two criteria of using
either truncated or exact solutions are both acceptable.

To summarize: since in the singlet case is not possible
to write down a solution of the DGLAP equation in a
closed exponential form because of the non-commutativity
of the operators $\hat{R}_i$, the best thing we can do is to arrange
the singlet DGLAP equation in the truncated form,
as in eqs.~(\ref{NLOsinglet2}) and (\ref{NNLOsinglet2}).
The truncated vector solutions
(\ref{NNLOtrsolsing}) and (\ref{NLOtrsolsin}) are equivalent to those obtained using the
vector recursion relations at NLO/NNLO.

Now, working at NNLO, we will show how the basic NNLO solution
can be improved and the higher truncates identified.
Clearly, it is important to show explicitely that these truncates,
generated after solving the recursion relations,
can be rewritten exactly in the form previously known from Mellin space.
We are going to show here that this
is in fact the case, although some of the explicit expressions for the higher order
coefficient functions $C_n, D_n'$s  will
be given explicitely only in part. The expressions are in fact slightly lengthy
\footnote{They will be included in
a file that will be made available in the same distribution of our code, which is in preparation.}.
Therefore, here we will just outline the procedure and
illustrate the proof up to the second truncate of the NNLO singlet only for the sake
of clarity.

The exact singlet NNLO equation in Mellin space is given by
\beqa
\frac{\partial\vec{f}(N,\alpha_s)}{\partial\alpha_s} &=&
-\frac{\left(\frac{\alpha_{s}}{2\pi}\right)\hat{P}^{(0)}(N)+
\left(\frac{\alpha_{s}}{2\pi}\right)^{2}\hat{P}^{(1)}(N)+
\left(\frac{\alpha_{s}}{2\pi}\right)^{3}
\hat{P}^{(2)}(N)}{\frac{\beta_{0}}{4\pi}
\alpha_{s}^{2}+\frac{\beta_{1}}{16\pi^{2}}\alpha_{s}^{3}+
\frac{\beta_{2}}{64\pi^{3}}\alpha_{s}^{4}}\vec{f}(N,\alpha_{s}) \nonumber \\
&=&
\frac{\hat{P}^{NNLO}(N,\alpha_s)}{\beta^{NNLO}(\alpha_s)}\vec{f}(N,\alpha_{s}),
\nonumber\\
\ea
where we have introduced the singlet kernels. 
After a Taylor expansion of $\hat{P}^{NNLO}(N,\alpha_s)/\beta^{NNLO}(\alpha_s)$
up to $\alpha_s^3$ it becomes
\ba
\frac{\partial\vec{f}(N,\alpha_s)}{\partial\alpha_s}=
\frac{1}{\alpha_s}\left[\hat{R}_0-\frac{b_2}{(4\pi)^2}\hat{R}_1
+\alpha_s\hat{R}_1+\alpha_s^2\hat{R}_2-\frac{b_1}{4\pi}\alpha_s^3\hat{R}_2
\right]\vec{f}(N,\alpha_{s}),\,
\ea
which is the truncated equation of order $\alpha_s^3$. The $\hat{R}_i$ $(i=0,1,2)$ operators are listed below
\ba
&&\hat{R}_0=-\frac{2}{\beta_0}\hat{P}^{(0)},
\nonumber\\
&&\hat{R}_1=-\frac{\hat{P}^{(1)}}{\pi\beta_0}+\frac{b_1}{2\pi\beta_0}\hat{P}^{(0)},
\nonumber\\
&&\hat{R}_2=-\frac{\hat{P}^{(2)}}{2\pi^2\beta_0}+\frac{b_1}{4\pi^2\beta_0}\hat{P}^{(1)}
+\left[-\frac{b_1^2}{8\pi^2\beta_0}+\frac{b_2}{8\pi^2\beta_0}\right]\hat{P}^{(0)}.
\ea
The formal solution of this equation can be written as \cite{Buras},\cite{kosower}
\ba
\vec{f}(N,\alpha_s)&=&T_{\alpha}\left[\exp\left\{\int_{\alpha_0}^{\alpha_s} d\,\alpha_s'\,\frac{1}{\alpha_s'}\left(\hat{R}_0
-\frac{b_2}{(4\pi)^2}\hat{R}_1+\alpha_s'\hat{R}_1+{\alpha_s'}^2\hat{R}_2
-\frac{b_1}{4\pi}{\alpha_s'}^3\hat{R}_2\right)\right\}\right]\vec{f}(N,\alpha_0)\nonumber\\
&=&\hat{U}(N,\alpha_s)\hat{L}(\alpha_s,\alpha_0)
\hat{U}^{-1}(N,\alpha_0)\vec{f}(N,\alpha_0),
\ea
where the $T_{\alpha}$ operator acts on the exponential similarly to a 
time-ordered product, but this time in the space of the 
couplings. Again, expanding the $\hat{U}(N,\alpha_s)$ and
$\hat{U}^{-1}(N,\alpha_0)$ operators in the formal solution 
around $\alpha_s=0$ and $\alpha_0=0$, we have
\ba
\label{NNLOhigher1}
&&\vec{f}(N,\alpha_s)=\left[\hat{L}+\alpha_s\hat{U}_1\hat{L}
-\alpha_0\hat{L}\hat{U}_1+\alpha_s^2\hat{U}_2\hat{L}
-\alpha_s\alpha_0\hat{U}_1\hat{L}\hat{U}_1
+\alpha_0^2\hat{L}\left(\hat{U}_1^2-\hat{U}_2\right)\right.
\nonumber\\
&&\hspace{2cm}\left.+\alpha_s^3\hat{U}_3\hat{L}
+\alpha_s\alpha_0^2\hat{U}_1\hat{L}\left(\hat{U}_1^2-\hat{U}_2\right)
-\alpha_s^2\alpha_0\hat{U}_2\hat{L}\hat{U}_1
\right.\nonumber\\
&&\hspace{2cm}\left.-\alpha_0^3\hat{L}\left(\hat{U}_1^3
-\hat{U}_1\hat{U}_2-\hat{U}_2\hat{U}_1+\hat{U}_3\right)
\right]\vec{f}(N,\alpha_0).
\ea
Inserting the expanded solution into eq.~(\ref{NNLOhigher1}) and equating the various power
of $\alpha_s$ we arrive at the following chain of commutation relations
\ba
&&\left[\hat{R}_0,\hat{U}_1\right]=\hat{U}_1-\hat{R}_1,
\nonumber\\
&&\left[\hat{R}_0,\hat{U}_2\right]=-\hat{R}_2-\hat{R}_1\hat{U}_1+2\hat{U}_2,
\nonumber\\
&&\left[\hat{R}_0,\hat{U}_3\right]=\frac{b_2}{(4\pi)^2}\hat{R}_1
+\frac{b_1}{(4\pi)}\hat{R}_2-\hat{R}_1\hat{U}_2-\hat{R}_2\hat{U}_1
+3\hat{U}_3.
\ea
Removing the commutators by the $e_\pm$ projectors one obtains
\ba
&&\hat{U}_1^{++}=\hat{R}_1^{++},\nonumber\\
&&\hat{U}_1^{--}=\hat{R}_1^{--},\nonumber\\
&&\hat{U}_1^{+-}=-\frac{\hat{R}_1^{+-}}{r_+ -r_- -1},\nonumber\\
&&\hat{U}_1^{-+}=-\frac{\hat{R}_1^{-+}}{r_- -r_+ -1},\nonumber\\
&&\hat{U}_2^{++}=\frac{1}{2}\left[\hat{R}_1^{++}\hat{R}_1^{++}
+\hat{R}_2^{++}-\frac{\hat{R}_1^{+-}
\hat{R}_1^{-+}}{r_- -r_+ -1}\right],\nonumber\\
&&\hat{U}_2^{--}=\frac{1}{2}\left[\hat{R}_1^{--}\hat{R}_1^{--}
+\hat{R}_2^{--}-\frac{\hat{R}_1^{-+}
\hat{R}_1^{+-}}{r_+ -r_- -1}\right],\nonumber\\
&&\hat{U}_2^{+-}=\frac{1}{r_+ -r_- -2}\left[-\hat{R}_1^{+-}\hat{R}_1^{--}
-\hat{R}_2^{+-}+\frac{\hat{R}_1^{++}\hat{R}_1^{+-}}{r_+ -r_- -1}
\right],\nonumber\\
&&\hat{U}_2^{-+}=\frac{1}{r_- -r_+ -2}\left[-\hat{R}_1^{-+}\hat{R}_1^{++}
-\hat{R}_2^{-+}+\frac{\hat{R}_1^{--}\hat{R}_1^{-+}}{r_- -r_+ -1}
\right],\nonumber\\
&&\hat{U}_3^{++}=\frac{1}{3}\left[-\frac{b_1}{(4\pi)}\hat{R}_1^{++}
-\frac{b_2}{(4\pi)^2}\hat{R}_2^{++}+\hat{R}_1^{+-}\hat{U}_2^{-+}
+\hat{R}_1^{++}\hat{U}_2^{++}+\hat{R}_2^{+-}\hat{U}_1^{-+}
+\hat{R}_2^{++}\hat{U}_1^{++}\right],\nonumber\\
&&\hat{U}_3^{--}=\frac{1}{3}\left[-\frac{b_1}{(4\pi)}\hat{R}_1^{--}
-\frac{b_2}{(4\pi)^2}\hat{R}_2^{--}+\hat{R}_1^{-+}\hat{U}_2^{+-}
+\hat{R}_1^{--}\hat{U}_2^{--}+\hat{R}_2^{-+}\hat{U}_1^{+-}
+\hat{R}_2^{--}\hat{U}_1^{--}\right],
\nonumber\\
&&\hat{U}_3^{+-}=\frac{1}{r_+ -r_- -3}\left[-\frac{b_1}{(4\pi)}\hat{R}_1^{+-}
-\frac{b_2}{(4\pi)^2}\hat{R}_2^{+-}-\hat{R}_1^{+-}\hat{U}_2^{--}
-\hat{R}_1^{++}\hat{U}_2^{+-}-\hat{R}_2^{+-}\hat{U}_1^{--}
-\hat{R}_2^{++}\hat{U}_1^{+-}\right],
\nonumber\\
&&\hat{U}_3^{-+}=\frac{1}{r_- -r_+ -3}\left[-\frac{b_1}{(4\pi)}\hat{R}_1^{-+}
-\frac{b_2}{(4\pi)^2}\hat{R}_2^{-+}-\hat{R}_1^{-+}\hat{U}_2^{++}
-\hat{R}_1^{--}\hat{U}_2^{-+}-\hat{R}_2^{-+}\hat{U}_1^{++}
-\hat{R}_2^{--}\hat{U}_1^{-+}\right].\,
\nonumber\\
\ea
As one can see, the $\hat{U}_{3}$ operator is expressed in terms 
of the kernels $P^{(0)},\hat{P}^{(1)}$ and $\hat{P}^{(2)}$.
One can prove that by imposing a higher order ansatz in Mellin space
of the form
\ba
\vec{\tilde{f}}(N,\alpha_s)=\sum_{n=0}^{\infty}\frac{L^n}{n!}
\left[\vec{A}_n(N)+\alpha_s\vec{B}_n(N)+\alpha_s^2\vec{C}_n(N)
+\alpha_s^3 \vec{D}_n(N)\right]\,,
\ea
 the solution (\ref{NNLOhigher1}) is generated. 
The vector recursion relations in this case become 
\ba
&&\vec{A}_{n+1}=-\frac{2}{\beta_0}\hat{P}^{(0)}\vec{A}_{n},\nonumber\\
&&\vec{B}_{n+1}=-\vec{B}_{n}-\frac{1}{\pi\beta_0}\hat{P}^{(1)}\vec{A}_{n}-
\frac{\beta_1}{\pi\beta_0}\vec{A}_{n+1}-\frac{2}{\beta_0}\hat{P}^{(0)}\vec{B}_{n},
\nonumber\\
&&\vec{C}_{n+1}=-\frac{1}{2\pi^2\beta_0}\hat{P}^{(2)}\vec{A}_{n}
-\frac{\beta_2}{(4\pi)^2\beta_0}\vec{A}_{n+1}
-\frac{1}{\pi\beta_0}\hat{P}^{(1)}\vec{B}_{n},\nonumber\\
&&\hspace{1.4cm}-\frac{\beta_1}{4\pi\beta_0}\vec{B}_{n}
-\frac{\beta_1}{4\pi\beta_0}\vec{B}_{n+1}
-\frac{2}{\beta_0}\hat{P}^{(0)}\vec{C}_{n}
-2\vec{C}_{n},\nonumber\\
&&\vec{D}_{n+1}=-\frac{1}{2\pi^2\beta_0}\hat{P}^{(2)}\vec{B}_{n}
-\frac{\beta_2}{(4\pi)^2\beta_0}\vec{B}_{n}
-\frac{\beta_2}{(4\pi)^2\beta_0}\vec{B}_{n+1}
-\frac{1}{\pi\beta_0}\hat{P}^{(1)}\vec{C}_{n}\nonumber\\
&&\hspace{1.4cm}-\frac{\beta_1}{2\pi\beta_0}\vec{C}_{n}
-\frac{\beta_1}{(4\pi)\beta_0}\vec{C}_{n+1}
-\frac{2}{\beta_0}\hat{P}^{(0)}\vec{D}_{n}
-3\vec{D}_{n}.\,
\ea
Applying the properties of the $e_\pm $ operators, 
we project out the $\vec{A}_{n+1}^\pm $, $\vec{B}_{n+1}^\pm\dots$ components of these relations, 
and imposing the initial conditions $\vec{B}_0=\vec{C}_0=\vec{D}_0=0$ together with
\ba
\vec{f}(N,\alpha_0)=\vec{A}_0\,,
\ea
we obtain the explicit form of $\vec{\tilde{f}}(N,\alpha_s)$.
In order to construct the $\vec{\tilde{f}}(N,\alpha_s)$ solution,
the four ${\pm}$ projections of $\vec{D}_{n+1}$ must be solved
with respect to $\vec{A}_0(N)$, since the other projections
$\vec{A}_{n}^{\pm}$, $\vec{B}_{n}^{\pm}$, $\vec{C}_{n}^{\pm}$
are already known. A direct computation shows that the structure of the solution can be organized as follows in terms of the components of $R_i$ 
\ba
&&\vec{D}_{n}^{++}(N)=
{\bf W}_1(\hat{R}_1^{++\,3},r_+,r_-,N,\vec{A}_0)
+{\bf W}_2(\hat{R}_1^{+-}\hat{R}_1^{-+}\hat{R}_1^{++},r_+,r_-,N,\vec{A}_0)
\nonumber\\
&&\hspace{1.7cm}
+{\bf W}_3(\hat{R}_1^{+-}\hat{R}_1^{--}\hat{R}_1^{-+},r_+,r_-,N,\vec{A}_0)
+{\bf W}_4(\hat{R}_1^{++}\hat{R}_1^{+-}\hat{R}_1^{-+},r_+,r_-,N,\vec{A}_0)
\nonumber\\
&&\hspace{1.7cm}
+{\bf W}_5(\hat{R}_1^{++}\hat{R}_2^{++},r_+,r_-,N,\vec{A}_0)
+{\bf W}_6(\hat{R}_2^{++}\hat{R}_1^{++},r_+,r_-,N,\vec{A}_0)
\nonumber\\
&&\hspace{1.7cm}
+{\bf W}_7(\hat{R}_1^{+-}\hat{R}_2^{-+},r_+,r_-,N,\vec{A}_0)
+{\bf W}_8(\hat{R}_2^{+-}\hat{R}_1^{-+},r_+,r_-,N,\vec{A}_0)
\nonumber\\
&&\hspace{1.7cm}
+{\bf W}_9(\hat{R}_1^{++},r_+,r_-,N,\vec{A}_0)
+{\bf W}_{10}(\hat{R}_2^{++},r_+,r_-,N,\vec{A}_0),
\nonumber\\
\nonumber\\
&&\vec{D}_{n}^{--}(N)=
{\bf W}_1(\hat{R}_1^{--\,3},r_+,r_-,N,\vec{A}_0)
+{\bf W}_2(\hat{R}_1^{-+}\hat{R}_1^{+-}\hat{R}_1^{--},r_+,r_-,N,\vec{A}_0)
\nonumber\\
&&\hspace{1.7cm}
+{\bf W}_3(\hat{R}_1^{-+}\hat{R}_1^{++}\hat{R}_1^{+-},r_+,r_-,N,\vec{A}_0)
+{\bf W}_4(\hat{R}_1^{--}\hat{R}_1^{-+}\hat{R}_1^{+-},r_+,r_-,N,\vec{A}_0)
\nonumber\\
&&\hspace{1.7cm}
+{\bf W}_5(\hat{R}_1^{--}\hat{R}_2^{--},r_+,r_-,N,\vec{A}_0)
+{\bf W}_6(\hat{R}_2^{--}\hat{R}_1^{--},r_+,r_-,N,\vec{A}_0)
\nonumber\\
&&\hspace{1.7cm}
+{\bf W}_7(\hat{R}_1^{-+}\hat{R}_2^{+-},r_+,r_-,N,\vec{A}_0)
+{\bf W}_8(\hat{R}_2^{-+}\hat{R}_1^{+-},r_+,r_-,N,\vec{A}_0)
\nonumber\\
&&\hspace{1.7cm}
+{\bf W}_9(\hat{R}_1^{--},r_+,r_-,N,\vec{A}_0)
+{\bf W}_{10}(\hat{R}_2^{--},r_+,r_-,N,\vec{A}_0),
\nonumber\\
\nonumber\\
&&\vec{D}_{n}^{+-}(N)=
{\bf Z}_1(\hat{R}_1^{++}\hat{R}_1^{+-}\hat{R}_1^{--},r_+,r_-,N,\vec{A}_0)
+{\bf Z}_2(\hat{R}_1^{++}\hat{R}_1^{++}\hat{R}_1^{+-},r_+,r_-,N,\vec{A}_0)
\nonumber\\
&&\hspace{1.7cm}
+{\bf Z}_3(\hat{R}_1^{+-}\hat{R}_1^{--}\hat{R}_1^{--},r_+,r_-,N,\vec{A}_0)
+{\bf Z}_4(\hat{R}_1^{+-}\hat{R}_1^{-+}\hat{R}_1^{+-},r_+,r_-,N,\vec{A}_0)
\nonumber\\
&&\hspace{1.7cm}
+{\bf Z}_5(\hat{R}_1^{+-}\hat{R}_2^{--},r_+,r_-,N,\vec{A}_0)
+{\bf Z}_6(\hat{R}_2^{+-}\hat{R}_1^{--},r_+,r_-,N,\vec{A}_0)
\nonumber\\
&&\hspace{1.7cm}
+{\bf Z}_7(\hat{R}_1^{++}\hat{R}_2^{+-},r_+,r_-,N,\vec{A}_0)
+{\bf Z}_8(\hat{R}_2^{++}\hat{R}_1^{+-},r_+,r_-,N,\vec{A}_0)
\nonumber\\
&&\hspace{1.7cm}
+{\bf Z}_9(\hat{R}_1^{+-},r_+,r_-,N,\vec{A}_0)
+{\bf Z}_{10}(\hat{R}_2^{+-},r_+,r_-,N,\vec{A}_0),
\nonumber\\
\nonumber\\
&&\vec{D}_{n}^{-+}(N)=
{\bf Z}_1(\hat{R}_1^{--}\hat{R}_1^{-+}\hat{R}_1^{++},r_+,r_-,N,\vec{A}_0)
+{\bf Z}_2(\hat{R}_1^{--}\hat{R}_1^{--}\hat{R}_1^{-+},r_+,r_-,N,\vec{A}_0)
\nonumber\\
&&\hspace{1.7cm}
+{\bf Z}_3(\hat{R}_1^{-+}\hat{R}_1^{++}\hat{R}_1^{++},r_+,r_-,N,\vec{A}_0)
+{\bf Z}_4(\hat{R}_1^{-+}\hat{R}_1^{+-}\hat{R}_1^{-+},r_+,r_-,N,\vec{A}_0)
\nonumber\\
&&\hspace{1.7cm}
+{\bf Z}_5(\hat{R}_1^{-+}\hat{R}_2^{++},r_+,r_-,N,\vec{A}_0)
+{\bf Z}_6(\hat{R}_2^{-+}\hat{R}_1^{++},r_+,r_-,N,\vec{A}_0)
\nonumber\\
&&\hspace{1.7cm}
+{\bf Z}_7(\hat{R}_1^{--}\hat{R}_2^{-+},r_+,r_-,N,\vec{A}_0)
+{\bf Z}_8(\hat{R}_2^{--}\hat{R}_1^{-+},r_+,r_-,N,\vec{A}_0)
\nonumber\\
&&\hspace{1.7cm}
+{\bf Z}_9(\hat{R}_1^{-+},r_+,r_-,N,\vec{A}_0)
+{\bf Z}_{10}(\hat{R}_2^{-+},r_+,r_-,N,\vec{A}_0),
\nonumber\\
\ea
where we used the notation 
$\hat{R}_1^{++\,3}=\hat{R}_1^{++}\hat{R}_1^{++}\hat{R}_1^{++}$. The expressions of the functions W and Z's can 
be extracted by a symbolic manipulation of the coefficients $\vec{D}_{n}^{++}$. 
We have included one of the projections for completeness in an appendix for the interested reader. 

In $x$-space, the structure of the (2nd) truncated (or $\alpha_s^3$) NNLO solution can be expressed as a sequence of convolution products of the form  

\ba
&&\vec{f}_{O(\alpha_s^3)}^{NNLO}(x,\alpha_s)=\left[\hat{L}(x)+
\alpha_s\hat{U}_1(x)\otimes\hat{L}(x)
-\alpha_0\hat{L}(x)\otimes\hat{U}_1(x)+
\alpha_s^2\hat{U}_2(x)\otimes\hat{L}(x)
\right.\nonumber\\
&&\hspace{2cm}\left.
-\alpha_s\alpha_0\hat{U}_1(x)\otimes\hat{L}(x)\otimes\hat{U}_1(x)
+\alpha_0^2\hat{L}(x)\otimes\left(\hat{U}_1(x)\otimes
\hat{U}_1(x)-\hat{U}_2(x)\right)
\right.\nonumber\\
&&\hspace{2cm}\left.
+\alpha_s^3\hat{U}_3(x)\otimes\hat{L}(x)
+\alpha_s\alpha_0^2\hat{U}_1(x)\otimes\hat{L}(x)\otimes\left(\hat{U}_1(x)
\otimes\hat{U}_1(x)-\hat{U}_2(x)\right)
\right.\nonumber\\
&&\hspace{2cm}\left.
-\alpha_s^2\alpha_0\hat{U}_2(x)\otimes\hat{L}(x)\otimes\hat{U}_1(x)
-\alpha_0^3\hat{L}(x)\otimes
\right.\nonumber\\
&&\hspace{2cm}\left.
\left(\hat{U}_1(x)\otimes\hat{U}_1(x)\otimes\hat{U}_1(x)
-\hat{U}_1(x)\otimes\hat{U}_2(x)-\hat{U}_2(x)\otimes\hat{U}_1(x)
+\hat{U}_3(x)\right)
\right]\otimes\vec{f}(x,\alpha_0),
\nonumber\\
\ea
and is reproduced by the $\alpha_s^3$ logarithmic expansion
\ba
\vec{\tilde{f}}(x,\alpha_s)=\sum_{n=0}^{\infty}\frac{L^n}{n!}
\left[\vec{A}_n(x)+\alpha_s\vec{B}_n(x)+\alpha_s^2\vec{C}_n(x)
+\alpha_s^3 \vec{D}_n(x)\right]\,,
\ea
with the initial condition
\ba
\vec{f}(x,\alpha_0)=\vec{A}_0(x).
\ea
The study of higher order truncates is performed numerically with the implementation of the
generalized recursion relations (\ref{genrecnnlo}) given in the previous sections.

\section{Comparison with existing programs}

In this section we present a numerical test of our solution algorithm.
For this aim, we compare the results of a computer program that implements our method 
with the results of QCD-Pegasus \cite{Vogt3}, a PDF evolution program based on Mellin-space inversion, 
which has been used by the QCD Working Group to set some
benchmark results \cite{leshouches,parton2005}. In the following, we refer to these results as to the \emph{benchmark}.

In the tables we are going to show in this paper, we set the renormalization and
factorization scales to be equal, and we adopt the fixed flavor number scheme. The
final evolution scale is $\mu_{F}^{2}=10^{4}\,\textrm{GeV}^{2}$.
We limit ourselves to this case because it is enough to test the reliability of our method.
A more lenghty and detailed analysis,
which will take into account many other cases with renormalization scale dependence
and variable flavor number scheme, will be presented elsewhere.

As in the published benchmarks, we start the evolution at $\mu_{F,0}^{2}=2\,\textrm{GeV}^{2}$,
where the test input distributions, regardless of the perturbative order, are parametrized
by the following toy model\begin{eqnarray}
xu_{v}(x) & = & 5.107200x^{0.8}(1-x)^{3}\nonumber \\
xd_{v}(x) & = & 3.064320x^{0.8}(1-x)^{4}\nonumber \\
xg(x) & = & 1.700000x^{-0.1}(1-x)^{5}\nonumber \\
x\bar{d}(x) & = & 0.1939875x^{-0.1}(1-x)^{6}\nonumber \\
x\bar{u}(x) & = & (1-x)x\bar{d}(x)\nonumber \\
xs(x)=x\bar{s}(x) & = & 0.2x(\bar{u}+\bar{d})(x)\end{eqnarray}
and the running coupling has the value\begin{equation}
\alpha_{s}(\mu_{R,0}^{2}=2\,\textrm{GeV}^{2})=0.35.\end{equation}
We remind that $q_{v}=q-\bar{q}$, $q_{+}=q+\bar{q}$,
$L_{\pm}=\bar{d}\pm\bar{u}$. Our results are obtained using the exact solution method for the nonsinglet and the
LO singlet, and the $\kappa$-th truncate method for the NLO and NNLO singlet, with $\kappa=10$.
In each entry in the tables, the first number is our result and the second is the difference
between our results and the benchmark.

In Table \ref{LOtable} we compare our results at leading order with the results 
reported in Table 2 of \cite{leshouches}. The agreement is excellent for any
value of $x$ for the nonsinglet ($xu_{v}$ and $xd_{v}$); regarding the singlet
($xg$ column), the agreement is excellent except at very high $x$: we have a sizeable
difference at $x=0.9$. In Table \ref{NLOtable} we analyze the next-to-leading order evolution; the results for the 
proposed benchmarks are reported in Table 3 of \cite{leshouches}. The agreement is very good
for any value of $x$ for the nonsinglet; for the singlet the agreement is good,
except at very high $x$ ($x=0.9$).

Moving to the NNLO case (Table \ref{NNLOtable}, the benchmarks are shown in Table 14 of \cite{parton2005}), 
some comments are in order.
We don't solve the non-singlet equation as in PEGASUS, 
since (\ref{eqnn}) admits an exact solution 
(\ref{nnlotest}), which in PEGASUS is obtained only by iteration of truncated solutions. 
Our implementation is based on the exact solution presented in this work.
The discrepancy between our results and PEGASUS are of the order of few percent,
and they become large in the gluon case at $x=0.9$, as for the lower orders. Another comments 
should be made for the sea asimmetry of the $s$ quark, which is nonvanishing at NNLO. 
In this case we have a sizeable
relative discrepancy for any value of $x$ (reported in the column $xs_{v}$), but it is evident 
that this asymmetry is quite small, especially if compared with the column $xs_{+}$, whose entries 
are several orders of magnitude larger. This means that $xs$ and $x\bar{s}$ should be comparable and their 
difference therefore pretty small.
We have to notice that $xs_{v}$
 cannot be computed directly by a single evolution equation.
Indeed, it is computed by a difference between very close numbers, a procedure
that amplifies the relative error.

\begin{sidewaystable}
\begin{small}
\begin{center}
\begin{tabular}{|c||c|c|c|c|c|c|c|c|}
\hline
\multicolumn{8}{|c|}{LO, $n_{f}=4$, $\mu_{F}^{2}=\mu_{R}^{2}=10^{4}\,\textrm{GeV}^{2}$}
\tabularnewline
\hline
$x$&
$xu_{v}$&
$xd_{v}$&
$xL_{-}$&
$2 xL_{+}$&
$xs_{+}$&
$xc_{+}$&
$xg$\tabularnewline
\hline
\hline
$10^{-7}$&
$5.7722\cdot10^{-5}$&
$3.4343\cdot10^{-5}$&
$7.6527\cdot10^{-7}$&
$9.9465\cdot10^{+1}$&
$4.8642\cdot10^{+1}$&
$4.7914\cdot10^{+1}$&
$1.3162\cdot10^{+3}$\tabularnewline
&
$0.0000\cdot10^{-5}$&
$0.0000\cdot10^{-5}$&
$0.0000\cdot10^{-7}$&
$0.0000\cdot10^{+1}$&
$0.0000\cdot10^{+1}$&
$0.0000\cdot10^{+1}$&
$0.0000\cdot10^{+3}$\tabularnewline
\hline
$10^{-6}$&
$3.3373\cdot10^{-4}$&
$1.9800\cdot10^{-4}$&
$5.0137\cdot10^{-6}$&
$5.0259\cdot10^{+1}$&
$2.4263\cdot10^{+1}$&
$2.3685\cdot10^{+1}$&
$6.0008\cdot10^{+2}$\tabularnewline
&
$0.0000\cdot10^{-4}$&
$0.0000\cdot10^{-4}$&
$0.0000\cdot10^{-6}$&
$0.0000\cdot10^{+1}$&
$0.0000\cdot10^{+1}$&
$0.0000\cdot10^{+1}$&
$0.0000\cdot10^{+2}$\tabularnewline
\hline
$10^{-5}$&
$1.8724\cdot10^{-3}$&
$1.1065\cdot10^{-3}$&
$3.1696\cdot10^{-5}$&
$2.4378\cdot10^{+1}$&
$1.1501\cdot10^{+1}$&
$1.1042\cdot10^{+1}$&
$2.5419\cdot10^{+2}$\tabularnewline
&
$0.0000\cdot10^{-3}$&
$0.0000\cdot10^{-3}$&
$0.0000\cdot10^{-5}$&
$0.0000\cdot10^{+1}$&
$0.0000\cdot10^{+1}$&
$0.0000\cdot10^{+1}$&
$0.0000\cdot10^{+2}$\tabularnewline
\hline
$10^{-4}$&
$1.0057\cdot10^{-2}$&
$5.9076\cdot10^{-3}$&
$1.9071\cdot10^{-4}$&
$1.1323\cdot10^{+1}$&
$5.1164\cdot10^{+0}$&
$4.7530\cdot10^{+0}$&
$9.7371\cdot10^{+1}$\tabularnewline
&
$0.0000\cdot10^{-2}$&
$0.0000\cdot10^{-3}$&
$0.0000\cdot10^{-4}$&
$0.0000\cdot10^{+1}$&
$0.0000\cdot10^{+0}$&
$0.0000\cdot10^{+0}$&
$0.0000\cdot10^{+1}$\tabularnewline
\hline
$10^{-3}$&
$5.0392\cdot10^{-2}$&
$2.9296\cdot10^{-2}$&
$1.0618\cdot10^{-3}$&
$5.0324\cdot10^{+0}$&
$2.0918\cdot10^{+0}$&
$1.8089\cdot10^{+0}$&
$3.2078\cdot10^{+1}$\tabularnewline
&
$0.0000\cdot10^{-2}$&
$0.0000\cdot10^{-2}$&
$0.0000\cdot10^{-3}$&
$0.0000\cdot10^{+0}$&
$0.0000\cdot10^{+0}$&
$0.0000\cdot10^{+0}$&
$0.0000\cdot10^{+1}$\tabularnewline
\hline
$10^{-2}$&
$2.1955\cdot10^{-1}$&
$1.2433\cdot10^{-1}$&
$4.9731\cdot10^{-3}$&
$2.0433\cdot10^{+0}$&
$7.2814\cdot10^{-1}$&
$5.3247\cdot10^{-1}$&
$8.0546\cdot10^{+0}$\tabularnewline
&
$0.0000\cdot10^{-1}$&
$0.0000\cdot10^{-1}$&
$0.0000\cdot10^{-3}$&
$0.0000\cdot10^{+0}$&
$0.0000\cdot10^{-1}$&
$0.0000\cdot10^{-1}$&
$0.0000\cdot10^{+0}$\tabularnewline
\hline
$0.1$&
$5.7267\cdot10^{-1}$&
$2.8413\cdot10^{-1}$&
$1.0470\cdot10^{-2}$&
$4.0832\cdot10^{-1}$&
$1.1698\cdot10^{-1}$&
$5.8864\cdot10^{-2}$&
$8.8766\cdot10^{-1}$\tabularnewline
&
$0.0000\cdot10^{-1}$&
$0.0000\cdot10^{-1}$&
$0.0000\cdot10^{-2}$&
$0.0000\cdot10^{-1}$&
$0.0000\cdot10^{-1}$&
$0.0000\cdot10^{-2}$&
$0.0000\cdot10^{-1}$\tabularnewline
\hline
$0.3$&
$3.7925\cdot10^{-1}$&
$1.4186\cdot10^{-1}$&
$3.3029\cdot10^{-3}$&
$4.0165\cdot10^{-2}$&
$1.0516\cdot10^{-2}$&
$4.1380\cdot10^{-3}$&
$8.2676\cdot10^{-2}$\tabularnewline
&
$0.0000\cdot10^{-1}$&
$0.0000\cdot10^{-1}$&
$0.0000\cdot10^{-3}$&
$0.0000\cdot10^{-2}$&
$0.0000\cdot10^{-2}$&
$+0.0001\cdot10^{-3}$&
$0.0000\cdot10^{-2}$\tabularnewline
\hline
$0.5$&
$1.3476\cdot10^{-1}$&
$3.5364\cdot10^{-2}$&
$4.2815\cdot10^{-4}$&
$2.8624\cdot10^{-3}$&
$7.3137\cdot10^{-4}$&
$2.6481\cdot10^{-4}$&
$7.9242\cdot10^{-3}$\tabularnewline
&
$0.0000\cdot10^{-1}$&
$0.0000\cdot10^{-2}$&
$0.0000\cdot10^{-4}$&
$0.0000\cdot10^{-3}$&
$-0.0001\cdot10^{-4}$&
$0.0000\cdot10^{-4}$&
$+0.0002\cdot10^{-3}$\tabularnewline
\hline
$0.7$&
$2.3123\cdot10^{-2}$&
$3.5943\cdot10^{-3}$&
$1.5868\cdot10^{-5}$&
$6.8970\cdot10^{-5}$&
$1.7730\cdot10^{-5}$&
$6.5593\cdot10^{-6}$&
$3.7301\cdot10^{-4}$\tabularnewline
&
$0.0000\cdot10^{-2}$&
$0.0000\cdot10^{-3}$&
$0.0000\cdot10^{-5}$&
$+0.0009\cdot10^{-5}$&
$+0.0005\cdot10^{-5}$&
$+0.0044\cdot10^{-6}$&
$-0.0010\cdot10^{-4}$\tabularnewline
\hline
$0.9$&
$4.3443\cdot10^{-4}$&
$2.2287\cdot10^{-5}$&
$1.1042\cdot10^{-8}$&
$3.3030\cdot10^{-8}$&
$8.5607\cdot10^{-9}$&
$3.2577\cdot10^{-9}$&
$1.3887\cdot10^{-6}$\tabularnewline
&
$0.0000\cdot10^{-4}$&
$0.0000\cdot10^{-5}$&
$0.0000\cdot10^{-8}$&
$-0.3263\cdot10^{-8}$&
$-0.1631\cdot10^{-8}$&
$-1.6316\cdot10^{-9}$&
$+0.2969\cdot10^{-6}$\tabularnewline
\hline
\end{tabular}
\end{center}

\caption{Comparison between our and the benchmark results at LO. In each entry, the first number is our result and the second one is the difference between our result and the benchmark. The benchmark values are reported in Table 2 of \cite{leshouches}}
\label{LOtable}
\end{small}
\end{sidewaystable}

\begin{sidewaystable}
\begin{small}
\begin{center}
\begin{tabular}{|c||c|c|c|c|c|c|c|c|}
\hline
\multicolumn{8}{|c|}{NLO, $n_{f}=4$, $\mu_{F}^{2}=\mu_{R}^{2}=10^{4}\,\textrm{GeV}^{2}$}
\tabularnewline
\hline
$x$&
$xu_{v}$&
$xd_{v}$&
$xL_{-}$&
$2 xL_{+}$&
$xs_{+}$&
$xc_{+}$&
$xg$\tabularnewline
\hline
\hline
$10^{-7}$&
$1.0620\cdot10^{-4}$&
$6.2353\cdot10^{-5}$&
$4.2455\cdot10^{-6}$&
$1.3710\cdot10^{+2}$&
$6.7469\cdot10^{+1}$&
$6.6750\cdot10^{+1}$&
$1.1517\cdot10^{+3}$\tabularnewline
&
$+0.0004\cdot10^{-4}$&
$+0.0025\cdot10^{-5}$&
$+0.0015\cdot10^{-6}$&
$+0.0112\cdot10^{+2}$&
$+0.0556\cdot10^{+1}$&
$+0.0555\cdot10^{+1}$&
$+0.0034\cdot10^{+3}$\tabularnewline
\hline
$10^{-6}$&
$5.4196\cdot10^{-4}$&
$3.1730\cdot10^{-4}$&
$1.9247\cdot10^{-5}$&
$6.8896\cdot10^{+1}$&
$3.3592\cdot10^{+1}$&
$3.3021\cdot10^{+1}$&
$5.4048\cdot10^{+2}$\tabularnewline
&
$+0.0019\cdot10^{-4}$&
$+0.0011\cdot10^{-4}$&
$+0.0006\cdot10^{-5}$&
$+0.0500\cdot10^{+1}$&
$+0.0250\cdot10^{+1}$&
$+0.0250\cdot10^{+1}$&
$+0.0137\cdot10^{+2}$\tabularnewline
\hline
$10^{-5}$&
$2.6878\cdot10^{-3}$&
$1.5682\cdot10^{-3}$&
$8.3598\cdot10^{-5}$&
$3.2936\cdot10^{+1}$&
$1.5788\cdot10^{+1}$&
$1.5335\cdot10^{+1}$&
$2.3578\cdot10^{+2}$\tabularnewline
&
$+0.0008\cdot10^{-3}$&
$+0.0005\cdot10^{-3}$&
$+0.0023\cdot10^{-5}$&
$+0.0208\cdot10^{+1}$&
$+0.0103\cdot10^{+1}$&
$+0.0104\cdot10^{+1}$&
$+0.0050\cdot10^{+2}$\tabularnewline
\hline
$10^{-4}$&
$1.2844\cdot10^{-2}$&
$7.4576\cdot10^{-3}$&
$3.4919\cdot10^{-4}$&
$1.4824\cdot10^{+1}$&
$6.8744\cdot10^{+0}$&
$6.5156\cdot10^{+0}$&
$9.3026\cdot10^{+1}$\tabularnewline
&
$+0.0003\cdot10^{-2}$&
$+0.0018\cdot10^{-3}$&
$+0.0008\cdot10^{-4}$&
$+0.0078\cdot10^{+1}$&
$+0.0389\cdot10^{+0}$&
$+0.0387\cdot10^{+0}$&
$+0.0154\cdot10^{+1}$\tabularnewline
\hline
$10^{-3}$&
$5.7937\cdot10^{-2}$&
$3.3343\cdot10^{-2}$&
$1.4164\cdot10^{-3}$&
$6.1899\cdot10^{+0}$&
$2.6783\cdot10^{+0}$&
$2.4001\cdot10^{+0}$&
$3.1540\cdot10^{+1}$\tabularnewline
&
$+0.0011\cdot10^{-2}$&
$+0.0006\cdot10^{-2}$&
$+0.0002\cdot10^{-3}$&
$+0.0251\cdot10^{+0}$&
$+0.0124\cdot10^{+0}$&
$+0.0123\cdot10^{+0}$&
$+0.0038\cdot10^{+1}$\tabularnewline
\hline
$10^{-2}$&
$2.3029\cdot10^{-1}$&
$1.2930\cdot10^{-1}$&
$5.3258\cdot10^{-3}$&
$2.2587\cdot10^{+0}$&
$8.4518\cdot10^{-1}$&
$6.5540\cdot10^{-1}$&
$8.1120\cdot10^{+0}$\tabularnewline
&
$+0.0003\cdot10^{-1}$&
$+0.0002\cdot10^{-1}$&
$+0.0007\cdot10^{-3}$&
$+0.0060\cdot10^{+0}$&
$+0.0298\cdot10^{-1}$&
$+0.0294\cdot10^{-1}$&
$+0.0054\cdot10^{+0}$\tabularnewline
\hline
$0.1$&
$5.5456\cdot10^{-1}$&
$2.7338\cdot10^{-1}$&
$1.0012\cdot10^{-2}$&
$3.9392\cdot10^{-1}$&
$1.1517\cdot10^{-1}$&
$6.0619\cdot10^{-2}$&
$8.9872\cdot10^{-1}$\tabularnewline
&
$+0.0004\cdot10^{-1}$&
$+0.0002\cdot10^{-1}$&
$+0.0001\cdot10^{-2}$&
$+0.0056\cdot10^{-1}$&
$+0.0028\cdot10^{-1}$&
$+0.0268\cdot10^{-2}$&
$+0.0005\cdot10^{-1}$\tabularnewline
\hline
$0.3$&
$3.5395\cdot10^{-1}$&
$1.3158\cdot10^{-1}$&
$3.0363\cdot10^{-3}$&
$3.5884\cdot10^{-2}$&
$9.2210\cdot10^{-3}$&
$3.4066\cdot10^{-3}$&
$8.3415\cdot10^{-2}$\tabularnewline
&
$+0.0002\cdot10^{-1}$&
$0.0000\cdot10^{-1}$&
$+0.0001\cdot10^{-3}$&
$+0.0036\cdot10^{-2}$&
$+0.0180\cdot10^{-3}$&
$+0.0176\cdot10^{-3}$&
$-0.0036\cdot10^{-2}$\tabularnewline
\hline
$0.5$&
$1.2271\cdot10^{-1}$&
$3.1968\cdot10^{-2}$&
$3.8266\cdot10^{-4}$&
$2.4149\cdot10^{-3}$&
$5.8539\cdot10^{-4}$&
$1.7068\cdot10^{-4}$&
$8.0412\cdot10^{-3}$\tabularnewline
&
$0.0000\cdot10^{-1}$&
$+0.0001\cdot10^{-2}$&
$+0.0001\cdot10^{-4}$&
$+0.0023\cdot10^{-3}$&
$+0.0115\cdot10^{-4}$&
$+0.0113\cdot10^{-4}$&
$-0.0061\cdot10^{-3}$\tabularnewline
\hline
$0.7$&
$2.0429\cdot10^{-2}$&
$3.1474\cdot10^{-3}$&
$1.3701\cdot10^{-5}$&
$5.3703\cdot10^{-5}$&
$1.2432\cdot10^{-5}$&
$2.8201\cdot10^{-6}$&
$3.8654\cdot10^{-4}$\tabularnewline
&
$0.0000\cdot10^{-2}$&
$+0.0001\cdot10^{-3}$&
$0.0000\cdot10^{-5}$&
$+0.0081\cdot10^{-5}$&
$+0.0039\cdot10^{-5}$&
$+0.0394\cdot10^{-6}$&
$-0.0067\cdot10^{-4}$\tabularnewline
\hline
$0.9$&
$3.6097\cdot10^{-4}$&
$1.8317\cdot10^{-5}$&
$8.9176\cdot10^{-9}$&
$1.6196\cdot10^{-8}$&
$1.6717\cdot10^{-9}$&
$-2.6084\cdot10^{-9}$&
$1.8308\cdot10^{-6}$\tabularnewline
&
$+0.0001\cdot10^{-4}$&
$0.0000\cdot10^{-5}$&
$-0.0054\cdot10^{-9}$&
$-0.4724\cdot10^{-8}$&
$-2.3673\cdot10^{-9}$&
$-2.3681\cdot10^{-9}$&
$+0.6181\cdot10^{-6}$\tabularnewline
\hline
\end{tabular}
\end{center}

\caption{Same as in Table \ref{LOtable} in the NLO case. The benchmark values are reported in Table 3 of \cite{leshouches}}
\label{NLOtable}
\end{small}
\end{sidewaystable}

\begin{sidewaystable}
\begin{small}
\begin{center}
\begin{tabular}{|c||c|c|c|c|c|c|c|c|}
\hline
\multicolumn{9}{|c|}{NNLO, $n_{f}=4$, $\mu_{F}^{2}=\mu_{R}^{2}=10^{4}\,\textrm{GeV}^{2}$}
\tabularnewline
\hline
$x$&
$xu_{v}$&
$xd_{v}$&
$xL_{-}$&
$2 xL_{+}$&
$xs_{v}$&
$xs_{+}$&
$xc_{+}$&
$xg$\tabularnewline
\hline
\hline
$10^{-7}$&
$1.4069\cdot10^{-4}$&
$9.0435\cdot10^{-5}$&
$5.5759\cdot10^{-6}$&
$1.4184\cdot10^{+2}$&
$1.9569\cdot10^{-5}$&
$6.9844\cdot10^{+1}$&
$6.9127\cdot10^{+1}$&
$1.0519\cdot10^{+3}$\tabularnewline
&
$-0.1218\cdot10^{-4}$&
$-0.1201\cdot10^{-4}$&
$-0.1259\cdot10^{-6}$&
$+0.0994\cdot10^{+2}$&
$-1.1868\cdot10^{-5}$&
$+0.4967\cdot10^{+1}$&
$+0.4966\cdot10^{+1}$&
$+0.0543\cdot10^{+3}$\tabularnewline
\hline
$10^{-6}$&
$6.5756\cdot10^{-4}$&
$4.0826\cdot10^{-4}$&
$2.4722\cdot10^{-5}$&
$7.1794\cdot10^{+1}$&
$5.8862\cdot10^{-5}$&
$3.5043\cdot10^{+1}$&
$3.4474\cdot10^{+1}$&
$5.1093\cdot10^{+2}$\tabularnewline
&
$-0.3420\cdot10^{-4}$&
$-0.3458\cdot10^{-4}$&
$-0.0688\cdot10^{-5}$&
$+0.3295\cdot10^{+1}$&
$-3.5417\cdot10^{-5}$&
$+0.1646\cdot10^{+1}$&
$+0.1646\cdot10^{+1}$&
$+0.1969\cdot10^{+2}$\tabularnewline
\hline
$10^{-5}$&
$3.0260\cdot10^{-3}$&
$1.8199\cdot10^{-3}$&
$1.0393\cdot10^{-4}$&
$3.4373\cdot10^{+1}$&
$1.4230\cdot10^{-4}$&
$1.6509\cdot10^{+1}$&
$1.6057\cdot10^{+1}$&
$2.2899\cdot10^{+2}$\tabularnewline
&
$-0.0721\cdot10^{-3}$&
$-0.0775\cdot10^{-3}$&
$-0.0326\cdot10^{-4}$&
$+0.0902\cdot10^{+1}$&
$-0.8560\cdot10^{-4}$&
$+0.0450\cdot10^{+1}$&
$+0.0450\cdot10^{+1}$&
$+0.0602\cdot10^{+2}$\tabularnewline
\hline
$10^{-4}$&
$1.3656\cdot10^{-2}$&
$8.0052\cdot10^{-3}$&
$4.1299\cdot10^{-4}$&
$1.5403\cdot10^{+1}$&
$2.2837\cdot10^{-4}$&
$7.1661\cdot10^{+0}$&
$6.8085\cdot10^{+0}$&
$9.2125\cdot10^{+1}$\tabularnewline
&
$-0.0066\cdot10^{-2}$&
$-0.0967\cdot10^{-3}$&
$-0.1259\cdot10^{-4}$&
$+0.0199\cdot10^{+1}$&
$-1.3807\cdot10^{-4}$&
$+0.0991\cdot10^{+0}$&
$+0.0988\cdot10^{+0}$&
$+0.1457\cdot10^{+1}$\tabularnewline
\hline
$10^{-3}$&
$5.9360\cdot10^{-2}$&
$3.4135\cdot10^{-2}$&
$1.5650\cdot10^{-3}$&
$6.3657\cdot10^{+0}$&
$8.9572\cdot10^{-5}$&
$2.7684\cdot10^{+0}$&
$2.4913\cdot10^{+0}$&
$3.1592\cdot10^{+1}$\tabularnewline
&
$+0.0200\cdot10^{-2}$&
$+0.0085\cdot10^{-2}$&
$-0.0358\cdot10^{-3}$&
$+0.0427\cdot10^{+0}$&
$-0.5522\cdot10^{-4}$&
$+0.0210\cdot10^{+0}$&
$+0.0209\cdot10^{+0}$&
$+0.0243\cdot10^{+1}$\tabularnewline
\hline
$10^{-2}$&
$2.3139\cdot10^{-1}$&
$1.2958\cdot10^{-1}$&
$5.5064\cdot10^{-3}$&
$2.2868\cdot10^{+0}$&
$-3.5702\cdot10^{-4}$&
$8.6094\cdot10^{-1}$&
$6.7224\cdot10^{-1}$&
$8.1503\cdot10^{+0}$\tabularnewline
&
$+0.0061\cdot10^{-1}$&
$+0.0039\cdot10^{-1}$&
$-0.0624\cdot10^{-3}$&
$+0.0116\cdot10^{+0}$&
$+2.1611\cdot10^{-4}$&
$+0.0592\cdot10^{-1}$&
$+0.0601\cdot10^{-1}$&
$+0.0122\cdot10^{+0}$\tabularnewline
\hline
$0.1$&
$5.5125\cdot10^{-1}$&
$2.7142\cdot10^{-1}$&
$9.9834\cdot10^{-3}$&
$3.9119\cdot10^{-1}$&
$-1.9045\cdot10^{-4}$&
$1.1453\cdot10^{-1}$&
$6.0520\cdot10^{-2}$&
$8.9909\cdot10^{-1}$\tabularnewline
&
$-0.0052\cdot10^{-1}$&
$-0.0023\cdot10^{-1}$&
$-0.0040\cdot10^{-2}$&
$+0.0100\cdot10^{-1}$&
$+1.1582\cdot10^{-4}$&
$+0.0067\cdot10^{-1}$&
$+0.0747\cdot10^{-2}$&
$-0.0654\cdot10^{-1}$\tabularnewline
\hline
$0.3$&
$3.5017\cdot10^{-1}$&
$1.3005\cdot10^{-1}$&
$3.0025\cdot10^{-3}$&
$3.4975\cdot10^{-2}$&
$-1.9830\cdot10^{-5}$&
$8.8758\cdot10^{-3}$&
$3.1421\cdot10^{-3}$&
$8.3041\cdot10^{-2}$\tabularnewline
&
$-0.0054\cdot10^{-1}$&
$-0.0020\cdot10^{-1}$&
$-0.0073\cdot10^{-3}$&
$-0.0383\cdot10^{-2}$&
$+1.2061\cdot10^{-5}$&
$-0.1722\cdot10^{-3}$&
$-0.1640\cdot10^{-3}$&
$-0.1145\cdot10^{-2}$\tabularnewline
\hline
$0.5$&
$1.2099\cdot10^{-1}$&
$3.1485\cdot10^{-2}$&
$3.7667\cdot10^{-4}$&
$2.1876\cdot10^{-3}$&
$-1.6924\cdot10^{-6}$&
$4.8155\cdot10^{-4}$&
$7.4120\cdot10^{-5}$&
$7.9784\cdot10^{-3}$\tabularnewline
&
$-0.0018\cdot10^{-1}$&
$-0.0043\cdot10^{-2}$&
$-0.0075\cdot10^{-4}$&
$-0.1991\cdot10^{-3}$&
$+1.0291\cdot10^{-6}$&
$-0.9810\cdot10^{-4}$&
$-0.9758\cdot10^{-4}$&
$-0.1342\cdot10^{-3}$\tabularnewline
\hline
$0.7$&
$2.0052\cdot10^{-2}$&
$3.0849\cdot10^{-3}$&
$1.3411\cdot10^{-5}$&
$1.5984\cdot10^{-5}$&
$-6.2854\cdot10^{-8}$&
$-6.1500\cdot10^{-6}$&
$-1.5545\cdot10^{-5}$&
$3.8226\cdot10^{-4}$\tabularnewline
&
$-0.0025\cdot10^{-2}$&
$-0.0037\cdot10^{-3}$&
$-0.0023\cdot10^{-5}$&
$-3.8260\cdot10^{-5}$&
$+0.3821\cdot10^{-7}$&
$-1.9084\cdot10^{-5}$&
$-1.9075\cdot10^{-5}$&
$-0.0722\cdot10^{-4}$\tabularnewline
\hline
$0.9$&
$3.5078\cdot10^{-4}$&
$1.7767\cdot10^{-5}$&
$8.6326\cdot10^{-9}$&
$-6.4293\cdot10^{-7}$&
$-9.1828\cdot10^{-11}$&
$-3.2776\cdot10^{-7}$&
$-3.3190\cdot10^{-7}$&
$1.9280\cdot10^{-6}$\tabularnewline
&
$-0.0033\cdot10^{-4}$&
$-0.0016\cdot10^{-5}$&
$-0.0184\cdot10^{-9}$&
$-6.6986\cdot10^{-7}$&
$+0.5579\cdot10^{-10}$&
$-3.3487\cdot10^{-7}$&
$-3.3487\cdot10^{-7}$&
$+0.7144\cdot10^{-6}$\tabularnewline
\hline
\end{tabular}
\end{center}

\caption{Same as in Table \ref{LOtable} in the NNLO case. The benchmark values are reported in Table 14 of \cite{parton2005}}
\label{NNLOtable}
\end{small}
\end{sidewaystable}

\section{Conclusions}

We have shown that logarithmic expansions identified in $x$-space and
implemented in this space carry the same information as the solution
of evolution equations in Mellin space. This has been obtained by the introduction of new and
generalized expansions that we expect to be very useful in order to establish benchmarks for the evolution of the pdf's at the
LHC. Our analysis has been presented up to NNLO.
We have also shown how exact expansions can be derived. We have presented analytical proofs of the equivalence and clarified the role of previous similar analysis which were quite limited in their reach. The numerical implementation of our results will be illustrated in
a forthcoming paper where various comparisons between our approach and other approaches will be analized
thoroughly. There are in fact several issues which are still unclear in this area and concern the role
of the NNLO effects in the evolution and in the hard scatterings, the role of the theoretical errors
in the determination of the pdf's, whether they dominate over the NNLO effects or not, and the
impact of the choices of various truncations in the determination of the numerical solution of the pdf's, along
the lines of our work. Similar analysis can be performed by other methods, but we think
that it is important, in the search for precise determination of cross sections at the LHC, to state clearly
which algorithm is implemented and what accuracy is retained, with a particular attention to the issues connected 
to the resummation of the perturbative expansion \cite{Catani}. Our work, here, has been limited to a (fixed order) NNLO 
analysis. We hope that our analysis has shown conclusively that x-space approaches have a very solid base and provide a simple view on the structure
of the solutions of the DGLAP equations, valid to all orders. We have also shown a numerical comparison 
of our results against those obtained using PEGASUS, for a specific setting. The overall agreement, as we have seen, 
is very good down to very small x-values. We will return on this and other related points in future work.

\centerline{\bf Acknowledgements}
C.C. thanks A. Faraggi, N. Irges, T. Tomaras  and K. Tamvakis and the theory groups at the University of
Liverpool, Crete and Ioannina for hospitality. The work of A.C. is supported by the University of Crete.
The work of C.C. is partly supported by INFN of Italy and the work of M.G. by MIUR of Italy.

\section{Appendix A. Derivation of the recursion relations at NNLO}
As an illustration we have included here a derivation of the recursion relations
for the first truncated ansatz of $O(\alpha_s^2)$ that appears at NNLO.

Inserting the NNLO truncated ansatz for the solution into the
DGLAP equation  we get at the left-hand-side of the defining equation
\begin{eqnarray}
&&\sum_{n=1}^{\infty}\left\{ \frac{A_{n}(x)}{n!}nL^{n-1}
\frac{\beta(\alpha_{s})}{\alpha_{s}}+\alpha_{s}\frac{B_{n}(x)}{n!}nL^{n-1}
\frac{\beta(\alpha_{s})}{\alpha_{s}}\right.\nonumber \\
&&\left.\qquad+\alpha_{s}^{2}\frac{C_{n}(x)}{n!}nL^{n-1}
\frac{\beta(\alpha_{s})}{\alpha_{s}}\right\} \nonumber \\
&&+\sum_{n=0}^{\infty}\left\{ \beta(\alpha_{s})\frac{B_{n}(x)}{n!}L^{n}+
2\alpha_{s}\beta(\alpha_{s})\frac{C_{n}(x)}{n!}L^{n}\right\} .
\label{eq:ansatz2}
\end{eqnarray}
Note that the first sum starts at $n=1$, because the $n=0$ term
in (\ref{eq:ansatz2}) does not have a $Q^{2}$ dependence. Sending
$n\rightarrow n+1$ in the first sum, using the three-loop expansion
of the beta function (\ref{eq:beta_exp}) and neglecting all the terms
of order $\alpha_{s}^{4}$ or higher, the previous formula becomes
\begin{eqnarray}
&&\sum_{n=0}^{\infty}\left\{ \frac{A_{n+1}(x)}{n!}L^{n}
\left(-\frac{\beta_{0}}{4\pi}\alpha_{s}-\frac{\beta_{1}}{16\pi^{2}}\alpha_{s}^{2}-
\frac{\beta_{2}}{64\pi^{3}}\alpha_{s}^{3}\right)\right.\nonumber \\
&&+\frac{B_{n+1}(x)}{n!}L^{n}\left(-\frac{\beta_{0}}{4\pi}\alpha_{s}^{2}-
\frac{\beta_{1}}{16\pi^{2}}\alpha_{s}^{3}\right)+\frac{C_{n+1}(x)}{n!}L^{n}
\left(-\frac{\beta_{0}}{4\pi}\alpha_{s}^{3}\right)\nonumber \\
&&\left.+\frac{B_{n}(x)}{n!}L^{n}\left(-\frac{\beta_{0}}{4\pi}\alpha_{s}^{2}-
\frac{\beta_{1}}{16\pi^{2}}\alpha_{s}^{3}\right)+2\frac{C_{n}(x)}{n!}L^{n}
\left(-\frac{\beta_{0}}{4\pi}\alpha_{s}^{3}\right)\right\} .\label{eq:recrelLHS}
\end{eqnarray}
At this point we use the NNLO expansion of the kernels.
We get at the right-hand-side of the defining equation
\begin{eqnarray}
&&\sum_{n=0}^{\infty}\frac{L^{n}}{n!}\left\{ \frac{\alpha_{s}}{2\pi}\left[P^{(0)}
\otimes A_{n}\right](x)+\frac{\alpha_{s}^{2}}{4\pi^{2}}\left[P^{(1)}\otimes A_{n}\right](x)\right.\nonumber \\
&&\quad\quad\quad+\frac{\alpha_{s}^{3}}{8\pi^{3}}\left[P^{(2)}
\otimes A_{n}\right](x)+\frac{\alpha_{s}^{2}}{2\pi}\left[P^{(0)}
\otimes B_{n}\right](x)\nonumber \\
&&\left.\quad\quad\quad+\frac{\alpha_{s}^{3}}{4\pi^{2}}\left[P^{(1)}
\otimes B_{n}\right](x)+\frac{\alpha_{s}^{3}}{2\pi}\left[P^{(0)}
\otimes C_{n}\right](x)\right\}.
\label{eq:recrelRHS}
\end{eqnarray}
Equating (\ref{eq:recrelLHS}) and (\ref{eq:recrelRHS}) term by term
and grouping the terms proportional respectively to $\alpha_{s}$,
$\alpha_{s}^{2}$ and $\alpha_{s}^{3}$ we get the three desired recursion
relations (\ref{NNLOrec_nonsing}).
Setting $Q=Q_{0}$ in (\ref{nnloans}) we get
\begin{equation}
f(x,Q_{0}^{2})=A_{0}(x)+\alpha_{s}(Q_{0}^{2})B_{0}(x)+
\left(\alpha_{s}(Q^{2})\right)^{2}C_{0}(x).
\label{eq:boundary2}
\end{equation}
We have seen that the initial conditions should be chosen as
\begin{equation}
B_{0}(x)=C_{0}(x)=0,\qquad f(x,Q_{0}^{2})=A_{0}(x)
\end{equation}
in order to reproduce the moments of the truncated solution of the DGLAP 
equation.

\section{Appendix B. NNLO singlet truncated solution.}

Putting all the projections of the coefficients $\vec{C}_n$
into the NNLO singlet ansatz we get
\ba
&&\vec{f}(N,\alpha_s)=\sum_{n=0}^{\infty}\frac{L^n}{n!}
\left[\vec{A}_{n}^{++}+\vec{A}_{n}^{--}+\alpha_s
\left(\vec{B}_{n}^{++}+\vec{B}_{n}^{--}
+\vec{B}_{n}^{-+}+\vec{B}_{n}^{+-}\right)\right.\nonumber\\
&&\hspace{3.5cm}\left.+\alpha_s^2
\left(\vec{C}_{n}^{++}+\vec{C}_{n}^{--}
+\vec{C}_{n}^{-+}+\vec{C}_{n}^{+-}\right)\right]\,,
\ea
and exponentiating
\ba
&&\vec{f}(N,\alpha_s)= e_{+}\vec{A}_0 \left(\frac{\alpha_s}{\alpha_0}\right)^{r_+}+
e_{-}\vec{A}_0 \left(\frac{a_s}{\alpha_0}\right)^{r_-}+\nonumber\\
&&\hspace{1.5cm}\alpha_s\left\{
e_{+}\hat{R}_1e_{+}
\left(\frac{\alpha_s}{\alpha_0}\right)^{r_+}-
e_{+}\hat{R}_1e_{+}
\left(\frac{\alpha_s}{\alpha_0}\right)^{(r_+ -1)}+\right.\nonumber\\
&&\hspace{2cm}\left.
e_{-}\hat{R}_1e_{-}
\left(\frac{\alpha_s}{\alpha_0}\right)^{r_-}-
e_{-}\hat{R}_1e_{-}
\left(\frac{\alpha_s}{\alpha_0}\right)^{(r_- -1)}+\right.\nonumber\\
&&\hspace{2cm}\left.\frac{1}{(r_+ -r_- -1)}\left[
-e_{+}\hat{R}_1e_{-}
\left(\frac{\alpha_s}{\alpha_0}\right)^{r_-}+
e_{+}\hat{R}_1e_{-}
\left(\frac{\alpha_s}{\alpha_0}\right)^{(r_+ -1)}\right]+\right.\nonumber\\
&&\hspace{2cm}\left.\frac{1}{(r_- -r_+ -1)}\left[
-e_{-}\hat{R}_1e_{+}
\left(\frac{\alpha_s}{\alpha_0}\right)^{r_+}+
e_{-}\hat{R}_1e_{+}
\left(\frac{\alpha_s}{\alpha_0}\right)^{(r_- -1)}\right]\right\}\vec{A}_0
\nonumber\\
&&\hspace{1.5cm}+\alpha_s^2\left\{
\left(\frac{\alpha_s}{\alpha_0}\right)^{r_+}\left[
\frac{1}{2\alpha_s^2}
\left(e_{+}\hat{R}_1e_{+}\hat{R}_1e_{+}(\alpha_0-\alpha_s)^2
-e_{+}\hat{R}_2e_{+}(\alpha_0^2-\alpha_s^2)\right)
\right.\right.\nonumber\\
&&\hspace{4cm}\left.\left.
+\alpha_s\alpha_0\frac{e_{+}\hat{R}_1e_{-}\hat{R}_1e_{+}}
{\alpha_s^2((r_- -r_+)^2 -1)}
\left(\frac{\alpha_s}{\alpha_0}\right)^{r_--r_+}
\right.\right.\nonumber\\
&&\hspace{4cm}\left.\left.
+e_{+}\hat{R}_1e_{-}\hat{R}_1e_{+}\frac{\left((r_--r_+-1)\alpha_0^2
+(r_+-r_- -1)\alpha_s^2\right)}{2\alpha_s^2((r_- -r_+)^2 -1)}
\right]\right\}\vec{A}_0
\nonumber\\
&&\hspace{1.5cm}+\alpha_s^2\left\{
\left(\frac{\alpha_s}{\alpha_0}\right)^{r_-}\left[
\frac{1}{2\alpha_s^2}
\left(e_{-}\hat{R}_1e_{-}\hat{R}_1e_{-}(\alpha_0-\alpha_s)^2
-e_{-}\hat{R}_2e_{-}(\alpha_0^2-\alpha_s^2)\right)
\right.\right.\nonumber\\
&&\hspace{4cm}\left.\left.
+\alpha_s\alpha_0\frac{e_{-}\hat{R}_1e_{+}\hat{R}_1e_{-}}
{\alpha_s^2((r_- -r_+)^2 -1)}
\left(\frac{\alpha_s}{\alpha_0}\right)^{r_+-r_-}
\right.\right.\nonumber\\
&&\hspace{4cm}\left.\left.
+e_{-}\hat{R}_1e_{+}\hat{R}_1e_{-}\frac{\left((r_+-r_--1)\alpha_0^2
+(r_--r_+ -1)\alpha_s^2\right)}{2\alpha_s^2((r_- -r_+)^2 -1)}
\right]\right\}\vec{A}_0
\nonumber\\
&&\hspace{1.5cm}+\alpha_s^2\left\{
\left(\frac{\alpha_s}{\alpha_0}\right)^{r_+}
\left[
\frac{e_{+}\hat{R}_1e_{-}\hat{R}_1e_{-}\alpha_0^2}
{\alpha_s^2(1+r_--r_+)(2+r_- -r_+)}
+\frac{\left(e_{+}\hat{R}_1e_{+}\hat{R}_1e_{-}
-e_{+}\hat{R}_2e_{-}\right)\alpha_0^2}
{\alpha_s^2(2+r_- -r_+)}
\right.\right.\nonumber\\
&&\hspace{4cm}\left.\left.+
\frac{e_{+}\hat{R}_1e_{+}\hat{R}_1e_{-}\alpha_0}
{\alpha_s(1+r_- -r_+)}
\right]+
\right.\nonumber\\
&&\hspace{2.6cm}\left.
\left(\frac{\alpha_s}{\alpha_0}\right)^{r_-}
\left[
\frac{e_{+}\hat{R}_1e_{+}\hat{R}_1e_{-}}
{(1+r_--r_+)(2+r_- -r_+)}+
\frac{\left(e_{+}\hat{R}_2e_{-}
+e_{+}\hat{R}_1e_{-}\hat{R}_1e_{-}\right)}
{(2+r_- -r_+)}
\right.\right.\nonumber\\
&&\hspace{4cm}\left.\left.
-\frac{e_{+}\hat{R}_1e_{-}\hat{R}_1e_{-}\alpha_0}
{\alpha_s(1+r_--r_+)}\right]
\right\}\vec{A}_0
\nonumber\\
&&\hspace{1.5cm}+\alpha_s^2\left\{
\left(\frac{\alpha_s}{\alpha_0}\right)^{r_-}
\left[
\frac{e_{-}\hat{R}_1e_{+}\hat{R}_1e_{+}\alpha_0^2}
{(r_- -r_+-1)(r_--r_+-2)\alpha_s^2}
+\frac{\left(e_{-}\hat{R}_2e_{+}-
e_{-}\hat{R}_1e_{-}\hat{R}_1e_{+}\right)\alpha_0^2}
{(r_--r_+-2)\alpha_s^2}
\right.\right.\nonumber\\
&&\hspace{4cm}\left.\left.
+\frac{e_{-}\hat{R}_1e_{-}\hat{R}_1e_{+}\alpha_0}
{(r_--r_+-1)\alpha_s}
\right]
\right.\nonumber\\
&&\hspace{2.6cm}\left.
\left(\frac{\alpha_s}{\alpha_0}\right)^{r_+}
\left[
\frac{e_{-}\hat{R}_1e_{-}\hat{R}_1e_{+}}
{(r_--r_+-1)(r_--r_+-2)}
-\frac{\left(e_{-}\hat{R}_1e_{+}\hat{R}_1e_{+}+e_{-}\hat{R}_2e_{+}
\right)}{(r_--r_+-2)}
\right.\right.\nonumber\\
&&\hspace{4cm}\left.\left.
+\frac{e_{-}\hat{R}_1e_{+}\hat{R}_1e_{+}\alpha_0}
{(r_--r_+-1)\alpha_s}\right]
\right\}\vec{A}_0
\nonumber\\
\ea

This expression is equivalent to that obtained from eq. (\ref{NNLOtrsolsing}).
Projecting over all the $\pm$ components we can write
\ba
\label{eqlhs}
&&{\vec{q}(N,\alpha_s)}^{++}=\left(\frac{\alpha_s}{\alpha_0}\right)^{r_+}
\left\{
e_+ + (\alpha_s -\alpha_0)e_+\hat{R}_1 e_+
+\alpha_s^2\frac{1}{2}\left[e_+\hat{R}_1 e_+\hat{R}_1 e_+
+e_+\hat{R}_2 e_+ -\frac{e_+\hat{R}_1 e_-\hat{R}_1 e_+}{(r_--r_+-1)}\right]
\right.\nonumber\\
&&\hspace{3.5cm}\left.-\alpha_s \alpha_0\left[
e_+\hat{R}_1e_+\hat{R}_1 e_+ +
\frac{e_+\hat{R}_1e_-\hat{R}_1 e_+}{(r_+-r_--1)(r_--r_+-1)}
\left(\frac{\alpha_s}{\alpha_0}\right)^{r_--r_+}\right]
\right.\nonumber\\
&&\hspace{4cm}\left.+\alpha_0^2 \left[
\frac{1}{2}e_+\hat{R}_1e_+\hat{R}_1 e_+ +
\frac{e_+\hat{R}_1e_-\hat{R}_1 e_+}{(r_+-r_--1)(r_--r_+-1)}
\right.\right.\nonumber\\
&&\hspace{5cm}\left.\left.
-\frac{1}{2}e_+\hat{R}_2 e_+ +\frac{1}{2}\frac{e_+\hat{R}_1e_-\hat{R}_1 e_+}{(r_--r_+-1)}
\right]\right\}\vec{q}(N,\alpha_0)\,,
\nonumber\\\\
&&{\vec{q}(N,\alpha_s)}^{--}=\left(\frac{\alpha_s}{\alpha_0}\right)^{r_-}
\left\{
e_- + (\alpha_s -\alpha_0)e_-\hat{R}_1 e_-
+\alpha_s^2\frac{1}{2}\left[e_-\hat{R}_1e_-\hat{R}_1 e_-
+e_-\hat{R}_2e_- -\frac{e_-\hat{R}_1 e_+\hat{R}_1 e_-}{(r_+-r_--1)}\right]
\right.\nonumber\\
&&\hspace{3.5cm}\left.-\alpha_s \alpha_0\left[
e_-\hat{R}_1e_-\hat{R}_1e_- +
\frac{e_-\hat{R}_1e_+\hat{R}_1 e_-}{(r_--r_+-1)(r_+-r_--1)}
\left(\frac{\alpha_s}{\alpha_0}\right)^{r_+-r_-}\right]
\right.\nonumber\\
&&\hspace{4cm}\left.+\alpha_0^2 \left[
\frac{1}{2}e_-\hat{R}_1e_-\hat{R}_1 e_- +
\frac{e_-\hat{R}_1e_+\hat{R}_1 e_-}{(r_--r_+-1)(r_+-r_--1)}
\right.\right.\nonumber\\
&&\hspace{5cm}\left.\left.
-\frac{1}{2}e_-\hat{R}_2 e_- +\frac{1}{2}\frac{e_-\hat{R}_1e_+\hat{R}_1 e_-}{(r_+-r_--1)}
\right]\right\}\vec{q}(N,\alpha_0)\,,
\nonumber\\\\
&&{\vec{q}(N,\alpha_s)}^{+-}=\left\{
-\alpha_s\frac{e_+\hat{R}_1e_-}{(r_+-r_--1)}\left(\frac{\alpha_s}{\alpha_0}\right)^{r_-}
+\alpha_0\frac{e_+\hat{R}_1e_-}{(r_+-r_--1)}\left(\frac{\alpha_s}{\alpha_0}\right)^{r_+}
\right.\nonumber\\
&&\hspace{2.5cm}\left.+\frac{\alpha_s^2}{(r_+-r_--2)}\left[
-e_+\hat{R}_1e_-\hat{R}_1e_- - e_+\hat{R}_2e_- +\frac{e_+\hat{R}_1e_+\hat{R}_1e_-}
{(r_+-r_--1)}\right]\left(\frac{\alpha_s}{\alpha_0}\right)^{r_-}
\right.\nonumber\\
&&\hspace{2.5cm}\left.
-\alpha_s\alpha_0\left[-\frac{e_+\hat{R}_1e_+\hat{R}_1e_-}{(r_+-r_--1)}
\left(\frac{\alpha_s}{\alpha_0}\right)^{r_+}
-\frac{e_+\hat{R}_1e_-\hat{R}_1e_-}{(r_+-r_--1)}
\left(\frac{\alpha_s}{\alpha_0}\right)^{r_-}\right]
\right.\nonumber\\
&&\hspace{2.5cm}\left.+\alpha_0^2\left(\frac{\alpha_s}{\alpha_0}\right)^{r_+}
\left[\left(-\frac{e_+\hat{R}_1e_+\hat{R}_1e_-}{(r_+-r_--1)}
-\frac{e_+\hat{R}_1e_-\hat{R}_1e_-}{(r_+-r_--1)}\right)
\right.\right.\nonumber\\
&&\hspace{4.8cm}\left.\left.
-\left(-\frac{e_+\hat{R}_1e_-\hat{R}_1e_-}{(r_+-r_--2)}
-\frac{e_+\hat{R}_2e_-}{(r_+-r_--2)}
\right.\right.\right.\nonumber\\
&&\hspace{5cm}\left.\left.\left.
+\frac{e_+\hat{R}_1e_+\hat{R}_1e_-}{(r_+-r_--2)(r_+-r_--1)}\right)
\right]\right\}\vec{q}(N,\alpha_0)\,,
\nonumber\\\\
&&{\vec{q}(N,\alpha_s)}^{-+}=\left\{
-\alpha_s\frac{e_-\hat{R}_1e_+}{(r_--r_+ -1)}\left(\frac{\alpha_s}{\alpha_0}\right)^{r_+}
+\alpha_0\frac{e_-\hat{R}_1e_+}{(r_--r_+ -1)}\left(\frac{\alpha_s}{\alpha_0}\right)^{r_-}
\right.\nonumber\\
&&\hspace{2.5cm}\left.+\frac{\alpha_s^2}{(r_--r_+-2)}\left[
-e_-\hat{R}_1e_+\hat{R}_1e_+ - e_-\hat{R}_2e_+ +\frac{e_-\hat{R}_1e_-\hat{R}_1e_+}
{(r_--r_+-1)}\right]\left(\frac{\alpha_s}{\alpha_0}\right)^{r_+}
\right.\nonumber\\
&&\hspace{2.5cm}\left.
-\alpha_s\alpha_0\left[-\frac{e_-\hat{R}_1e_-\hat{R}_1e_+}{(r_--r_+-1)}
\left(\frac{\alpha_s}{\alpha_0}\right)^{r_-}
-\frac{e_-\hat{R}_1e_+\hat{R}_1e_+}{(r_--r_+-1)}
\left(\frac{\alpha_s}{\alpha_0}\right)^{r_+}\right]
\right.\nonumber\\
&&\hspace{2.5cm}\left.+\alpha_0^2\left(\frac{\alpha_s}{\alpha_0}\right)^{r_-}
\left[\left(-\frac{e_-\hat{R}_1e_-\hat{R}_1e_+}{(r_--r_+-1)}
-\frac{e_-\hat{R}_1e_+\hat{R}_1e_+}{(r_--r_+-1)}\right)
\right.\right.\nonumber\\
&&\hspace{4.8cm}\left.\left.
-\left(-\frac{e_-\hat{R}_1e_+\hat{R}_1e_+}{(r_--r_+-2)}
-\frac{e_-\hat{R}_2e_+}{(r_--r_+ -2)}
\right.\right.\right.\nonumber\\
&&\hspace{5cm}\left.\left.\left.
+\frac{e_-\hat{R}_1e_-\hat{R}_1e_+}{(r_--r_+-2)(r_--r_+-1)}\right)
\right]\right\}\vec{q}(N,\alpha_0)\,.
\nonumber\\
\ea
Using $\vec{A}_0=\vec{f}(N,\alpha_0)=\vec{q}(N,\alpha_0)$
we obtain
\ba
&&{\vec{q}(N,\alpha_s)}^{++}=\sum_{n=0}^{\infty}\frac{L^n}{n!}
\left[\vec{A}_{n}^{++}+\alpha_s
\vec{B}_{n}^{++}+\alpha_s^2\vec{C}_{n}^{++}\right]
\nonumber\\
&&{\vec{q}(N,\alpha_s)}^{--}=\sum_{n=0}^{\infty}\frac{L^n}{n!}
\left[\vec{A}_{n}^{--}+\alpha_s
\vec{B}_{n}^{--}+\alpha_s^2\vec{C}_{n}^{--}\right]
\nonumber\\
&&{\vec{q}(N,\alpha_s)}^{+-}=\sum_{n=0}^{\infty}\frac{L^n}{n!}
\left[\alpha_s
\vec{B}_{n}^{+-}+\alpha_s^2\vec{C}_{n}^{+-}\right]
\nonumber\\
&&{\vec{q}(N,\alpha_s)}^{-+}=\sum_{n=0}^{\infty}\frac{L^n}{n!}
\left[\alpha_s
\vec{B}_{n}^{-+}+\alpha_s^2\vec{C}_{n}^{-+}\right]\,.
\ea
For example we can check the first of the relations above,
gives
\ba
&&\hspace{-1cm}\sum_{n=0}^{\infty}\frac{L^n}{n!}
\left[\vec{A}_{n}^{++}+\alpha_s
\vec{B}_{n}^{++}+\alpha_s^2\vec{C}_{n}^{++}\right]=
\left(\frac{\alpha_s}{\alpha_0}\right)^{r_+}
\left\{e_{+}+(\alpha_s-\alpha_0)e_{+}\hat{R}_1e_{+}
\right\}\vec{f}(N,\alpha_0)
\nonumber\\
&&\hspace{5cm}+\left\{
\left(\frac{\alpha_s}{\alpha_0}\right)^{r_+}\left[
\frac{1}{2}
\left(e_{+}\hat{R}_1e_{+}\hat{R}_1e_{+}(\alpha_0-\alpha_s)^2
-e_{+}\hat{R}_2e_{+}(\alpha_0^2-\alpha_s^2)\right)
\right.\right.\nonumber\\
&&\hspace{7.5cm}\left.\left.
+\alpha_s\alpha_0\frac{e_{+}\hat{R}_1e_{-}\hat{R}_1e_{+}}
{(r_- -r_+)^2 -1}
\left(\frac{\alpha_s}{\alpha_0}\right)^{r_--r_+}
\right.\right.\nonumber\\
&&\hspace{5cm}\left.\left.
+e_{+}\hat{R}_1e_{-}\hat{R}_1e_{+}\frac{\left((r_--r_+-1)\alpha_0^2
+(r_+-r_- -1)\alpha_s^2\right)}{2((r_- -r_+)^2 -1)}
\right]\right\}\vec{f}(N,\alpha_0)\,.
\nonumber\\
\ea
Factorizing $(\alpha_s/\alpha_0)^{r_+}$ and expanding the power of $\alpha_s$
the previous expression becomes
\ba
&&{\vec{f}(N,\alpha_s)}^{++}=\left(\frac{\alpha_s}{\alpha_0}\right)^{r_+}ba
\left\{e_+ + (\alpha_s-\alpha_0)e_+\hat{R}_1 e_+
+\frac{1}{2}\alpha_s^2 \left[e_+\hat{R}_1 e_+\hat{R}_1 e_+
+e_+\hat{R}_2 e_+ -\frac{e_+\hat{R}_1 e_-\hat{R}_1 e_+}{(r_--r_+-1)}\right]
\right.\nonumber\\
&&\hspace{2cm}\left.+\frac{1}{2}\alpha_0^2
\left[-\frac{e_+\hat{R}_1 e_-\hat{R}_1 e_+}
{(r_+-r_--1)}+e_+\hat{R}_1 e_+\hat{R}_1 e_+ -
e_+\hat{R}_2 e_+\right]
\right.\nonumber\\
&&\hspace{2cm}\left.-\alpha_s\alpha_0\left[
e_+\hat{R}_1e_+\hat{R}_1 e_+ +
\frac{e_+\hat{R}_1e_-\hat{R}_1 e_+}{(r_+-r_--1)(r_--r_+-1)}
\left(\frac{\alpha_s}{\alpha_0}\right)^{r_--r_+}\right]
\right\}\vec{f}(N,\alpha_0)\,.
\ea
which agrees with the left hand side of eq.~(\ref{eqlhs}).

\section{Appendix B. Calculation of $\vec{D}_{n}^{++}$}

An explicit calculation of the vector coefficient $\vec{D}_{n}^{++}$ of the
$\kappa=4$ (4th truncated) solution of the NNLO singlet equation has been done
in this section. Since the expressions of the coefficients
$\vec{D}_{n}^{+-}$, $\vec{D}_{n}^{-+}$ and $\vec{D}_{n}^{--}$ have a 
structure similar to $\vec{D}_{n}^{++}$, we omit them and
give only the explicit form of this one 
\begin{small}
\ba
&&\vec{D}_{n}^{++}={\left( -3 + r_+ \right) }^n\,\left( 4\,b_2\,\hat{R}_1^{++}
- 32\,{\pi }^2\,\hat{R}_1^{++\,3} +
16\,b_1\,\pi \,\hat{R}_2^{++} - 5\,b_2\,\hat{R}_1^{++}\,{r_-}^2
+ 40\,{\pi }^2\,\hat{R}_1^{++\,3}\,{r_-}^2
\right.\nonumber\\
&&\left.
-20\,b_1\,\pi \,\hat{R}_2^{++}\,{r_-}^2 + b_2\,\hat{R}_1^{++}\,{r_-}^4 -
8\,{\pi }^2\,\hat{R}_1^{++\,3}\,{r_-}^4 + 4\,b_1\,\pi \,\hat{R}_2^{++}\,{r_-}^4 +
96\,{\pi }^2\,\hat{R}_1^{++\,3}\,{\left( \frac{-2 + r_+}{-3 + r_+} \right) }^n -
\right.\nonumber\\
&&\left.
120\,{\pi }^2\,\hat{R}_1^{++\,3}\,{r_-}^2\,{\left( \frac{-2 + r_+}{-3 + r_+} \right) }^n +
24\,{\pi }^2\,\hat{R}_1^{++\,3}\,{r_-}^4\,{\left( \frac{-2 + r_+}{-3 + r_+} \right) }^n -
96\,{\pi }^2\,\hat{R}_1^{++\,3}\,{\left( \frac{-1 + r_+}{-3 + r_+} \right) }^n +
\right.\nonumber\\
&&\left.
120\,{\pi }^2\,\hat{R}_1^{++\,3}\,{r_-}^2\,{\left( \frac{-1 + r_+}{-3 + r_+} \right) }^n -
24\,{\pi }^2\,\hat{R}_1^{++\,3}\,{r_-}^4\,{\left( \frac{-1 + r_+}{-3 + r_+} \right) }^n +
10\,b_2\,\hat{R}_1^{++}\,r_-\,r_+
- 80\,{\pi }^2\,\hat{R}_1^{++\,3}\,r_-\,r_+ +
\right.\nonumber\\
&&\left.
40\,b_1\,\pi \,\hat{R}_2^{++}\,r_-\,r_+ - 4\,b_2\,\hat{R}_1^{++}\,{r_-}^3\,r_+ +
32\,{\pi }^2\,\hat{R}_1^{++\,3}\,{r_-}^3\,r_+ - 16\,b_1\,\pi \,\hat{R}_2^{++}\,{r_-}^3\,r_+ +
\right.\nonumber\\
&&\left.
240\,{\pi }^2\,\hat{R}_1^{++\,3}\,r_-\,{\left( \frac{-2 + r_+}{-3 + r_+} \right) }^n\,r_+
-96\,{\pi }^2\,\hat{R}_1^{++\,3}\,{r_-}^3\,{\left( \frac{-2 + r_+}{-3 + r_+} \right) }^n\,r_+ -
240\,{\pi }^2\,\hat{R}_1^{++\,3}\,r_-\,{\left( \frac{-1 + r_+}{-3 + r_+} \right) }^n\,r_+
\right.\nonumber\\
&&\left.
+96\,{\pi }^2\,\hat{R}_1^{++\,3}\,{r_-}^3\,{\left( \frac{-1 + r_+}{-3 + r_+} \right) }^n\,r_+
-5\,b_2\,\hat{R}_1^{++}\,{r_+}^2 + 40\,{\pi }^2\,\hat{R}_1^{++\,3}\,{r_+}^2 -
20\,b_1\,\pi \,\hat{R}_2^{++}\,{r_+}^2 + 6\,b_2\,\hat{R}_1^{++}\,{r_-}^2\,{r_+}^2
\right.\nonumber\\
&&\left.
-48\,{\pi }^2\,\hat{R}_1^{++\,3}\,{r_-}^2\,{r_+}^2 +
24\,b_1\,\pi \,\hat{R}_2^{++}\,{r_-}^2\,{r_+}^2 -
120\,{\pi }^2\,\hat{R}_1^{++\,3}\,{\left( \frac{-2 + r_+}{-3 + r_+} \right) }^n\,{r_+}^2
\right.\nonumber\\
&&\left.
+144\,{\pi }^2\,\hat{R}_1^{++\,3}\,{r_-}^2\,{\left( \frac{-2 + r_+}{-3 + r_+} \right) }^n\,{r_+}^2 +
120\,{\pi }^2\,\hat{R}_1^{++\,3}\,{\left( \frac{-1 + r_+}{-3 + r_+} \right) }^n\,{r_+}^2 -
144\,{\pi }^2\,\hat{R}_1^{++\,3}\,{r_-}^2\,{\left( \frac{-1 + r_+}{-3 + r_+} \right) }^n\,{r_+}^2
\right.\nonumber\\
&&\left.
-4\,b_2\,\hat{R}_1^{++}\,r_-\,{r_+}^3 + 32\,{\pi }^2\,\hat{R}_1^{++\,3}\,r_-\,{r_+}^3 -
16\,b_1\,\pi \,\hat{R}_2^{++}\,r_-\,{r_+}^3
-96\,{\pi }^2\,\hat{R}_1^{++\,3}\,r_-\,{\left( \frac{-2 + r_+}{-3 + r_+} \right) }^n\,{r_+}^3 +
\right.\nonumber\\
&&\left.
96\,{\pi }^2\,\hat{R}_1^{++\,3}\,r_-\,{\left( \frac{-1 + r_+}{-3 + r_+} \right) }^n\,{r_+}^3 +
b_2\,\hat{R}_1^{++}\,{r_+}^4 - 8\,{\pi }^2\,\hat{R}_1^{++\,3}\,{r_+}^4 +
4\,b_1\,\pi \,\hat{R}_2^{++}\,{r_+}^4 +
\right.\nonumber\\
&&\left.24\,{\pi }^2\,\hat{R}_1^{++\,3}\,{\left( \frac{-2 + r_+}{-3 + r_+} \right) }^n\,{r_+}^4 -
24\,{\pi }^2\,\hat{R}_1^{++\,3}\,{\left( \frac{-1 + r_+}{-3 + r_+} \right) }^n\,{r_+}^4 -
4\,b_2\,\hat{R}_1^{++}\,{\left( \frac{r_+}{-3 + r_+} \right) }^n
\right.\nonumber\\
&&\left.
+32\,{\pi }^2\,\hat{R}_1^{++\,3}\,{\left( \frac{r_+}{-3 + r_+} \right) }^n -
16\,b_1\,\pi \,\hat{R}_2^{++}\,{\left( \frac{r_+}{-3 + r_+} \right) }^n +
5\,b_2\,\hat{R}_1^{++}\,{r_-}^2\,{\left( \frac{r_+}{-3 + r_+} \right) }^n 
\right.\nonumber\\
&&\left.
-40\,{\pi }^2\,\hat{R}_1^{++\,3}\,{r_-}^2\,{\left( \frac{r_+}{-3 + r_+} \right) }^n +
20\,b_1\,\pi \,\hat{R}_2^{++}\,{r_-}^2\,{\left( \frac{r_+}{-3 + r_+} \right) }^n -
b_2\,\hat{R}_1^{++}\,{r_-}^4\,{\left( \frac{r_+}{-3 + r_+} \right) }^n +
\right.\nonumber\\
&&\left.
8\,{\pi }^2\,\hat{R}_1^{++\,3}\,{r_-}^4\,{\left( \frac{r_+}{-3 + r_+} \right) }^n -
4\,b_1\,\pi \,\hat{R}_2^{++}\,{r_-}^4\,{\left( \frac{r_+}{-3 + r_+} \right) }^n -
10\,b_2\,\hat{R}_1^{++}\,r_-\,r_+\,{\left( \frac{r_+}{-3 + r_+} \right) }^n
\right.\nonumber\\
&&\left.
+80\,{\pi }^2\,\hat{R}_1^{++\,3}\,r_-\,r_+\,{\left( \frac{r_+}{-3 + r_+} \right) }^n -
40\,b_1\,\pi \,\hat{R}_2^{++}\,r_-\,r_+\,{\left( \frac{r_+}{-3 + r_+} \right) }^n +
4\,b_2\,\hat{R}_1^{++}\,{r_-}^3\,r_+\,{\left( \frac{r_+}{-3 + r_+} \right) }^n
\right.\nonumber\\
&&\left.
-32\,{\pi }^2\,\hat{R}_1^{++\,3}\,{r_-}^3\,r_+\,{\left( \frac{r_+}{-3 + r_+} \right) }^n +
16\,b_1\,\pi \,\hat{R}_2^{++}\,{r_-}^3\,r_+\,{\left( \frac{r_+}{-3 + r_+} \right) }^n +
5\,b_2\,\hat{R}_1^{++}\,{r_+}^2\,{\left( \frac{r_+}{-3 + r_+} \right) }^n -
\right.\nonumber\\
&&\left.
40\,{\pi }^2\,\hat{R}_1^{++\,3}\,{r_+}^2\,{\left( \frac{r_+}{-3 + r_+} \right) }^n +
20\,b_1\,\pi \,\hat{R}_2^{++}\,{r_+}^2\,{\left( \frac{r_+}{-3 + r_+} \right) }^n -
6\,b_2\,\hat{R}_1^{++}\,{r_-}^2\,{r_+}^2\,{\left( \frac{r_+}{-3 + r_+} \right) }^n
\right.\nonumber\\
&&\left.
+48\,{\pi }^2\,\hat{R}_1^{++\,3}\,{r_-}^2\,{r_+}^2\,{\left( \frac{r_+}{-3 + r_+} \right) }^n -
24\,b_1\,\pi \,\hat{R}_2^{++}\,{r_-}^2\,{r_+}^2\,{\left( \frac{r_+}{-3 + r_+} \right) }^n +
4\,b_2\,\hat{R}_1^{++}\,r_-\,{r_+}^3\,{\left( \frac{r_+}{-3 + r_+} \right) }^n 
\right.\nonumber\\
&&\left.
-32\,{\pi }^2\,\hat{R}_1^{++\,3}\,r_-\,{r_+}^3\,{\left( \frac{r_+}{-3 + r_+} \right) }^n +
16\,b_1\,\pi \,\hat{R}_2^{++}\,r_-\,{r_+}^3\,{\left( \frac{r_+}{-3 + r_+} \right) }^n -
b_2\,\hat{R}_1^{++}\,{r_+}^4\,{\left( \frac{r_+}{-3 + r_+} \right) }^n
\right.\nonumber\\
&&\left.
+8\,{\pi}^2\,\hat{R}_1^{++\,3}\,{r_+}^4\,{\left( \frac{r_+}{-3 + r_+} \right) }^n -
4\,b_1\,\pi \,\hat{R}_2^{++}\,{r_+}^4\,{\left( \frac{r_+}{-3 + r_+} \right) }^n
\right.\nonumber\\
&&\left.
+16\,{\pi }^2 \left( -2 + r_- + {r_-}^2 - r_+ - 2\,r_-\,r_+ + {r_+}^2 \right)\times
\right.\nonumber\\
&&\left.
\left( -2 + r_- + 3\,{\left(\frac{-2 + r_-}{-3 + r_+} \right)}^n -
{\left( \frac{r_+}{-3 + r_+} \right) }^n - r_-\,{\left( \frac{r_+}{-3 + r_+} \right) }^n
+r_+\,\left( -1 + {\left( \frac{r_+}{-3 + r_+} \right) }^n \right)  \right)
\,\hat{R}_1^{+-}\,\hat{R}_2^{-+}
\right.\nonumber\\
&&\left.
-8\,{\pi }^2\,\left( -2 + 3\,{\left( \frac{-2 + r_+}{-3 + r_+} \right) }^n -
{\left( \frac{r_+}{-3 + r_+} \right) }^n \right) \times\,
\right.\nonumber\\
&&\left.
\left( 4 + {r_-}^4 - 4\,{r_-}^3\,r_+ - 5\,{r_+}^2 + {r_+}^4 +
{r_-}^2\,\left( -5 + 6\,{r_+}^2 \right)  + r_-\,\left( 10\,r_+ - 4\,{r_+}^3 \right)  \right) \,
\hat{R}_1^{++}\, \hat{R}_2^{++}
\right.\nonumber\\
&&\left.
+ 32\,{\pi }^2\,\hat{R}_2^{+-}\, \hat{R}_1^{-+} -
16\,{\pi }^2\,r_-\,\hat{R}_2^{+-}\, \hat{R}_1^{-+}
- 32\,{\pi }^2\,{r_-}^2\,\hat{R}_2^{+-}\,\hat{R}_1^{-+} +
16\,{\pi }^2\,{r_-}^3\,\hat{R}_2^{+-}\, \hat{R}_1^{-+}
\right.\nonumber\\
&&\left.
-96\,{\pi }^2\,{\left( \frac{-1 + r_-}{-3 + r_+} \right) }^n\,\hat{R}_2^{+-}\, \hat{R}_1^{-+} -
48\,{\pi }^2\,r_-\,{\left( \frac{-1 + r_-}{-3 + r_+} \right) }^n\,\hat{R}_2^{+-}\, \hat{R}_1^{-+} +
48\,{\pi }^2\,{r_-}^2\,{\left( \frac{-1 + r_-}{-3 + r_+} \right) }^n\,\hat{R}_2^{+-}\, \hat{R}_1^{-+}
\right.\nonumber\\
&&\left.
+16\,{\pi }^2\,r_+\,\hat{R}_2^{+-}\, \hat{R}_1^{-+} +
64\,{\pi }^2\,r_-\,r_+\,\hat{R}_2^{+-}\, \hat{R}_1^{-+}
-48\,{\pi }^2\,{r_-}^2\,r_+\,\hat{R}_2^{+-}\, \hat{R}_1^{-+}
\right.\nonumber\\
&&\left.
+48\,{\pi }^2\,{\left( \frac{-1 + r_-}{-3 + r_+} \right) }^n\,r_+\,\hat{R}_2^{+-}\, \hat{R}_1^{-+}
-96\,{\pi }^2\,r_-\,{\left( \frac{-1 + r_-}{-3 + r_+} \right) }^n\,r_+\,\hat{R}_2^{+-}\, \hat{R}_1^{-+}
\right.\nonumber\\
&&\left.
-32\,{\pi }^2\,{r_+}^2\,\hat{R}_2^{+-}\, \hat{R}_1^{-+} +
48\,{\pi }^2\,r_-\,{r_+}^2\,\hat{R}_2^{+-}\, \hat{R}_1^{-+}
\right.\nonumber\\
&&\left.
+48\,{\pi }^2\,{\left( \frac{-1 + r_-}{-3 + r_+} \right) }^n\,{r_+}^2\,\hat{R}_2^{+-}\, \hat{R}_1^{-+} -
16\,{\pi }^2\,{r_+}^3\,\hat{R}_2^{+-}\, \hat{R}_1^{-+}
\right.\nonumber\\
&&\left.
+64\,{\pi }^2\,{\left( \frac{r_+}{-3 + r_+} \right) }^n\,\hat{R}_2^{+-}\, \hat{R}_1^{-+} +
64\,{\pi }^2\,r_-\,{\left( \frac{r_+}{-3 + r_+} \right) }^n\,\hat{R}_2^{+-}\, \hat{R}_1^{-+} -
16\,{\pi }^2\,{r_-}^2\,{\left( \frac{r_+}{-3 + r_+} \right) }^n\,\hat{R}_2^{+-}\, \hat{R}_1^{-+}
\right.\nonumber\\
&&\left.
-16\,{\pi }^2\,{r_-}^3\,{\left( \frac{r_+}{-3 + r_+} \right) }^n\,\hat{R}_2^{+-}\, \hat{R}_1^{-+} -
64\,{\pi }^2\,r_+\,{\left( \frac{r_+}{-3 + r_+} \right) }^n\,\hat{R}_2^{+-}\, \hat{R}_1^{-+}
\right.\nonumber\\
&&\left.
+32\,{\pi }^2\,r_-\,r_+\,{\left( \frac{r_+}{-3 + r_+} \right) }^n\,\hat{R}_2^{+-}\, \hat{R}_1^{-+}
+48\,{\pi }^2\,{r_-}^2\,r_+\,{\left( \frac{r_+}{-3 + r_+} \right) }^n\,\hat{R}_2^{+-}\, \hat{R}_1^{-+}
\right.\nonumber\\
&&\left.
-16\,{\pi }^2\,{r_+}^2\,{\left( \frac{r_+}{-3 + r_+} \right) }^n\,\hat{R}_2^{+-}\, \hat{R}_1^{-+} -
48\,{\pi }^2\,r_-\,{r_+}^2\,{\left( \frac{r_+}{-3 + r_+} \right) }^n\,\hat{R}_2^{+-}\, \hat{R}_1^{-+}
\right.\nonumber\\
&&\left.
+16\,{\pi }^2\,{r_+}^3\,{\left( \frac{r_+}{-3 + r_+} \right) }^n\,\hat{R}_2^{+-}\, \hat{R}_1^{-+} +
32\,{\pi }^2\,\hat{R}_2^{++}\, \hat{R}_1^{++} - 40\,{\pi }^2\,{r_-}^2\,\hat{R}_2^{++}\, \hat{R}_1^{++} +
8\,{\pi }^2\,{r_-}^4\,\hat{R}_2^{++}\, \hat{R}_1^{++}
\right.\nonumber\\
&&\left.
-96\,{\pi }^2\,{\left( \frac{-1 + r_+}{-3 + r_+} \right) }^n\,\hat{R}_2^{++}\, \hat{R}_1^{++} +
120\,{\pi }^2\,{r_-}^2\,{\left( \frac{-1 + r_+}{-3 + r_+} \right) }^n\,\hat{R}_2^{++}\, \hat{R}_1^{++} -
24\,{\pi }^2\,{r_-}^4\,{\left( \frac{-1 + r_+}{-3 + r_+} \right) }^n\,\hat{R}_2^{++}\, \hat{R}_1^{++}
\right.\nonumber\\
&&\left.
+80\,{\pi }^2\,r_-\,r_+\,\hat{R}_2^{++}\, \hat{R}_1^{++} -
32\,{\pi }^2\,{r_-}^3\,r_+\,\hat{R}_2^{++}\, \hat{R}_1^{++} -
240\,{\pi }^2\,r_-\,{\left( \frac{-1 + r_+}{-3 + r_+} \right) }^n\,r_+\,\hat{R}_2^{++}\, \hat{R}_1^{++}
\right.\nonumber\\
&&\left.
+96\,{\pi }^2\,{r_-}^3\,{\left( \frac{-1 + r_+}{-3 + r_+} \right) }^n\,r_+\,
 \hat{R}_2^{++}\, \hat{R}_1^{++} - 40\,{\pi }^2\,{r_+}^2\,\hat{R}_2^{++}\, \hat{R}_1^{++} +
48\,{\pi }^2\,{r_-}^2\,{r_+}^2\,\hat{R}_2^{++}\, \hat{R}_1^{++}
\right.\nonumber\\
&&\left.
+120\,{\pi }^2\,{\left( \frac{-1 + r_+}{-3 + r_+} \right) }^n\,{r_+}^2\,\hat{R}_2^{++}\, \hat{R}_1^{++} -
144\,{\pi }^2\,{r_-}^2\,{\left( \frac{-1 + r_+}{-3 + r_+} \right) }^n\,{r_+}^2\,
\hat{R}_2^{++}\, \hat{R}_1^{++} - 32\,{\pi }^2\,r_-\,{r_+}^3\,\hat{R}_2^{++}\, \hat{R}_1^{++}
\right.\nonumber\\
&&\left.
+96\,{\pi }^2\,r_-\,{\left( \frac{-1 + r_+}{-3 + r_+} \right) }^n\,{r_+}^3\,
 \hat{R}_2^{++}\, \hat{R}_1^{++} + 8\,{\pi }^2\,{r_+}^4\,\hat{R}_2^{++}\, \hat{R}_1^{++}
\right.\nonumber\\
&&\left.
-24\,{\pi }^2\,{\left( \frac{-1 + r_+}{-3 + r_+} \right) }^n\,{r_+}^4\,\hat{R}_2^{++}\, \hat{R}_1^{++} +
64\,{\pi }^2\,{\left( \frac{r_+}{-3 + r_+} \right) }^n\,\hat{R}_2^{++}\, \hat{R}_1^{++}
\right.\nonumber\\
&&\left.
-80\,{\pi }^2\,{r_-}^2\,{\left( \frac{r_+}{-3 + r_+} \right) }^n\,\hat{R}_2^{++}\, \hat{R}_1^{++} +
16\,{\pi }^2\,{r_-}^4\,{\left( \frac{r_+}{-3 + r_+} \right) }^n\,\hat{R}_2^{++}\, \hat{R}_1^{++}
\right.\nonumber\\
&&\left.
+160\,{\pi }^2\,r_-\,r_+\,{\left( \frac{r_+}{-3 + r_+} \right) }^n\,\hat{R}_2^{++}\, \hat{R}_1^{++}
-64\,{\pi }^2\,{r_-}^3\,r_+\,{\left( \frac{r_+}{-3 + r_+} \right) }^n\,\hat{R}_2^{++}\, \hat{R}_1^{++}
\right.\nonumber\\
&&\left.
-80\,{\pi }^2\,{r_+}^2\,{\left( \frac{r_+}{-3 + r_+} \right) }^n\,\hat{R}_2^{++}\, \hat{R}_1^{++}
+96\,{\pi }^2\,{r_-}^2\,{r_+}^2\,{\left( \frac{r_+}{-3 + r_+} \right) }^n\,
\hat{R}_2^{++}\, \hat{R}_1^{++}
\right.\nonumber\\
&&\left.
- 64\,{\pi }^2\,r_-\,{r_+}^3\,{\left( \frac{r_+}{-3 + r_+} \right) }^n\,
\hat{R}_2^{++}\, \hat{R}_1^{++}
+ 16\,{\pi }^2\,{r_+}^4\,{\left( \frac{r_+}{-3 + r_+} \right) }^n\,
\hat{R}_2^{++}\, \hat{R}_1^{++} - 32\,{\pi }^2\,\hat{R}_1^{+-}\, \hat{R}_1^{--}\, \hat{R}_1^{-+}
\right.\nonumber\\
&&\left.
+48\,{\pi }^2\,r_-\,\hat{R}_1^{+-}\, \hat{R}_1^{--}\, \hat{R}_1^{-+}
-16\,{\pi }^2\,{r_-}^2\,\hat{R}_1^{+-}\, \hat{R}_1^{--}\, \hat{R}_1^{-+} +
96\,{\pi }^2\,{\left( \frac{-2 + r_-}{-3 + r_+} \right) }^n\,\hat{R}_1^{+-}\, \hat{R}_1^{--}\, \hat{R}_1^{-+}
\right.\nonumber\\
&&\left.
-48\,{\pi }^2\,r_-\,{\left( \frac{-2 + r_-}{-3 + r_+} \right) }^n\,
 \hat{R}_1^{+-}\, \hat{R}_1^{--}\, \hat{R}_1^{-+}
-48\,{\pi }^2\,{r_-}^2\,{\left( \frac{-2 + r_-}{-3 + r_+} \right) }^n\,
 \hat{R}_1^{+-}\, \hat{R}_1^{--}\, \hat{R}_1^{-+}
\right.\nonumber\\
&&\left.
-96\,{\pi }^2\,{\left( \frac{-1 + r_-}{-3 + r_+} \right) }^n\,\hat{R}_1^{+-}\, \hat{R}_1^{--}\, \hat{R}_1^{-+}
-48\,{\pi }^2\,r_-\,{\left( \frac{-1 + r_-}{-3 + r_+} \right) }^n\,
 \hat{R}_1^{+-}\, \hat{R}_1^{--}\, \hat{R}_1^{-+}
\right.\nonumber\\
&&\left.
+48\,{\pi }^2\,{r_-}^2\,{\left( \frac{-1 + r_-}{-3 + r_+} \right) }^n\,
 \hat{R}_1^{+-}\, \hat{R}_1^{--}\, \hat{R}_1^{-+}
-48\,{\pi }^2\,r_+\,\hat{R}_1^{+-}\, \hat{R}_1^{--}\, \hat{R}_1^{-+} +
\right.\nonumber\\
&&\left.
32\,{\pi }^2\,r_-\,r_+\,\hat{R}_1^{+-}\, \hat{R}_1^{--}\, \hat{R}_1^{-+} +
48\,{\pi }^2\,{\left( \frac{-2 + r_-}{-3 + r_+} \right) }^n\,r_+\,
 \hat{R}_1^{+-}\, \hat{R}_1^{--}\, \hat{R}_1^{-+} 
\right.\nonumber\\
&&\left.
+96\,{\pi }^2\,r_-\,{\left( \frac{-2 + r_-}{-3 + r_+} \right) }^n\,r_+\,
 \hat{R}_1^{+-}\, \hat{R}_1^{--}\, \hat{R}_1^{-+} +
48\,{\pi }^2\,{\left( \frac{-1 + r_-}{-3 + r_+} \right) }^n\,r_+\,
 \hat{R}_1^{+-}\, \hat{R}_1^{--}\, \hat{R}_1^{-+} 
\right.\nonumber\\
&&\left.
-96\,{\pi }^2\,r_-\,{\left( \frac{-1 + r_-}{-3 + r_+} \right) }^n\,r_+\,
 \hat{R}_1^{+-}\, \hat{R}_1^{--}\, \hat{R}_1^{-+} -
16\,{\pi }^2\,{r_+}^2\,\hat{R}_1^{+-}\, \hat{R}_1^{--}\, \hat{R}_1^{-+}
\right.\nonumber\\
&&\left.
-48\,{\pi }^2\,{\left( \frac{-2 + r_-}{-3 + r_+} \right) }^n\,{r_+}^2\,
 \hat{R}_1^{+-}\, \hat{R}_1^{--}\, \hat{R}_1^{-+} +
48\,{\pi }^2\,{\left( \frac{-1 + r_-}{-3 + r_+} \right) }^n\,{r_+}^2\,
 \hat{R}_1^{+-}\, \hat{R}_1^{--}\, \hat{R}_1^{-+}
\right.\nonumber\\
&&\left.
+32\,{\pi }^2\,{\left( \frac{r_+}{-3 + r_+} \right) }^n\,\hat{R}_1^{+-}\, \hat{R}_1^{--}\,
\hat{R}_1^{-+} +
48\,{\pi }^2\,r_-\,{\left( \frac{r_+}{-3 + r_+} \right) }^n\,
 \hat{R}_1^{+-}\, \hat{R}_1^{--}\, \hat{R}_1^{-+} 
\right.\nonumber\\
&&\left.
+16\,{\pi }^2\,{r_-}^2\,{\left( \frac{r_+}{-3 + r_+} \right) }^n\,
 \hat{R}_1^{+-}\, \hat{R}_1^{--}\, \hat{R}_1^{-+} -
48\,{\pi }^2\,r_+\,{\left( \frac{r_+}{-3 + r_+} \right) }^n\,
\hat{R}_1^{+-}\, \hat{R}_1^{--}\, \hat{R}_1^{-+}
\right.\nonumber\\
&&\left.
-32\,{\pi }^2\,r_-\,r_+\,{\left( \frac{r_+}{-3 + r_+} \right) }^n\,
 \hat{R}_1^{+-}\, \hat{R}_1^{--}\, \hat{R}_1^{-+} +
16\,{\pi }^2\,{r_+}^2\,{\left( \frac{r_+}{-3 + r_+} \right) }^n\,
 \hat{R}_1^{+-}\, \hat{R}_1^{--}\, \hat{R}_1^{-+} 
\right.\nonumber\\
&&\left.
- 32\,{\pi }^2\,\hat{R}_1^{+-}\,
 \hat{R}_1^{-+}\, \hat{R}_1^{++}
+32\,{\pi }^2\,r_-\,\hat{R}_1^{+-}\, \hat{R}_1^{-+}\, \hat{R}_1^{++} +
8\,{\pi }^2\,{r_-}^2\,\hat{R}_1^{+-}\, \hat{R}_1^{-+}\, \hat{R}_1^{++}
\right.\nonumber\\
&&\left.
-8\,{\pi }^2\,{r_-}^3\,\hat{R}_1^{+-}\, \hat{R}_1^{-+}\, \hat{R}_1^{++} +
96\,{\pi }^2\,{\left( \frac{-2 + r_-}{-3 + r_+} \right) }^n\,\hat{R}_1^{+-}\,
\hat{R}_1^{-+}\, \hat{R}_1^{++} + 48\,{\pi }^2\,r_-\,{\left( \frac{-2 + r_-}{-3 + r_+} \right) }^n\,
 \hat{R}_1^{+-}\, \hat{R}_1^{-+}\, \hat{R}_1^{++}
\right.\nonumber\\
&&\left.
-96\,{\pi }^2\,{\left( \frac{-1 + r_+}{-3 + r_+} \right) }^n\,\hat{R}_1^{+-}\,
\hat{R}_1^{-+}\, \hat{R}_1^{++} -
96\,{\pi }^2\,r_-\,{\left( \frac{-1 + r_+}{-3 + r_+} \right) }^n\,
 \hat{R}_1^{+-}\, \hat{R}_1^{-+}\, \hat{R}_1^{++}
\right.\nonumber\\
&&\left.
+24\,{\pi }^2\,{r_-}^2\,{\left( \frac{-1 + r_+}{-3 + r_+} \right) }^n\,
 \hat{R}_1^{+-}\, \hat{R}_1^{-+}\, \hat{R}_1^{++}
+24\,{\pi }^2\,{r_-}^3\,{\left( \frac{-1 + r_+}{-3 + r_+} \right) }^n\,
 \hat{R}_1^{+-}\, \hat{R}_1^{-+}\, \hat{R}_1^{++}
\right.\nonumber\\
&&\left.
-32\,{\pi }^2\,r_+\,\hat{R}_1^{+-}\, \hat{R}_1^{-+}\, \hat{R}_1^{++}
-16\,{\pi }^2\,r_-\,r_+\,\hat{R}_1^{+-}\, \hat{R}_1^{-+}\, \hat{R}_1^{++} +
24\,{\pi }^2\,{r_-}^2\,r_+\,\hat{R}_1^{+-}\, \hat{R}_1^{-+}\, \hat{R}_1^{++}
\right.\nonumber\\
&&\left.
-48\,{\pi }^2\,{\left( \frac{-2 + r_-}{-3 + r_+} \right) }^n\,r_+\,
 \hat{R}_1^{+-}\, \hat{R}_1^{-+}\, \hat{R}_1^{++}
+96\,{\pi }^2\,{\left( \frac{-1 + r_+}{-3 + r_+} \right) }^n\,r_+\,
 \hat{R}_1^{+-}\, \hat{R}_1^{-+}\, \hat{R}_1^{++}
\right.\nonumber\\
&&\left.
-48\,{\pi }^2\,r_-\,{\left( \frac{-1 + r_+}{-3 + r_+} \right) }^n\,r_+\,
 \hat{R}_1^{+-}\, \hat{R}_1^{-+}\, \hat{R}_1^{++}
-72\,{\pi }^2\,{r_-}^2\,{\left( \frac{-1 + r_+}{-3 + r_+} \right) }^n\,r_+\,
 \hat{R}_1^{+-}\, \hat{R}_1^{-+}\, \hat{R}_1^{++}
\right.\nonumber\\
&&\left.
+8\,{\pi }^2\,{r_+}^2\,\hat{R}_1^{+-}\, \hat{R}_1^{-+}\, \hat{R}_1^{++}
-24\,{\pi }^2\,r_-\,{r_+}^2\,\hat{R}_1^{+-}\, \hat{R}_1^{-+}\, \hat{R}_1^{++} 
\right.\nonumber\\
&&\left.
+24\,{\pi }^2\,{\left( \frac{-1 + r_+}{-3 + r_+} \right) }^n\,{r_+}^2\,
\hat{R}_1^{+-}\, \hat{R}_1^{-+}\, \hat{R}_1^{++} +
72\,{\pi }^2\,r_-\,{\left( \frac{-1 + r_+}{-3 + r_+} \right) }^n\,{r_+}^2\,
 \hat{R}_1^{+-}\, \hat{R}_1^{-+}\, \hat{R}_1^{++} 
\right.\nonumber\\
&&\left.
+8\,{\pi }^2\,{r_+}^3\,\hat{R}_1^{+-}\, \hat{R}_1^{-+}\, \hat{R}_1^{++} -
24\,{\pi }^2\,{\left( \frac{-1 + r_+}{-3 + r_+} \right) }^n\,{r_+}^3\,
 \hat{R}_1^{+-}\, \hat{R}_1^{-+}\, \hat{R}_1^{++} +
32\,{\pi }^2\,{\left( \frac{r_+}{-3 + r_+} \right) }^n\,\hat{R}_1^{+-}\,
\hat{R}_1^{-+}\, \hat{R}_1^{++}
\right.\nonumber\\
&&\left.
+16\,{\pi }^2\,r_-\,{\left( \frac{r_+}{-3 + r_+} \right) }^n\,
 \hat{R}_1^{+-}\, \hat{R}_1^{-+}\, \hat{R}_1^{++}
-32\,{\pi }^2\,{r_-}^2\,{\left( \frac{r_+}{-3 + r_+} \right) }^n\,
 \hat{R}_1^{+-}\, \hat{R}_1^{-+}\, \hat{R}_1^{++}
\right.\nonumber\\
&&\left.
 -16\,{\pi }^2\,{r_-}^3\,{\left( \frac{r_+}{-3 + r_+} \right) }^n\,
 \hat{R}_1^{+-}\, \hat{R}_1^{-+}\, \hat{R}_1^{++}
-16\,{\pi }^2\,r_+\,{\left( \frac{r_+}{-3 + r_+} \right) }^n\,
 \hat{R}_1^{+-}\, \hat{R}_1^{-+}\, \hat{R}_1^{++}
\right.\nonumber\\
&&\left.
+64\,{\pi }^2\,r_-\,r_+\,{\left( \frac{r_+}{-3 + r_+} \right) }^n\,
 \hat{R}_1^{+-}\, \hat{R}_1^{-+}\, \hat{R}_1^{++} +
48\,{\pi }^2\,{r_-}^2\,r_+\,{\left( \frac{r_+}{-3 + r_+} \right) }^n\,
 \hat{R}_1^{+-}\, \hat{R}_1^{-+}\, \hat{R}_1^{++}
\right.\nonumber\\
&&\left.
 -32\,{\pi }^2\,{r_+}^2\,{\left( \frac{r_+}{-3 + r_+} \right) }^n\,
 \hat{R}_1^{+-}\, \hat{R}_1^{-+}\, \hat{R}_1^{++} -
48\,{\pi }^2\,r_-\,{r_+}^2\,{\left( \frac{r_+}{-3 + r_+} \right) }^n\,
 \hat{R}_1^{+-}\, \hat{R}_1^{-+}\, \hat{R}_1^{++}
 \right.\nonumber\\
&&\left.
+16\,{\pi }^2\,{r_+}^3\,{\left( \frac{r_+}{-3 + r_+} \right) }^n\,
 \hat{R}_1^{+-}\, \hat{R}_1^{-+}\, \hat{R}_1^{++}
 - 32\,{\pi }^2\,\hat{R}_1^{++}\, \hat{R}_1^{+-}\,
 \hat{R}_1^{-+} +16\,{\pi }^2\,r_-\,\hat{R}_1^{++}\, \hat{R}_1^{+-}\, \hat{R}_1^{-+}
 \right.\nonumber\\
&&\left.
 +32\,{\pi }^2\,{r_-}^2\,\hat{R}_1^{++}\, \hat{R}_1^{+-}\, \hat{R}_1^{-+} -
16\,{\pi }^2\,{r_-}^3\,\hat{R}_1^{++}\, \hat{R}_1^{+-}\, \hat{R}_1^{-+} -
96\,{\pi }^2\,{\left( \frac{-1 + r_-}{-3 + r_+} \right) }^n\,\hat{R}_1^{++}\,
\hat{R}_1^{+-}\, \hat{R}_1^{-+}
 \right.\nonumber\\
&&\left.
+48\,{\pi }^2\,r_-\,{\left( \frac{-1 + r_-}{-3 + r_+} \right) }^n\,
 \hat{R}_1^{++}\, \hat{R}_1^{+-}\, \hat{R}_1^{-+} +
96\,{\pi }^2\,{\left( \frac{-2 + r_+}{-3 + r_+} \right) }^n\,\hat{R}_1^{++}\,
\hat{R}_1^{+-}\, \hat{R}_1^{-+} -
\right.\nonumber\\
&&\left.
96\,{\pi }^2\,r_-\,{\left( \frac{-2 + r_+}{-3 + r_+} \right) }^n\,
 \hat{R}_1^{++}\, \hat{R}_1^{+-}\, \hat{R}_1^{-+} -
24\,{\pi }^2\,{r_-}^2\,{\left( \frac{-2 + r_+}{-3 + r_+} \right) }^n\,
 \hat{R}_1^{++}\, \hat{R}_1^{+-}\, \hat{R}_1^{-+}
\right.\nonumber\\
&&\left.
 +24\,{\pi }^2\,{r_-}^3\,{\left( \frac{-2 + r_+}{-3 + r_+} \right) }^n\,
 \hat{R}_1^{++}\, \hat{R}_1^{+-}\, \hat{R}_1^{-+} -
16\,{\pi }^2\,r_+\,\hat{R}_1^{++}\, \hat{R}_1^{+-}\, \hat{R}_1^{-+} 
-64\,{\pi }^2\,r_-\,r_+\,\hat{R}_1^{++}\, \hat{R}_1^{+-}\, \hat{R}_1^{-+}
\right.\nonumber\\
&&\left.
+48\,{\pi }^2\,{r_-}^2\,r_+\,\hat{R}_1^{++}\, \hat{R}_1^{+-}\, \hat{R}_1^{-+} -
48\,{\pi }^2\,{\left( \frac{-1 + r_-}{-3 + r_+} \right) }^n\,r_+\,
 \hat{R}_1^{++}\, \hat{R}_1^{+-}\, \hat{R}_1^{-+} +
 \right.\nonumber\\
&&\left.
96\,{\pi }^2\,{\left( \frac{-2 + r_+}{-3 + r_+} \right) }^n\,r_+\,
 \hat{R}_1^{++}\, \hat{R}_1^{+-}\, \hat{R}_1^{-+} +
48\,{\pi }^2\,r_-\,{\left( \frac{-2 + r_+}{-3 + r_+} \right) }^n\,r_+\,
 \hat{R}_1^{++}\, \hat{R}_1^{+-}\, \hat{R}_1^{-+} 
\right.\nonumber\\
&&\left.
-72\,{\pi }^2\,{r_-}^2\,{\left( \frac{-2 + r_+}{-3 + r_+} \right) }^n\,r_+\,
 \hat{R}_1^{++}\, \hat{R}_1^{+-}\, \hat{R}_1^{-+} +
32\,{\pi }^2\,{r_+}^2\,\hat{R}_1^{++}\, \hat{R}_1^{+-}\, \hat{R}_1^{-+}
\right.\nonumber\\
&&\left.
-48\,{\pi }^2\,r_-\,{r_+}^2\,\hat{R}_1^{++}\, \hat{R}_1^{+-}\, \hat{R}_1^{-+} -
24\,{\pi }^2\,{\left( \frac{-2 + r_+}{-3 + r_+} \right) }^n\,{r_+}^2\,
 \hat{R}_1^{++}\, \hat{R}_1^{+-}\, \hat{R}_1^{-+} 
\right.\nonumber\\
&&\left.
+72\,{\pi }^2\,r_-\,{\left( \frac{-2 + r_+}{-3 + r_+} \right) }^n\,{r_+}^2\,
 \hat{R}_1^{++}\, \hat{R}_1^{+-}\, \hat{R}_1^{-+} +
16\,{\pi }^2\,{r_+}^3\,\hat{R}_1^{++}\, \hat{R}_1^{+-}\, \hat{R}_1^{-+}
 \right.\nonumber\\
&&\left.
-24\,{\pi }^2\,{\left( \frac{-2 + r_+}{-3 + r_+} \right) }^n\,{r_+}^3\,
 \hat{R}_1^{++}\, \hat{R}_1^{+-}\, \hat{R}_1^{-+} +
32\,{\pi }^2\,{\left( \frac{r_+}{-3 + r_+} \right) }^n\,\hat{R}_1^{++}\,
\hat{R}_1^{+-}\, \hat{R}_1^{-+}
 \right.\nonumber\\
&&\left.
+32\,{\pi }^2\,r_-\,{\left( \frac{r_+}{-3 + r_+} \right) }^n\,
 \hat{R}_1^{++}\, \hat{R}_1^{+-}\, \hat{R}_1^{-+} -
8\,{\pi }^2\,{r_-}^2\,{\left( \frac{r_+}{-3 + r_+} \right) }^n\,
 \hat{R}_1^{++}\, \hat{R}_1^{+-}\, \hat{R}_1^{-+}
 \right.\nonumber\\
&&\left.
 -8\,{\pi }^2\,{r_-}^3\,{\left( \frac{r_+}{-3 + r_+} \right) }^n\,
 \hat{R}_1^{++}\, \hat{R}_1^{+-}\, \hat{R}_1^{-+} -
32\,{\pi }^2\,r_+\,{\left( \frac{r_+}{-3 + r_+} \right) }^n\,
 \hat{R}_1^{++}\, \hat{R}_1^{+-}\, \hat{R}_1^{-+}
 \right.\nonumber\\
&&\left.
+16\,{\pi }^2\,r_-\,r_+\,{\left( \frac{r_+}{-3 + r_+} \right) }^n\,
 \hat{R}_1^{++}\, \hat{R}_1^{+-}\, \hat{R}_1^{-+}
 24\,{\pi }^2\,{r_-}^2\,r_+\,{\left( \frac{r_+}{-3 + r_+} \right) }^n\,
 \hat{R}_1^{++}\, \hat{R}_1^{+-}\, \hat{R}_1^{-+}
 \right.\nonumber\\
&&\left.
-8\,{\pi }^2\,{r_+}^2\,{\left( \frac{r_+}{-3 + r_+} \right) }^n\,
 \hat{R}_1^{++}\, \hat{R}_1^{+-}\, \hat{R}_1^{-+} -
24\,{\pi }^2\,r_-\,{r_+}^2\,{\left( \frac{r_+}{-3 + r_+} \right) }^n\,
 \hat{R}_1^{++}\, \hat{R}_1^{+-}\, \hat{R}_1^{-+}
 \right.\nonumber\\
&&\left.
 +8\,{\pi }^2\,{r_+}^3\,{\left( \frac{r_+}{-3 + r_+} \right) }^n\,
\hat{R}_1^{++}\, \hat{R}_1^{+-}\, \hat{R}_1^{-+} \right)\vec{A}_0
\nonumber\\
&&\left[48\,{\pi }^2\,\left( 4 + {r_-}^4 - 4\,{r_-}^3\,r_+ - 5\,{r_+}^2 + {r_+}^4 +
{r_-}^2\,\left( -5 + 6\,{r_+}^2 \right)  + r_-\,\left( 10\,r_+ - 4\,{r_+}^3 
\right)\right) \right]^{-1}\,.
\nonumber\\
\ea
\end{small}

\end{document}